\documentclass[12pt,a4paper]{article}

\usepackage{colortbl}
\usepackage{ascmac}
\usepackage{bm}
\usepackage[pdftex]{graphicx}

\usepackage{amsmath}
\usepackage{amsfonts}
\usepackage{amssymb}

\newcommand{\CC}{\mathbb{C}}
\newcommand{\ZZ}{\mathbb{Z}}
\newcommand{\RR}{\mathbb{R}}

\newcommand{\ol}{\overline}
\newcommand{\wt}{\widetilde}

\newcommand{\wh}{\widehat}

\newcommand{\sX}{\mathsf{X}}
\newcommand{\sY}{\mathsf{Y}}
\newcommand{\sZ}{\mathsf{Z}}
\newcommand{\sI}{\mathsf{I}}

\usepackage{ascmac}
\usepackage{fancybox}

\newcommand{\nn}{\nonumber}

\def\Sym{\mathop{\rm Sym}\nolimits}
\def\tr{\mathop{\rm tr}\nolimits}
\def\Pexp{\mathop{\rm Pexp}\nolimits}

\makeatletter

\@addtoreset{equation}{section}
\makeatother

\begin{document}
\begin{titlepage}
\title{
\vspace{-1.5cm}
\begin{flushright}
{\normalsize TIT/HEP-671\\ April 2019}
\end{flushright}
\vspace{3.5cm}
\LARGE{Finite $N$ Corrections to the Superconformal Index of S-fold Theories}}
\author{\large{
Reona {\scshape Arai\footnote{E-mail: r.arai@th.phys.titech.ac.jp}}\ 
and 
Yosuke {\scshape Imamura\footnote{E-mail: imamura@phys.titech.ac.jp}}
}\\
\\
{\itshape Department of Physics, Tokyo Institute of Technology}, \\ {\itshape Tokyo 152-8551, Japan}}

\maketitle
\thispagestyle{empty}

\begin{abstract}
We study the superconformal index of S-fold theories
by using AdS/CFT correspondence.
It has been known that the index in the large $N$ limit 
is reproduced as the contribution of bulk Kaluza-Klein
modes.
For finite $N$ D3-branes
wrapped around the non-trivial cycle
in $\bm{S}^5/\ZZ_k$, which corresponds
to Pfaffian-like operators,
give the corrections of order $q^N$ to
the index.
We calculate the finite $N$ corrections
by analyzing the fluctuations of
wrapped D3-branes.
Comparisons to known results
show that our formula correctly reproduces
the corrections up to errors of order $q^{2N}$.
\end{abstract}
\end{titlepage}
\section{Introduction}

The superconformal index \cite{Kinney:2005ej} and its various limits \cite{Gadde:2011uv}
are useful observables of quantum field theories to study
non-perturbative properties like dualities among them.
Recent progress in supersymmetric gauge theories
owes a lot to the development of methods of calculating the indices.
If a Lagrangian description is known, we can in principle calculate the index
by using the localization method.
However, it has been known that there are many types of theories
that do not have Lagrangian descriptions.
In such a case we need to rely on some duality
which connects the target theory to another
calculable system.

In this paper we investigate such a class of theories,
S-fold theories \cite{Garcia-Etxebarria:2015wns}.
Each of them is defined as the theory on the worldvolume of D3-branes
in an S-fold background of type IIB string theory.
An S-fold theory is specified by three numbers:
the order of the S-fold group $k$,
the number of D3-branes $N$,
and $p=0,1$ associated with the three-form
discrete torsion.
We denote the theory by $S(k,N,p)$.
For the consistency with $SL(2,\ZZ)$ symmetry
$k$ takes only values $1,2,3,4,6$.
If $k\geq3$, they are strongly-coupled and
no Lagrangian description have been known.
An S-fold with $k=2$ is the orientifold with
a fixed three-plane, and the S-fold theory
is the ${\cal N}=4$ SYM with the gauge group $G=SO(2N)$
for $p=0$ or $G=SO(2N+1)$ for $p=1$.
$k=1$ gives the background without any S-folding.

In the large $N$ limit, we can analyze the S-fold theories by
using AdS/CFT correspondence \cite{Maldacena:1997re}.
The gravity dual is type IIB string theory
in the $AdS_5\times\bm{S}^5/\ZZ_k$ background.
The superconformal index for $N=\infty$ was calculated
as the index of the Kaluza-Klein excitation of the
massless fields in the type IIB supergravity
in \cite{Imamura:2016abe}.

It is known that for finite $N$ the Kaluza-Klein analysis fails to give
correct index.
We give the index as a series expansion with respect to $q$.
Roughly speaking, $q^n$ terms correspond to
operators with dimension $n$.
(See (\ref{defsc}) for the precise definition.)
In the case of $U(N)$ SYM, which can be regarded as the
S-fold theory $S(1,N,0)$, we have
corrections starting from $q^N$.
On the gravity side this is interpreted as the
effect of giant gravitons \cite{McGreevy:2000cw,Grisaru:2000zn,Hashimoto:2000zp}.
For angular momenta of order $N$
the Kaluza-Klein analysis becomes invalid,
and we should treat the excitations as giant gravitons.
For S-fold theories with $k\geq2$, because the
flux in the covering space is $kN$, the
correction by giant gravitons starts from $q^{kN}$.

In addition,
because the internal space
$\bm{S}^5/\ZZ_k$ has a non-trivial third homology
\begin{align}
H_3(\bm{S}^5/\ZZ_k)=\ZZ_k,
\end{align}
we also have corrections due to D3-branes wrapped
around topologically non-trivial cycles.
In the case of an S-fold theory with $p=1$,
which is associated with background with a non-trivial discrete torsion,
the wrapping of a D3-brane is not allowed because the interaction between
the non-trivial torsion flux prevents the existence of a consistent gauge bundle on
the worldvolume \cite{Witten:1998xy,Aharony:2016kai}.
In such a case the finite $N$ correction starts around $q^{kN}$,
and there is no correction starting from $q^N$.
By this reason we focus mainly on theories with $p=0$.

The theory $S(k,N,0)$ has a discrete $\ZZ_k$ symmetry.
A D3-brane with the wrapping number $m$ corresponds to
an operator carrying $\ZZ_k$ charge $m$ \cite{Aharony:2016kai}.
Such operators are
called Pfaffian-like operators because
typical examples of such operators are
the Pfaffian of scalar fields in $SO(2N)$ SYM \cite{Witten:1998xy}.
A Pfaffian-like operator
with charge $m\in\ZZ_k$ has the dimension $\gtrsim|m|N$
where $|m|$ for $m\in\ZZ_k$ is defined by
\begin{align}
|m|=\min |k\ZZ +m|.
\end{align}
They give finite $N$ corrections of order $q^{|m|N}$ to the index \cite{Imamura:2016abe}.

The corrections are summarized as follows.
We show only the $q$-dependence and
all coefficients are omitted.
\begin{align}
\mbox{Kaluza-Klein modes}\quad & 1+\cdots + q^N+\cdots +q^{2N}+\cdots +q^{kN}+\cdots,\nonumber\\
\mbox{Giant gravitons}\quad & \phantom{1+\cdots + q^N+\cdots +q^{2N}+\cdots +}q^{kN}+\cdots,\nonumber\\
\mbox{Wrapped D3 with $|m|=1$}\quad&\phantom{1+\cdots + }q^N+\cdots +q^{2N}+\cdots +q^{kN}+\cdots,\nonumber\\
\mbox{Wrapped D3 with $|m|=2$}\quad&\phantom{1+\cdots +q^N+\cdots +}q^{2N}+\cdots +q^{kN}+\cdots,\nonumber\\
\vdots
\end{align}
where we included $1$, the contribution of the ground state,
in the first line.
In this paper we are interested in the leading corrections starting from $q^N$.
We calculate the contribution of wrapped D3-branes with $|m|=1$ to the index.
It is expected that this calculation,
(combined with the Kaluza-Klein analysis) gives the index
up to errors starting from $q^{2N}$.
For theories with (manifest or hidden) the ${\cal N}=4$ supersymmetry
we confirm that this is actually the case by comparing
to the results of numerical calculation by localization.

We calculate the contribution of wrapped D3-branes by quantizing the fluctuations
of D3-branes around ground state configurations.
The worldvolume of a ground state configuration is
given by the intersection of a four-dimensional plane and $\bm{S}^5$ with
$\ZZ_k$ identification.
In general there is a continuous family of such ground state configurations,
and to cover fluctuations around all of them we adopt a patch-wise method.
Namely, we choose a few ground state configurations and quantize the fluctuations around each of them.
Then we collect the results in all patches to construct the index.

This paper is organized as follows.
In the next section, as a preparation for the index calculation,
we establish the patch-wise method for the BPS partition function,
for which an exact formula is known \cite{Arai:2018utu}.
We derive a formula which relates the
single-particle partition functions
for the excitations on the wrapped D3-branes
to the finite $N$ correction to the BPS partition function.
In section \ref{index.sec} we define the superconformal index,
and summarize known results in the large $N$ limit.
In section \ref{d3.sec} we apply the patch-wise method
to the superconformal index,
and propose a formula which gives the correction starting from $q^N$.
In section \ref{checks.sec} we compare the results on the AdS side
to the results of the localization,
and confirm that our formula gives correct indices
up to errors of order $q^{2N}$.
We also show that the formula works even for the $k=1$ case.
The last section is devoted for conclusions and discussion.
Details of calculations, conventions, and
collection of results are provided in appendices.

\section{BPS partition function}\label{bps.sec}
Before we investigate the superconformal index
let us discuss the BPS partition function.
Similarly to the superconformal index
the BPS partition function is defined as the trace
of a symmetry group element over gauge invariant operators.
The difference from the index is that
only operators made of scalar fields are
taken into account, and this makes the calculation
much simpler than the index.
Indeed, an analytic formula is known for the
BPS partition function \cite{Arai:2018utu}.
Using the knowledge of the analytic formula
we will establish a method
to reproduce the finite $N$ corrections to the
partition function from the analysis of wrapped D3-branes.
The method, which we call ``the patch-wise calculation,''
will be used in the following sections
to calculate the finite $N$ corrections to
the superconformal index.

\subsection{BPS condition and S-folding}
Let $\mathsf{X}$, $\mathsf{Y}$, and $\mathsf{Z}$ be the three scalar fields
in the ${\cal N}=4$ vector multiplet.
When we discuss the BPS partition function
we are interested in BPS operators ${\cal O}$ made of scalar fields satisfying
\begin{align}
[{\cal Q},{\cal O}]=0,
\end{align}
where ${\cal Q}$ is one of the eight supercharges $Q^I$ and $\ol Q_I$ ($I=1,2,3,4$).
Operators satisfying this ${\cal Q}$-closedness condition 
saturate at least one of the BPS bounds
\begin{align}
H\geq \pm R_x\pm R_y\pm R_z,
\end{align}
where $H$ is the dilatation (or, Hamiltonian in the radial quantization),
and $R_x$, $R_y$, and $R_z$ are the Cartan generators of $SU(4)_R$
each of which acts on the scalar field $\mathsf{X}$, $\mathsf{Y}$, and $\mathsf{Z}$, respectively.
The eight combinations of the three signs correspond to eight supercharges
$Q^I$ and $\ol Q_I$,
and once we choose the signs, the supercharge ${\cal Q}$ is uniquely fixed.
In this paper we always use the ``all plus'' convention.
Namely, BPS operators satisfy
\begin{align}
H=R_x+R_y+R_z.
\label{ppp}
\end{align}
With our convention $\sX$, $\sY$, and $\sZ$ satisfy (\ref{ppp}).
In the D3-brane construction of the ${\cal N}=4$ SYM
the scalar fields correspond to $\CC^3$ transverse to the D3-brane worldvolume.
Let $X$, $Y$, and $Z$ be the complex coordinates corresponding to the scalar fields
$\sX$, $\sY$, and $\sZ$, respectively.
The above choice of the signs fixes a natural complex structure in $\CC^3$.
Namely, the coordinates $X$, $Y$, and $Z$ are treated as holomorphic coordinates.

The S-folding group $\ZZ_k$ is generated by
\begin{align}
{\cal R}=\exp\left(\frac{2\pi i}{k}\left(S-\frac{A}{2}\right)\right) ,
\end{align}
where $S$ is defined by
\begin{align}
S=\pm R_x\pm R_y\pm R_z,
\label{sdef}
\end{align}
and $A$ is the $U(1)_A$ charge which will be defined later.
$A$ acts on the scalar fields trivially
and we can neglect it in this section.
When $k\geq 3$ the projection by ${\cal R}$ reduces the ${\cal N}=4$
supersymmetry to ${\cal N}=3$.
Again we have eight combinations of signs in (\ref{sdef}).
Up to overall the sign that does not affect the definition of the S-fold,
we have four possible choices of $S$.
These correspond to four possibilities of
the eliminated supercharge in the reduction from ${\cal N}=4$ to ${\cal N}=3$.
The symmetric choice $S=R_x+R_y+R_z$
is not allowed
because the associated projection eliminates ${\cal Q}$
and is incompatible with the definition of BPS operators.
Therefore, we need to use an asymmetric choice.
We take $S=-R_x+R_y+R_z$
which gives the inhomogeneous action
\begin{align}
(X,Y,Z)\stackrel{\cal R}{\rightarrow}
(\omega_k^{-1}X,\omega_kY,\omega_kZ),
\label{omegampp}
\end{align}
where $\omega_k=\exp(2\pi i/k)$.

\subsection{Geometric quantization}
The BPS partition function is given as the sum
\begin{align}
Z(x,y,z,q;N)=\sum_{m=0}^{k-1}Z_m(x,y,z,q;N).
\end{align}
$Z_m$ is the BPS partition function of the sector $m$ defined by
\begin{align}
Z_m(x,y,z,q;N)=\tr(x^{R_x}y^{R_y}z^{R_z}q^H),
\label{zmdef}
\end{align}
where the trace is taken over gauge invariant operators
with the $\ZZ_k$ charge $m$ made of scalar fields.
In the context of the boundary SCFT $R_x$, $R_y$, and $R_z$ are
Cartan generators of the $SU(4)_R$ symmetry,
and $H$ is the dilatation operator.
They are interpreted in the holographic description
as the angular momenta in $\bm{S}^5$ or $\bm{S}^5/\ZZ_k$ and the Hamiltonian
normalized by the inverse of the AdS radius.
The definition (\ref{zmdef})
is redundant because
BPS operators satisfy (\ref{ppp}),
and $H$ is not independent of the R-charges.
$Z_m$ depends only on the three combinations of fugacities $qx$, $qy$, and $qz$.
Although we can set one of the fugacities to be $1$ without loosing information
we leave the redundancy for later convenience.

A simple way to derive the BPS partition function is
to use the geometric quantization of BPS D3-branes
in $\bm{S}^5/\ZZ_k$ \cite{Arai:2018utu}
following the similar analysis of
giant gravitons in \cite{Biswas:2006tj}.
(For an $\mathcal{N}=4$ SYM there is a complementary way
to reproduce the same result \cite{Mandal:2006tk} .
We can also calculate the exact BPS partition function
by using dual giant gravitons, D3-branes expanded in AdS \cite{Grisaru:2000zn,Hashimoto:2000zp}.
In this work, although it may be possible and interesting,
we will not discuss the derivation of the BPS partition functions
for S-folds and its extension to the index
using dual giant gravitons.)
Let us represent $\bm{S}^5$ as the subspace of $\CC^3$
defined by
\begin{align}
|X|^2+|Y|^2+|Z|^2=1.
\end{align}
As is shown in \cite{Mikhailov:2000ya}
the worldvolume of BPS D3-branes in $\bm{S}^5$
is given as the intersection of
$\bm{S}^5$ and a holomorphic surface in $\CC^3$
given by
\begin{align}
f(X,Y,Z)=0,
\label{surface}
\end{align}
where $f(X,Y,Z)$ is an arbitrary holomorphic function.
In the case of S-fold we need to impose the condition of the $\ZZ_k$ invariance of 
the surface (\ref{surface}).
This requires the function $f(X,Y,Z)$ to satisfy
\begin{align}
{\cal R}f(X,Y,Z)=\omega_k^m f(X,Y,Z).
\label{mdef}
\end{align}
$m$ is a $\ZZ_k$-valued quantity interpreted as the wrapping number of the surface
around the topologically non-trivial $3$-cycle in $\bm{S}^3/\ZZ_k$.
This is confirmed by considering simple examples of the function $f$ satisfying (\ref{mdef}).
\begin{align}
f=X^{k-m},\quad
f=Y^m,\quad
f=Z^m.
\label{threef}
\end{align}
The latter two give $m$ coincident planes in $\CC^3$,
and the intersection with $\bm{S}^5$ gives $m$ branes wrapped
over the large $\bm{S}^3$ specified by $Y=0$ and $Z=0$, respectively.
The first one looks different from the latter two if $k-m\neq m$.
This is because we use asymmetric definition of ${\cal R}$ and
the wrapping over $X=0$ plane must be counted with the opposite sign.
Because continuous deformations of the function $f$ that
keep the condition
(\ref{mdef}) satisfied
do not change the
homology class of the worldvolume in $\bm{S}^5/\ZZ_k$,
an arbitrary function satisfying 
(\ref{mdef}) has the wrapping number $m$.

The Taylor expansion of the function $f$ is
\begin{align}
f(X,Y,Z)=\sum_{u,v,w}c_{u,v,w}X^uY^vZ^w,
\end{align}
where the summation is taken over non-negative integers $(u,v,w)$
satisfying
\begin{align}
-u+v+w=m\mod k.
\label{pqrcond}
\end{align}
We treat the coefficients $c_{u,v,w}$ as dynamical variables.
Because overall factor is irrelevant to the worldvolume defined by
(\ref{surface}),
the configuration space ${\cal M}$ is the projective space $\CC\bm{P}^\infty$
with the homogeneous coordinates $c_{u,v,w}$.

There are two issues which make the problem complicated.
One is that different functions $f$ may give the same brane configuration, and we should remove the redundancy.
The other is that the surface $f=0$ may not intersect with $\bm{S}^5$,
and the parameter region giving such a surface should be removed from the configuration space.
The detailed analysis in \cite{Biswas:2006tj} shows that
even if we take account of these issues the result is the same as what we obtained by naive analysis
neglecting these issues.

A quantum state of D3-branes is specified by the wave-function
$\Psi(c_{u,v,w})$ over ${\cal M}$.
Due to the coupling to the background RR flux the wave-function is
not a function but a section of the line bundle ${\cal O}(N)$ over
$\CC\bm{P}^\infty$.
Therefore, the determination of the Fock space of D3-branes
reduces to the problem of finding global holomorphic sections
of ${\cal O}(N)$.

This is actually a very simple problem.
In terms of the homogeneous coordinates $c_{u,v,w}$,
holomorphic sections of ${\cal O}(N)$ are expressed
as homogeneous polynomials of degree $N$.
Therefore, we can identify a quantum state of D3-branes with
a collection of $N$ coefficients with duplication allowed.
It is convenient to treat each coefficient $c_{u,v,w}$ as a bosonic quantum with quantum numbers $(R_x,R_y,R_z)=(u,v,w)$.
$c_{u,v,w}$ contributes to the BPS partition function by
$x^uy^vz^wq^{u+v+w}$
and the single-particle partition function is given by
\begin{align}
I_m(x,y,z,q)=\sum_{u,v,w}x^uy^vz^wq^{u+v+w},
\end{align}
where the summation is taken over non-negative integers
satisfying
(\ref{pqrcond}).
The BPS partition function for a specific $N$ can be extracted from the grand partition function
$\Pexp(I_m(x,y,z,q)t)$
by picking up the $t^N$ term.
\begin{align}
Z_m(x,y,z,q;N)
=\left(\Pexp(I_m(x,y,z,q)t)\right)|_{t^N},
\label{ximdef}
\end{align}
where ``$\Pexp$'', the plethystic exponential, is defined by
\begin{align}
\Pexp f(x_i)=\exp\left(\sum_{m=1}^\infty\frac{1}{m}f(x_i^m)\right).
\end{align}
This is the operation which generates the multi-particle partition function
from the single-particle one.

\subsection{Large $N$ limit and finite $N$ corrections}
In this section we derive formulas for the large $N$ limit of
the BPS partition functions and the finite $N$ corrections.

Let us first derive a few formulas for the plethystic exponential.
We assume that a single-particle partition function is given by
$1+F$, where $F=\sum_i F_i$ includes monomials $F_i$ with
positive order.
``The order'' of $F_i$ may be defined as the exponent of $q$,
$H=R_x+R_y+R_z$.
For later convenience we generalize this to
a non-vanishing linear combination $R_*$ of the charges $R_x$, $R_y$, and $R_z$.
We assume $R_*\geq 1$ for all $F_i$.
We are interested in the $N$-particle partition function
given by
\begin{align}
\Pexp((1+F)t)|_{t^N}.
\label{nparticle}
\end{align}
The large $N$ limit of this partition function
can be extracted from the pole at $t=1$.
\begin{align}
\lim_{N\rightarrow\infty} \Pexp(1+F)|_{t^N}
&= \lim_{t\rightarrow 1}\left[\frac{1}{1-t}\Pexp((1+F)t)\right]
\nonumber\\
&=\lim_{t\rightarrow 1} \Pexp((1+F)t-t)
\nonumber\\
&= \Pexp F.
\label{PexplargeN}
\end{align}
For finite $N$ we use ``$\approx$'' to express the presence of
the finite $N$ correction:
\begin{align}
\Pexp((1+F)t)|_{t^N}\approx\Pexp F.
\label{laergsn}
\end{align}
The order of the error in (\ref{laergsn}) is
\begin{align}
(\mbox{error})={\cal O}(F)^{N+1}.
\label{finnerror}
\end{align}
This is shown as follows.
We first rewrite (\ref{nparticle}) as
\begin{align}
\Pexp((1+F)t)|_{t^N}
&= \left(\frac{1}{1-t}\prod_i \frac{1}{1-tF_i}\right)\Bigg|_{t^N}
\nonumber\\
&= \left(\prod_i\frac{1}{1-tF_i}\right)\Bigg|_{t^0,\cdots, t^N},
\end{align}
where $(\cdots)|_{t^0,\ldots,t^N}$ means the sum of the coefficients of
the terms with the indicated powers of $t$.
In the large $N$ limit this gives the formula (\ref{PexplargeN}).
The error for finite $N$ is
\begin{align}
\left(\prod_i\frac{1}{1-tF_i}\right)\Bigg|_{t^{N+1},t^{N+2},\cdots}={\cal O}(F)^{N+1},
\end{align}
as is shown in (\ref{laergsn}).

Now let us apply the formula 
(\ref{laergsn}) to $Z_m$ in (\ref{ximdef}).
$Z_0$ gives the contribution of D3-brane bubbles
in $\bm{S}^5/\ZZ_k$ without wrapping.
To emphasize this we use the notation $Z^{\rm KK}=Z_0$.
We also define the corresponding single-particle partition function $I^{\rm KK}=I_0-1$.
It is given by
\begin{align}
I^{\rm KK}&=-1+\sum_{-u+v+w=0\mod k}x^uy^vz^wq^{u+v+w}\quad
(u,v,w\geq0).
\label{ikksp}
\end{align}
The formula (\ref{laergsn}) gives
\begin{align}
Z^{\rm KK}\approx\Pexp I^{\rm KK}.
\label{z0largen}
\end{align}

A state contributing to $Z_m$ with $m\neq0$ in general includes not only D3-branes
wrapping around the non-trivial cycle in $\bm{S}^5/\ZZ_k$
but also disconnected components without wrapping.
We indicate this by using the notation $Z^{{\rm D3}+{\rm KK}}_m=Z_m$ for $m\neq0$.
The $q$ expansion of the corresponding single-particle partition function
$I_m$ ($m\neq0$) starts from
${\cal O}(q^{|m|})$.
Therefore, the Taylor expansion of $Z_m$
starts from ${\cal O}(q^{|m|N})$ and we cannot use (\ref{laergsn}) as it is.

To avoid this problem we choose
one of the ${\cal O}(q^{|m|})$ terms in $I_m$
and use it as a reference point.
Let $I^{\rm gr}$ be the chosen term.
(``gr'' stands for a ground state.)
Then the BPS partition function $Z_m$
includes the term $(I^{\rm gr})^N$ in the leading term in the
$q$ expansion.
This term corresponds to a D3-brane ground state configuration $C$.
In general we have several choices of $I^{\rm gr}$
corresponding to different $C$.
We denote $I^{\rm gr}$ associated with a specific
configuration $C$ by $I_C^{\rm gr}$.
In the following we are interested in the sectors with $|m|=1$, and
$C$ is one of $X=0$, $Y=0$, and $Z=0$.
$X=0$ has the wrapping number $m=-1$ and $Y=0$ and $Z=0$ have $m=+1$.
For each of them $I_C^{\rm gr}$ is given by
\begin{align}
I^{\rm gr}_{X=0}=qx,\quad
I^{\rm gr}_{Y=0}=qy,\quad
I^{\rm gr}_{Z=0}=qz.
\label{igrxyz}
\end{align}
(\ref{ximdef}) can be rewritten as
\begin{align}
Z_m^{{\rm D3}+{\rm KK}}=
(I_C^{\rm gr})^N\left.\left(
\Pexp\left(
(1+I_C^{{\rm D3}+{\rm KK}})
t\right)\right)\right|_{t^N},\quad
I_C^{{\rm D3}+{\rm KK}}=\frac{I_m}{I_C^{\rm gr}}-1.
\end{align}
The functions $I_C^{{\rm D3}+{\rm KK}}$
are given by
\begin{align}
I^{{\rm D3}+{\rm KK}}_{X=0}&=-1+\sum_{-u+v+w=0\mod k}x^uy^vz^wq^{u+v+w}\quad(u\geq-1,v\geq0,w\geq0),\nonumber\\
I^{{\rm D3}+{\rm KK}}_{Y=0}&=-1+\sum_{-u+v+w=0\mod k}x^uy^vz^wq^{u+v+w}\quad(u\geq0,v\geq-1,w\geq0),\nonumber\\
I^{{\rm D3}+{\rm KK}}_{Z=0}&=-1+\sum_{-u+v+w=0\mod k}x^uy^vz^wq^{u+v+w}\quad(u\geq0,v\geq0,w\geq-1).
\label{id3xyznc}
\end{align}
The formula (\ref{laergsn}) gives
\begin{align}
Z_m^{{\rm D3}+{\rm KK}}\approx
(I_C^{\rm gr})^N\Pexp I_C^{{\rm D3}+{\rm KK}}.
\label{zmlargen}
\end{align}

We also define the partition function
of connected configurations by
\begin{align}
Z_m^{\rm D3}=\frac{Z_m^{{\rm D3}+{\rm KK}}}{Z^{\rm KK}}.
\label{zmconn}
\end{align}
Combining (\ref{z0largen}) and (\ref{zmlargen}) we obtain
\begin{align}
Z_m^{\rm D3}
&\approx(I_C^{\rm gr})^N\Pexp I^{\rm D3}_C,
\label{zmlargencn}
\end{align}
where $I^{\rm D3}_C$ is defined by
\begin{align}
I^{\rm D3}_C=I_C^{{\rm D3}+{\rm KK}}-I^{\rm KK}.
\label{id3c}
\end{align}
The error of (\ref{zmlargencn}) is obtained
by combining (\ref{z0largen}) and (\ref{zmlargencn}) as
\begin{align}
(\mbox{error})
&=(I_C^{\rm gr})^N{\cal O}(I^{{\rm D3}+{\rm KK}}_C)^{N+1}.
\label{errformula}
\end{align}
For the three loci $I^{\rm D3}_C$ are given by
\begin{align}
I^{\rm D3}_{X=0}
&=\sum_{v+w=-1\mod k}q^{v+w-1}x^{-1}y^vz^w,\quad(v,w\geq0)\nonumber\\
I^{\rm D3}_{Y=0}&
=\sum_{-u+w=1\mod k}q^{u+w-1}x^uy^{-1}z^w,\quad(u,w\geq0)\nonumber\\
I^{\rm D3}_{Z=0}
&=\sum_{-u+v=1\mod k}q^{u+v-1}x^uy^vz^{-1},\quad(u,v\geq0)
\label{id3xyz}
\end{align}
Unlike (\ref{ikksp}) and (\ref{id3xyznc})
each function in (\ref{id3xyz}) is given as the sum over
two variables.
This strongly suggests that the relevant degrees of freedom
live in a lower dimensional space.
In the next subsection we will reproduce (\ref{id3xyz})
by the analysis of
the fluctuations of wrapped D3-branes.

The single-particle partition functions introduced above associated with different loci
are related among them by the permutations among fugacities
associated with the symmetry of the system.
For example,
three functions in
(\ref{id3xyz}) satisfy
\begin{align}
\sigma_{23}I^{\rm D3}_{X=0}=I^{\rm D3}_{X=0},\quad
\sigma_{23}I^{\rm D3}_{Y=0}=I^{\rm D3}_{Z=0}.
\end{align}
where $\sigma_{ij}$ is the operator which swaps the $i$-th and the $j$-th fugacities in
$(x,y,z)$.
In addition, in the $k=2$ case the following relations hold.
\begin{align}
\sigma_{123}I^{\rm D3}_{X=0}=I^{\rm D3}_{Y=0},\quad
\sigma_{123}^2I^{\rm D3}_{X=0}=I^{\rm D3}_{Z=0},
\end{align}
where $\sigma_{123}=\sigma_{12}\sigma_{23}$.

\subsection{D3-brane fluctuations}
Let us reproduce
(\ref{id3xyz}) as
the single-particle partition functions
of the D3-brane fluctuations.

\begin{figure}[htb]
\centering
\includegraphics[width=100mm, bb=86 420 376 540]{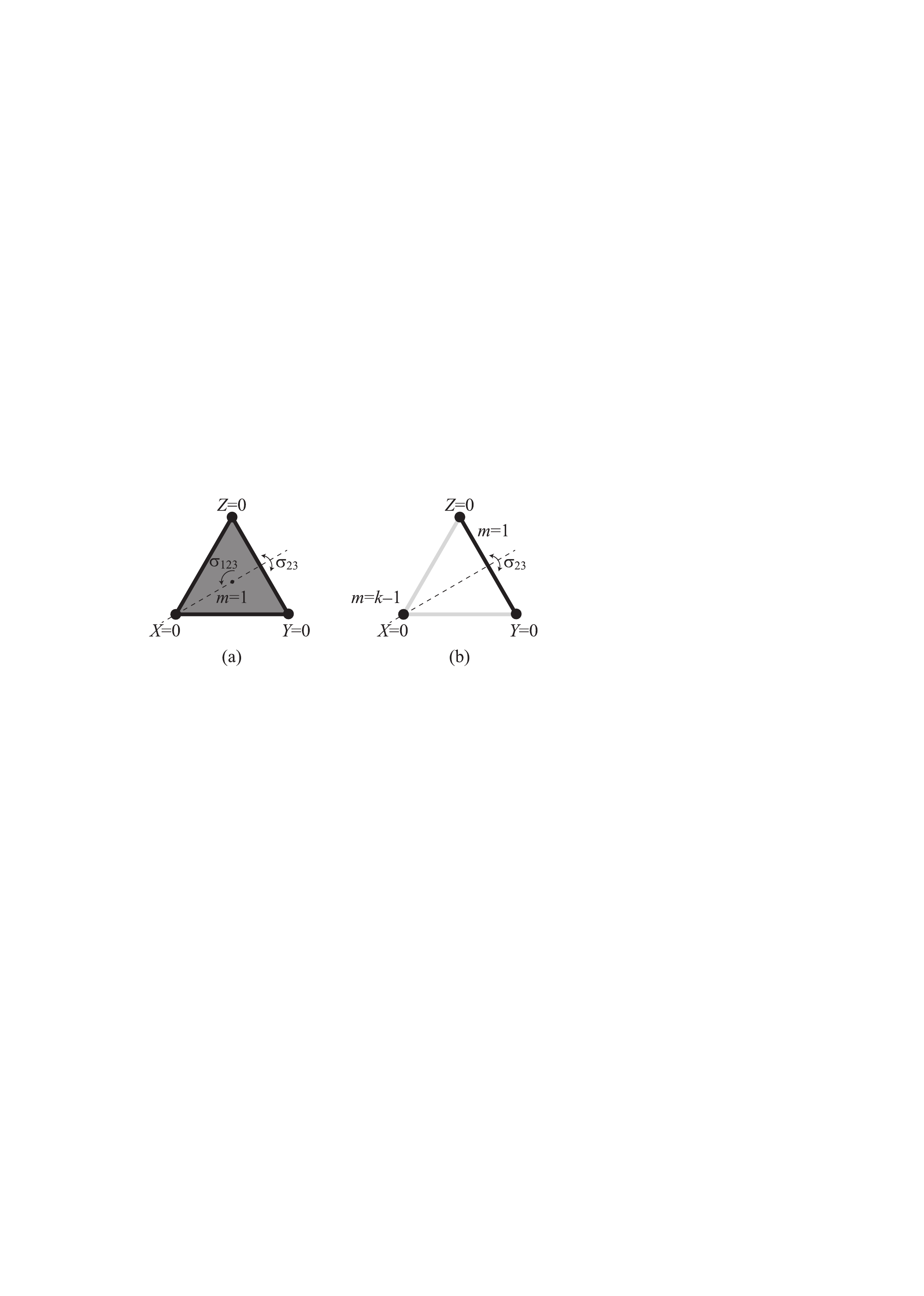}
\caption{The structure of the ground state configuration spaces for $k=2$ (a) and $k\geq3$ (b)
are shown.}\label{configspace.eps}
\end{figure}

We first summarize the structure of the
ground state configuration space of wrapped D3-branes.
We are interested in ${\cal O}(q^N)$ corrections, and
a wrapped D3-brane gives such a correction when $|m|=1$.
In the orientifold case with $k=2$ there is one such sector, $m=1$.
The ground state configuration is given by
\begin{align}
aX+bY+cZ=0,
\end{align}
and the coefficients $(a,b,c)$ are
the homogeneous coordinates of
the configuration space $\CC\bm{P}^2$.
(See (a) in Figure \ref{configspace.eps}.)
This structure is consistent with (\ref{id3xyz}).
If $k=2$ each of $I^{\rm D3}_C$ in (\ref{id3xyz})
includes two $q^0$ terms.
This implies that
each configuration is contained in a
two-dimensional configuration space.
For $k\geq3$, there are two sectors
with different wrapping numbers, $m=1$ and $m=k-1$.
(See (b) in Figure \ref{configspace.eps}.)
A configuration with $m=1$ is given by
\begin{align}
bY+cZ=0,
\end{align}
and the configuration space is $\CC\bm{P}^1$ with
the homogeneous coordinates $(b,c)$.
For $m=k-1$ there is only one configuration $X=0$.
Again, the structure is consistent with (\ref{id3xyz}).
If $k\geq3$ $I_{X=0}^{\rm D3}$ does not contain
the $q^0$ term.
This indicates the isolated configuration.
Each of $I_{Y=0}^{\rm D3}$ and $I_{Z=0}^{\rm D3}$
contains one $q^0$ term, and this is consistent with the
fact that the two configurations are contained in the
one-dimensional configuration space.

We can actually reproduce not only the $q^0$ terms but also
all terms in (\ref{id3xyz}).
Because the three loci can be treated in a parallel way
we here consider a D3-brane wrapped over $X=0$.
The fluctuations of this D3-brane is described by
giving $X$ as a function of $Y$ and $Z$.
As Mikhairov has shown in \cite{Mikhailov:2000ya} for the configuration to be BPS
this function should be holomorphic.
Let us consider a mode
\begin{align}
X=f(Y,Z)\propto Y^vZ^w.
\label{fxy}
\end{align}
This surface should be
compatible with the S-fold $\ZZ_k$
action (\ref{omegampp}).
This requires
\begin{align}
v+w=-1\mod k.
\label{pqcond}
\end{align}

A quantum of the excitation of this mode carries $(R_y,R_z)=(v,w)$.
Let us calculate the energy of the quantum
by using the D3-brane action.
We assume the fluctuation is small and
take only quadratic terms with respect to $X$.
The Born-Infeld action and the Chern-Simons action become
\begin{align}
S_{\rm BI}&=
\frac{kN}{2\pi^2}
\int dt\wedge\omega_3
\left(\frac{1}{2}|\dot X|^2-\frac{1}{2}|\nabla X|^2+\frac{3}{2}|X|^2\right),\nonumber\\
S_{\rm CS}
&=\frac{kN}{2\pi^2}\int dt\wedge\omega_3
(iX^*\dot X-iX\dot X^*),
\label{bics}
\end{align}
where the integral is taken over $\RR\times\bm{S}^3$
and $\omega_3$ is the volume form of $\bm{S}^3$.
$L$ is the AdS radius and we used
the D3-brane tension $T_{\rm D3}=kN/(2\pi^2L^4)$
to obtain the coefficient in $S_{\rm BI}$.
The equation of motion is
\begin{align}
-\ddot X+\nabla^2 X+3X+4i\dot X=0.
\end{align}
$\nabla^2$ is the Laplacian in $\bm{S}^3$ and its
eigenvalue associated with the mode (\ref{fxy}) is $-(v+w)(v+w+2)$.
The energy $E$ can be read off from the $t$ dependence of
a solution.
Let us assume the time dependence $X\propto e^{-iE t}$.
The equation of motion gives
\begin{align}
(E+1)(E+3)=(v+w)(v+w+2).
\end{align}
The non-negative solution corresponding to
the BPS excitation is
\begin{align}
E=v+w-1.
\label{sandpq}
\end{align}

The relation (\ref{sandpq}) holds even if we take account of higher order terms in the action.
Actually, (\ref{sandpq}) is easily derived by using the Mikhailov's solution.
The time dependence of the solution
in \cite{Mikhailov:2000ya}
is given by
\begin{align}
(X(t),Y(t),Z(t))
=(e^{it}X(0),e^{it}Y(0),e^{it}Z(0)).
\end{align}
Let us assume that the initial configuration is given by
\begin{align}
X(0)=f(Y(0),Z(0)).
\end{align}
Then the $Y$ and $Z$ dependence of $X$ at time $t$ is
\begin{align}
X(t)=e^{it}X(0)
=e^{it}f(e^{-it}Y(0),e^{-it}Z(0))
=e^{-i(v+w-1)t}f(Y(t),Z(t)),
\label{ztbyxtyt}
\end{align}
and the frequency agrees with (\ref{sandpq}).

Once we obtain the energy,
we obtain $R_z=-1$ from the BPS condition (\ref{ppp}).
A quantum of the mode (\ref{sandpq}) contributes to the BPS partition function by
$q^{v+w-1}x^{-1}y^vz^w$.
The single-particle partition function is obtained
by summing up this over $v$ and $w$ under the constraint (\ref{pqcond}),
and $I_{X=0}^{\rm D3}$ in (\ref{id3xyz}) is correctly reproduced.

\subsection{Patch-wise calculation}
The purpose of this subsection is to demonstrate
how we can reproduce the partition function $Z_m^{\rm D3}$
by using $I^{\rm D3}_C$.

Let us first consider the $k\geq3$ case.
The configuration space consists of two components with $m=1$ and $m=k-1$,
and we should calculate the contribution of each component
separately.
The component with $m=k-1$ includes one ground state configuration $X=0$,
and the fluctuations around it gives
the partition function
\begin{align}
Z_{k-1}^{\rm D3}
\approx
Z_{X=0}^{\rm D3}
\equiv
q^Nx^N\Pexp I^{\rm D3}_{X=0}.
\label{zd3nond}
\end{align}
The error is estimated by (\ref{finnerror}) as
\begin{align}
(\mbox{error})\sim
q^Nx^N{\cal O}(I^{{\rm D3}+{\rm KK}}_{X=0})^{N+1}
\lesssim
{\cal O}(q^{2N+1})
\end{align}
where we used the fact that all terms in $I^{{\rm D3}+{\rm KK}}_{X=0}$ have $H\geq1$.

The other component of
the ground state configuration space with $m=1$
is $\CC\bm{P}^1$, and we can cover it by two coordinate
patches each of which includes $Y=0$ or $Z=0$.
The fluctuations around $Y=0$ give
\begin{align}
Z_{Y=0}^{\rm D3}
\equiv
q^Ny^N\Pexp I_{Y=0}^{\rm D3},
\label{z1d3y}
\end{align}
and ones around $Z=0$ give
\begin{align}
Z_{Z=0}^{\rm D3}
\equiv
q^Nz^N\Pexp I_{Z=0}^{\rm D3}.
\label{z1d3z}
\end{align}
Two equations (\ref{z1d3y}) and (\ref{z1d3z})
give two ways of calculating the same partition function $Z_1^{\rm D3}$.
Both of them have errors,
and the ranges of validity are different from each other.
Namely,
$Z_1^{\rm D3}\approx Z_{Y=0}^{\rm D3}$ in a parameter region with small $z$,
while $Z_1^{\rm D3}\approx Z_{Z=0}^{\rm D3}$ in another region with small $y$.
Thanks to the symmetry of the system $Z_{Y=0}^{\rm D3}$ and $Z_{Z=0}^{\rm D3}$
are related by $\sigma_{23}Z_{Y=0}^{\rm D3}=Z_{Z=0}^{\rm D3}$.

Let us first look at (\ref{z1d3y}).
The $q$ expansion of the single-particle partition function
$I^{\rm D3}_{Y=0}$ includes a $q$-independent term.
We call such terms ``zero-mode terms.''
We divide $I_{Y=0}^{\rm D3}$ into the zero-mode term and the rest:
\begin{align}
I^{\rm D3}_{Y=0}=\frac{z}{y}+I'^{\rm D3}_{Y=0}.
\end{align}
$I'^{\rm D3}_{Y=0}$ includes terms with only positive powers of $q$.
If we assume $q$ is sufficiently small, $\Pexp I'^{\rm D3}_{Y=0}$
gives a convergent series,
while the convergence of the zero-mode contribution
requires the additional assumption $|\frac{z}{y}|\leq1$.
Then the plethystic exponential of the zero-mode term is
\begin{align}
\Pexp\left(\frac{z}{y}\right)=1+\frac{z}{y}+\left(\frac{z}{y}\right)^2+\cdots=\frac{1}{1-\frac{z}{y}},
\end{align}
and we obtain
\begin{align}
Z^{\rm D3}_{Y=0}=\frac{q^Ny^N\Pexp I'^{\rm D3}_{Y=0}}{1-\frac{z}{y}}.
\label{ik31}
\end{align}
We restrict ourselves to $k=3$ case for concreteness.
Then the explicit form of the numerator
of (\ref{ik31}) is
\begin{align}
(\mbox{num})
&=q^Ny^N+q^{N+1}y^{N-1}x^2+q^{N+2}(y^{N-1}z^2x+y^{N-2}x^4)+{\cal O}(q^{N+3}).
\label{ik31num}
\end{align}
Let us focus on the leading term $q^Ny^N$ in (\ref{ik31num}) for example.
Combining the factor $1/(1-\frac{z}{y})$ we obtain
the infinite series
starting from $q^Ny^N$:
\begin{align}
\frac{q^Ny^N}{1-\frac{z}{y}}
=q^N(y^N+y^{N-1}z+y^{N-2}z^2+\cdots).
\label{syn1}
\end{align}
Due to the error in (\ref{z1d3y}) only finite number of terms
in (\ref{syn1}) can be correct.
In addition,
negative powers of fugacities contradict
some of the BPS bounds, and
the series (\ref{syn1})
must terminate at some order.
This is the case, too, for the corresponding part of (\ref{z1d3z}):
\begin{align}
\sigma_{23}
\frac{q^Ny^N}{1-\frac{z}{y}}
=\frac{q^Nz^N}{1-\frac{y}{z}}
=q^N(z^N+yz^{N-1}+y^2z^{N-2}+\cdots).
\label{szn1}
\end{align}
Although neither (\ref{syn1}) nor (\ref{szn1}) is exact,
we can completely determine the order $q^N$ terms
by combining these two
as follows.
\begin{align}
q^N(y^N+y^{N-1}z+y^{N-2}z^2+\cdots+yz^{N-1}+z^N)=q^N\ol\chi_N(y,z).
\label{chinn}
\end{align}
($\ol\chi_n(y,z)$ is the $u(2)$ character defined in Appendix \ref{characters.sec}.)
In fact this is the sum of two functions (\ref{syn1}) and (\ref{szn1})
\begin{align}
\frac{q^Ny^N}{1-\frac{z}{y}}
+\frac{q^Nz^N}{1-\frac{y}{z}}
=q^N\frac{y^{N+1}-z^{N+1}}{y-z}
=q^N\ol\chi_N(y,z).
\label{leadingchu}
\end{align}
Remark that the ranges of convergence for (\ref{syn1}) and (\ref{szn1})
are different, and we need analytic continuation
for $\frac{z}{y}$ to sum up them.
The sum (\ref{chinn}) is a polynomial in $y$ and $z$, and we do not have to
care about convergence.
This procedure can be applied to other terms as follows.
Let us suppose there is a term $x^uy^vz^w$ in (\ref{ik31num}).
We temporarily omit the factor $q^{u+v+w}$, which is a spectator
in the following argument.
We combine this with the denominator
and express it as
\begin{align}
\frac{x^uy^vz^w}{1-\frac{z}{y}}
=\frac{\vec x^{\vec w}}{1-\vec x^{-\vec\alpha}},
\label{pqrterm}
\end{align}
where we introduced the vector notation such as
$\vec x=(x,y,z)$, $\vec w=(u,v,w)$, and $\vec\alpha=(0,1,-1)$.
Just like (\ref{leadingchu})
the sum of (\ref{pqrterm}) and the corresponding term in (\ref{z1d3z}) gives
the character of the $u(2)$ representation with highest weight $\vec w$:
\begin{align}
(1+\sigma_{23})\frac{\vec x^{\vec w}}{1-\vec x^{-\vec\alpha}}
=x^u(yz)^w\ol\chi_{v-w}(y,z).
\end{align}
Notice that this is nothing but the Weyl character formula for $u(2)$.
With our convention $\ol\chi_{v-w}(y,z)$ is the
character of the representation with highest weight
$(0,v-w,0)$, and the prefactor is needed to shift the weight.

After all, the sum of (\ref{z1d3y}) and (\ref{z1d3z})
\begin{align}
Z_{Y=0}^{\rm D3}
+Z_{Z=0}^{\rm D3}
=q^Ny^N\Pexp I_{Y=0}^{\rm D3}
+q^Nz^N\Pexp I_{Z=0}^{\rm D3}
\label{sumyz}
\end{align}
gives $Z_1^{\rm D3}$ up to some order of $q$.
We can obtain (\ref{sumyz})
from (\ref{ik31num})
by the replacement
\begin{align}
x^uy^vz^w\rightarrow x^u(yz)^w\ol\chi_{v-w}(y,z).
\label{replacementy}
\end{align}
We call this procedure
the Weyl completion.

Let us estimate the range of validity of (\ref{sumyz}).
All terms in $I^{{\rm D3}+{\rm KK}}_{Y=0}$
satisfy
$\frac{H}{2}+R_z+\frac{R_x}{2}\geq1$
and the error in (\ref{z1d3y}), $q^Ny^N{\cal O}(I^{{\rm D3}+{\rm KK}})^{N+1}$
does not affect terms satisfying
\begin{align}
\frac{H}{2}+R_z+\frac{R_y}{2}<\frac{3}{2}N+1
\label{error1}.
\end{align}
Similarly, the range of validity for
(\ref{z1d3z}) is
\begin{align}
\frac{H}{2}+R_y+\frac{R_x}{2}<\frac{3}{2}N+1.
\label{error2}
\end{align}
By summing up
(\ref{error1}) and (\ref{error2})
we obtain
\begin{align}
H<\frac{3}{2}N+1.
\label{su2invalid}
\end{align}
If this is satisfied, at least one of (\ref{error1}) and (\ref{error2})
holds, and (\ref{replacementy}) gives correct terms.
Namely, we can correctly determine
$Z_1^{\rm D3}$ for terms satisfying (\ref{su2invalid})
by the prescription above.

By summing up (\ref{zd3nond}) and (\ref{sumyz})
we obtain the formula
\begin{align}
Z_{k-1}^{\rm D3}+Z_1^{\rm D3}
\approx
Z_{X=0}^{\rm D3}
+Z_{Y=0}^{\rm D3}
+Z_{Z=0}^{\rm D3}.
\label{approxzformula}
\end{align}

We emphasize that we do not claim that
$Z_{Y=0}^{\rm D3}$ and $Z_{Z=0}^{\rm D3}$ give two independent contributions.
$Z_{Y=0}^{\rm D3}$ and $Z_{Z=0}^{\rm D3}$ are two approximate functions of $Z_1^{\rm D3}$,
and we should pick up
correct terms from each of them and combine the terms to obtain
$Z_1^{\rm D3}$.
Thanks to the
mathematical structure of the functions $Z_{Y=0}^{\rm D3}$ and $Z_{Z=0}^{\rm D3}$
this is accidentally equivalent to
simply summing up them.

The method of the Weyl completion works in the orientifold case, too.
In that case we have one sector with $m=1$.
The ground state configuration space is $\CC\bm{P}^2$, and we can cover it
by three coordinate patches each of which includes
$X=0$, $Y=0$, or $Z=0$.
(See (a) in Figure \ref{configspace.eps}).
$I^{\rm D3}_{X=0}$ includes two zero-mode terms
corresponding to the two-dimensional configuration space.
\begin{align}
I^{\rm D3}_{X=0}
=\frac{y}{x}+\frac{z}{x}+I'^{\rm D3}_{X=0}.
\label{id3x}
\end{align}
The corresponding partition function is
\begin{align}
Z_1^{\rm D3}\approx
&Z^{\rm D3}_{X=0}
=q^Nx^N\Pexp I^{\rm D3}_{X=0}
=\frac{(\mbox{num})}
{(1-\frac{y}{x})(1-\frac{z}{x})},
\label{z2excitation}
\end{align}
where the numerator is
\begin{align}
(\mbox{num})
&=q^Nx^N\Pexp  I'^{\rm D3}_{X=0}
\nonumber\\
&=q^Nx^N+q^{N+2}x^{N-1}\ol\chi_3
+q^{N+4}\left(x^{N-2}\ol\chi_6+x^{N-2}y^2z^2\ol\chi_2+x^{N-1}\ol\chi_5\right)
\nonumber\\&\hspace{2em}
+{\cal O}(q^{N+6}).
\end{align}

Thanks to the $SU(3)$ symmetry, the partition function for
the $Y=0$ and $Z=0$ loci are
\begin{align}
Z^{\rm D3}_{Y=0}
=\sigma_{123}Z^{\rm D3}_{X=0},\quad
Z^{\rm D3}_{Z=0}
=\sigma_{123}^2Z^{\rm D3}_{X=0}.
\end{align}
We can combine $Z_{X=0}^{\rm D3}$, $Z_{Y=0}^{\rm D3}$, and $Z_{Z=0}^{\rm D3}$
to obtain $Z_1^{\rm D3}$ up to some order of $q$
in a similar way  to the previous one.
The numerator of (\ref{z2excitation})
is a linear combination of $u(2)$ characters,
and we can rewrite it in the form of the Weyl character formula
\begin{align}
(\mbox{num})=(1+\sigma_{23})\frac{g(\vec x,q)}{1-\vec x^{-\vec\alpha}},
\end{align}
where $g$ is the function obtained by picking up
the highest weight term from each $u(2)$ character.
$Z_{X=0}^{\rm D3}$ is given by
\begin{align}
Z_{X=0}^{\rm D3}
=(1+\sigma_{23})\frac{g(\vec x,q)}
{(1-\vec x^{-\vec\alpha})(1-\vec x^{-\vec\alpha'})(1-\vec x^{-\vec\alpha''})},
\end{align}
where $\vec\alpha'=(1,0,-1)$ and $\vec\alpha''=(0,1,-1)$.
Notice that we can regard $\vec\alpha$, $\vec\alpha'$, and $\vec\alpha''$ as
positive roots of $su(3)$.
Just like (\ref{leadingchu})
we sum up three contributions:
\begin{align}
Z_1^{\rm D3}&\approx Z^{\rm D3}_{X=0}+Z^{\rm D3}_{Y=0}+Z^{\rm D3}_{Z=0}
\nonumber\\
&=(1+\sigma_{123}+\sigma_{123}^2)(1+\sigma_{23})
\frac{g(\vec x,q)}{(1-\vec x^{-\vec\alpha})(1-\vec x^{-\vec\alpha'})(1-\vec x^{-\vec\alpha''})}
\nonumber\\
&=\sum_{\sigma\in S_3}
\sigma\frac{g(\vec x,q)}{(1-\vec x^{-\vec\alpha})(1-\vec x^{-\vec\alpha'})(1-\vec x^{-\vec\alpha''})}.
\label{zd31}
\end{align}
Because the permutation group $S_3$ is the Weyl group of $u(3)$
the final expression of
(\ref{zd31}) has the form of the Weyl character formula
for $u(3)$, and gives $Z^{\rm D3}_1$ as a linear combination of
characters of $u(3)$ representations.
For example,
if the numerator in (\ref{z2excitation})
includes term $x^u(yz)^w\ol\chi_n(y,z)$,
the function $g$ includes the highest weight term
$x^uy^{w+n}z^w$,
and
(\ref{zd31})
gives the corresponding term $(xyz)^{w+n}\ol\chi_{(u-w-n,n)}$
in $Z_1^{\rm D3}$.
The rule to obtain $Z_1^{\rm D3}$ from the numerator
of (\ref{z2excitation}) is
\begin{align}
x^u(yz)^w\ol\chi_n(y,z)\rightarrow (xyz)^{w+n}\ol\chi_{(u-w-n,n)}.
\label{replso}
\end{align}
We give the result of this prescription for first few terms:
\begin{align}
Z_1^{\rm D3}
&=q^N\ol\chi_{(N,0)}
+q^{N+2}(xyz)^3\ol\chi_{(N-4,3)}
\nonumber\\&\hspace{2em}
+q^{N+4}((xyz)^5\ol\chi_{(N-6,5)}+(xyz)^6\ol\chi_{(N-8,6)}+(xyz)^4\ol\chi_{(N-6,2)})
\nonumber\\&\hspace{2em}
+{\cal O}(q^{N+6}).
\label{c012}
\end{align}

In summary, the contribution of wrapped D3-branes is given by
\begin{align}
Z_{X=0}^{\rm D3}+Z_{Y=0}^{\rm D3}+Z_{Z=0}^{\rm D3}
\label{zxyz}
\end{align}
for both $k=2$ and $k\geq3$.

\subsection{D3-brane wave function}\label{global.sec}

The patch-wise method in the previous subsections
may call the concept of fiber-bundles in mind.
Actually, the BPS partition function derived above
can be reproduced by counting global holomorphic sections
of a certain vector bundle.

Because the analysis in this subsection
will not be used in the following
we will not give a complete analysis.
We only discuss the orientifold case with the ground state configuration space $\CC\bm{P}^2$.
Generalization to the other cases is straightforward.

In the orientifold case, the ground state
configuration space is ${\cal M}_B=\CC\bm{P}^2$.
Let $O\in{\cal M}_B$ be the point corresponding to the worldvolume
$X=0$.
An excitation mode is specified by
$X=f(Y,Z)$ and
is regarded as an element of the vector
space associated with $O$.
If $f(Y,Z)$ is a degree $1$ polynomial,
the corresponding vector space is $T_O{\cal M}_B$,
the tangent space of ${\cal M}_B$ at $O$.
Therefore, fluctuations by degree $n$ polynomial
is associated with
the symmetric product
\begin{align}
\Sym^n(T_O{\cal M}_B),
\end{align}
and the space for small fluctuations
around $X=0$
is the direct sum
\begin{align}
\bigoplus_{n=1}^\infty\Sym^n(T_O{\cal M}_B),
\label{configs}
\end{align}
Because deformations by $n=1$ modes do not raise the energy,
it is natural to remove the restriction of the small fluctuation
and replace the $n=1$ factor in 
(\ref{configs}) by ${\cal M}_B$.
As the result we obtain the fiber bundle
\begin{align}
{\cal M}=\bigoplus_{n=2}^\infty\Sym^n(T{\cal M}_B)
\end{align}
over ${\cal M}_B$.
This is the configuration space for a wrapped D3-brane with small fluctuations.
The wave function $\Psi$ of the D3-brane
is a section
of a line bundle over ${\cal M}$.

Let us use the Born-Oppenheimer decomposition.
Namely, we treat the fluctuations along the fiber directions
as the fast variables,
and fix the functional dependence of $\Psi$ on the fiber coordinates.
Let $N_n$ for $n\geq2$ be the excitation number of quanta associated with
degree $n$ modes.
For each $P\in{\cal M}_B$
the wave function along $\Sym^n(T_P{\cal M}_B)$ is
(up to Gaussian factor we are not interested in)
a degree $N_n$ polynomial
of the fiber coordinates.
Such a function is an element of
\begin{align}
\Sym^{N_n}\left(\Sym^n(T_P^*{\cal M}_B)\right).
\end{align}
Therefore, the wave function $\Psi$ for a state with a particular set of excitation numbers
$N_n$ is a section of the vector bundle
\begin{align}
{\cal E}={\cal O}(N')\otimes\bigotimes_{n=2}^\infty\Sym^{N_n}\left(\Sym^n(T^*{\cal M}_B)\right).
\label{vbe}
\end{align}
The orbital factor ${\cal O}(N')$ can be determined as follows.
Because an element of the cotangent bundle
$T^*{\cal M}_B$ carries $R_x+R_y+R_z=1$,
the wave function in (\ref{vbe}) carries
\begin{align}
R_x+R_y+R_z=N'+\sum_{n=2}^\infty nN_n.
\label{eq96}
\end{align}
On the other hand,
because the ground state has the energy $N$ and an excitation of degree $n$ modes has
the energy $n-1$ (See (\ref{sandpq})),
the total energy is
\begin{align}
E=N+\sum_{n=2}^\infty(n-1)N_n.
\label{eq97}
\end{align}
By combining (\ref{eq96}), (\ref{eq97}), and the BPS condition (\ref{ppp}),
we obtain
\begin{align}
N'=N-\sum_{n=2}^\infty N_n.
\end{align}

The action of $R_x$, $R_y$, and $R_z$ on the base
${\cal M}_B$ can be naturally extended to the
vector bundle ${\cal E}$, and we can define
the character of the vector bundle $\chi({\cal E})$
as the trace of $x^{R_x}y^{R_y}z^{R_z}$
over the vector space of the global holomorphic sections.
We can reproduce $Z_1^{\rm D3}$ in (\ref{c012})
by calculating the characters
of the vector bundles (\ref{vbe}).

For the ground states with $N_n=0$, the wave function is
a section of
\begin{align}
{\cal E}_0^{(N)}\equiv {\cal O}(N),
\end{align}
and the character for the holomorphic section
is $\ol\chi_{(N,0)}$. (See (\ref{appeq1}).)
This gives the $q^N$ term in (\ref{c012}).

The $q^{N+2}$ term in (\ref{c012})
is the contribution of states with $N_3=1$ and $N_{n\neq3}=0$.
The associated bundle is
\begin{align}
{\cal E}_3^{(N-1)}\equiv {\cal O}(N-1)\otimes\Sym^3(T^*{\cal M}_B).
\end{align}
The formula (\ref{chichi}) shows that
the character of holomorphic
sections of this vector bundle is
\begin{align}
\chi({\cal E}_3^{(N-1)})=\ol\chi_{(N-1,0)}\ol\chi_{(3,0)}-\ol\chi_{(N,0)}\ol\chi_{(2,0)}
=(xyz)^3\ol\chi_{(N-4,3)},
\end{align}
for sufficiently large $N$.
This reproduces the $q^{N+2}$ term in (\ref{c012}).

There are two types of contributions
to the $q^{N+4}$ terms.
One is the contribution from
states with $N_5=1$ and $N_{n\neq5}=0$.
The associated vector bundle is
\begin{align}
{\cal E}_5^{(N-1)}\equiv{\cal O}(N-1)\otimes\Sym^5(T^*{\cal M}_B),
\end{align}
and (\ref{chichi}) gives the character
\begin{align}
\chi({\cal E}_5^{(N-1)})=
\ol\chi_{(N-1,0)}\ol\chi_{(5,0)}-\ol\chi_{(N,0)}\ol\chi_{(4,0)}
=(xyz)^5\ol\chi_{(N-6,5)},
\label{cont11}
\end{align}
for sufficiently large $N$.
The other is the contribution of
states with $N_3=2$ and $N_{n\neq3}=0$.
The wave function for such a state is a section of
\begin{align}
{\cal E}_{\{3,3\}}^{(N-2)}
\equiv{\cal O}(N-2)\otimes\Sym^2\left(\Sym^3(T^*{\cal M}_B)\right),
\end{align}
and
for sufficiently large $N$
(\ref{two1}) gives the character
\begin{align}
\chi({\cal E}_{\{3,3\}}^{(N-2)})
&=\ol\chi_{(N-2,0)}\frac{\ol\chi_{(3,0)}^2+\ol\chi_{(3,0)}^{(2)}}{2}
-\ol\chi_{(N-1,0)}\ol\chi_{(3,0)}\ol\chi_{(2,0)}
+\ol\chi_N\frac{\ol\chi_{(2,0)}^2-\ol\chi_{(2,0)}^{(2)}}{2}
\nonumber\\
&=(xyz)^4\ol\chi_{(N-6,2)}+(xyz)^6\ol\chi_{(N-8,6)},
\label{cont22}
\end{align}
where $\ol\chi_{(r_1,r_2)}^{(m)}\equiv\ol\chi_{(r_1,r_2)}(x^m,y^m,z^m)$.
The sum of (\ref{cont11}) and (\ref{cont22}) reproduces
the $q^{N+4}$ terms in (\ref{c012}).

\section{Superconformal index}\label{index.sec}
In the previous section we established a method to
calculate the contribution of wrapped D3-branes
to the BPS partition function
by using the single-particle partition functions $I_C^{\rm D3}$
of wrapped D3-branes.
In this and the following sections we follow the same strategy
by using the superconformal indices of wrapped D3-branes instead of
the BPS partition functions.

\subsection{Superconformal algebra}
Let us first summarize some notations and conventions
associated with
the superconformal algebra and the superconformal index.

The four-dimensional ${\cal N}=4$ superconformal group $PSU(2,2|4)$
is generated by the $30$ bosonic generators
\begin{align}
H,\quad
J^{(+)a}{}_b,\quad
J^{(-)\dot a}{}_{\dot b},\quad
P^{\dot a}{}_b,\quad
K^a{}_{\dot b},\quad
R^I{}_J,
\end{align}
and $32$ fermionic generators
\begin{align}
Q^I_a,\quad
\ol Q_I^{\dot a},\quad
S_I^a,\quad
\ol S^I_{\dot a}.
\end{align}
$a,b,\ldots=1,2$ ($\dot a,\dot b,\ldots=\dot1,\dot2$) are
spin indices associated with the group $SU(2)_J^{(+)}$ ($SU(2)_J^{(-)}$) generated by $J^{(+)a}{}_b$ ($J^{(-)\dot a}{}_{\dot b}$).
$I,J,\ldots=1,2,3,4$ are $SU(4)_R$ indices.
The anti-commutation relations among the fermionic generators are
\begin{align}
\{S_I^a,Q_b^J\}&=\tfrac{1}{2}\delta^a_b\delta_I^JH+\delta^J_IJ^{(+)a}{}_b+\delta^a_b\left(R^J{}_I-\tfrac{1}{4}\delta^J_IR^K{}_K\right),\nonumber\\
\{\ol Q_I^{\dot a},\ol S_{\dot b}^J\}&=\tfrac{1}{2}\delta^{\dot a}_{\dot b}\delta_I^JH-\delta^J_IJ^{(-)\dot a}{}_{\dot b}
-\delta^{\dot a}_{\dot b}\left(R^J{}_I-\tfrac{1}{4}\delta^J_IR^K{}_K\right),\nonumber\\
\{S_I^a,\ol S_{\dot b}^J\}&=\delta^J_IK^a{}_{\dot b},\nonumber\\
\{\ol Q_I^{\dot a},Q_b^J\}&=\delta^J_IP^{\dot a}{}_b.
\label{n4algebra}
\end{align}
See Appendix \ref{algebra.sec} for other commutation relations.

The superconformal index ${\cal I}$ receives the contribution
from BPS operators that are annihilated by ${\cal Q}$, one of the supercharges
$Q$ and $\ol Q$.
We will later also consider the Schur limit of ${\cal I}$, the Schur index $\wh{\cal I}$.
This receives the contribution from so-called Schur operators, which are annihilated by
both ${\cal Q}$ and ${\cal Q}'$,
where ${\cal Q}'$ is another one of $Q$ and $\ol Q$ that has opposite
chirality to ${\cal Q}$ and anti-commutes with ${\cal Q}$.
We choose ${\cal Q}$ and ${\cal Q}'$ as follows.
\begin{align}
{\cal Q}=\ol Q^{\dot 1}_1,\quad
{\cal Q}'=Q^4_2.
\label{qqp}
\end{align}
To describe the BPS bounds associated with these supercharges
it is convenient to define
\begin{align}
\Delta\equiv
2\{{\cal Q}'^\dagger,{\cal Q}'\}
=2\{S^2_4,Q_2^4\}
=H-2J^{(+)1}{}_1+2R^4{}_4-\tfrac{1}{2}R^I{}_I,\nonumber\\
\ol\Delta\equiv
2\{{\cal Q}^\dagger,{\cal Q}\}
=2\{\ol S^1_{\dot1},\ol Q^{\dot1}_1\}
=H-2J^{(-)\dot1}{}_{\dot1}-2R^1{}_1+\tfrac{1}{2}R^I{}_I.
\end{align}
By definition these satisfy
\begin{align}
\Delta\geq0,\quad
\ol\Delta\geq0.
\label{bounds}
\end{align}

In order to define the orientifold and S-folds, we introduce an extra symmetry $U(1)_A$,
which rotates the eight supercharges $Q^I_a$ by the same angle%
\footnote{
In the context of string theory the precise definition of the $U(1)_A$ symmetry is as follows.
The scalar fields in type IIB supergravity can be regarded as coordinates
of the coset space $SL(2,\RR)/U(1)_R$.
In the coset approach of type IIB supergravity \cite{Schwarz:1983qr}
the scalar fields are written as a matrix $V\in SL(2,\RR)$.
$SL(2,\RR)$ and $U(1)_R$ symmetries act on $V$ on the left and on the right, respectively.
$V$ has three independent components and one of them is an auxiliary field
introduced to realize the symmetries linearly.
We can eliminate the auxiliary field by imposing appropriate gauge fixing condition for $U(1)_R$
so that only the physical axio-dilaton field $\tau$ are left.
As the result, $SL(2,\RR)\times U(1)_R$ is broken to a diagonal subgroup
of $U(1)\subset SL(2,\RR)$ and $U(1)_R$,
and we call the diagonal group $U(1)_A$.
The $SL(2,\RR)$ symmetry of the supergravity is broken in string theory
to $SL(2,\ZZ)$ due to the brane charge quantization, and correspondingly the
$U(1)_A$ symmetry is broken to a discrete subgroup depending on the value of $\tau$.
}.
We denote the generator of this symmetry by $A$.
In the ${\cal N}=4$ SYM $U(1)_A$ acts on the electric and magnetic charges
in a non-trivial manner, and is broken to discrete subgroup
due to the discreteness of the charges.
For a generic value of the complex coupling $\tau$ $U(1)_A$ is broken down to $\ZZ_4$,
which is generated by $C=\exp(\frac{\pi i}{2}A)$.
$C$ flips the sign of the gauge fields, and is nothing but the charge conjugation.
$C^2$ acts only on fermions and flip their signs.
Namely, $C^2=(-1)^F$.
The orientifold action is defined by combining this $\ZZ_4\subset U(1)_A$ and
the center of the R-symmetry $\ZZ_4\subset SU(4)_R$.
For special values of the coupling, $\tau=e^{\frac{2\pi i}{k}}$ with $k=3,4,6$,
the $U(1)_A$ symmetry is broken to $\ZZ_{2k}$ (or $\ZZ_k$ up to $(-1)^F$).
The $\ZZ_k$ S-fold is defined by combining this $\ZZ_{2k}\subset U(1)_A$ and
$\ZZ_{2k}\subset SU(4)_R$.

$\ZZ_4\subset U(1)_A$ and $\ZZ_4\subset SU(4)_R$
act in the same way on the supercharges and
as far as the action on the supercharges concerned
$SU(4)_R$ and $U(1)_A$ are combined into
$U(4)=(SU(4)_R\times U(1)_A)/\ZZ_4$.
However, two $\ZZ_4$ act differently on the vector multiplet,
and the two groups cannot be combined to $U(4)$.

We define the superconformal index by
\begin{align}
{\cal I}(q,y,u,v)=
\tr[(-1)^F\ol x^{\ol\Delta}
q^{H+J^{(+)1}{}_1+J^{(-)\dot1}{}_{\dot1}}y^{2J^{(+)1}{}_1}u^{R^2{}_2-R^3{}_3}v^{R^3{}_3-R^4{}_4}].
\label{defsc}
\end{align}
Because only operators with $\ol\Delta=0$ contribute to ${\cal I}$, ${\cal I}$ is independent of $\ol x$.
The factor $u^{R^2{}_2-R^3{}_3}v^{R^3{}_3-R^4{}_4}$ gives the character for the unbroken $SU(3)\subset SU(4)_R$.
$J^{(+)1}{}_1$ appears in (\ref{defsc}) in the form
$(yq^{\frac{1}{2}})^{2J^{(+)1}{}_1}$,
and it is convenient to use $\wt y=yq^{\frac{1}{2}}$ instead of $y$
to see the $SU(2)_J^{(+)}$ multiplet structure.
We use $y$ and $\wt y$ interchangeably.

As we will explain in subsection \ref{subsec.Schur} in detail,
the Schur index is obtained by setting $v=y=1$ in (\ref{defsc}).
Note that the fugacity $q$ for the Schur index
is denoted by $q^{\frac{1}{2}}$ in the standard
reference \cite{Gadde:2011uv}.

\subsection{${\cal N}=4$ Maxwell theory}
The ${\cal N}=4$ Maxwell theory
consists of a single free vector multiplet.
The component fields are
\begin{align}
F_{ab},\quad
\lambda_{Ia},\quad
\phi_{IJ},\quad
\ol\lambda^{I\dot a},\quad
F^{\dot a\dot b}.
\end{align}
The scalar field $\phi_{IJ}$ has two anti-symmetric $SU(4)_R$ indices,
and satisfies the reality condition
$(\phi_{IJ})^*=-\frac{1}{2}\epsilon^{IJKL}\phi_{KL}$.
Therefore, $\phi_{IJ}$ contains three independent complex scalar fields.
The supersymmetry
transformation rules
for the scalar fields are
up to numerical coefficients
\begin{align}
[Q^I_a,\phi_{JK}]\propto\delta^I_{[J}\lambda_{K]a},\quad
[\ol Q_I^{\dot a},\phi_{JK}]\propto\epsilon_{IJKL}\ol\lambda^{L\dot a}.
\label{deltaphi}
\end{align}
With the latter transformation rule we can show that
the following three scalar fields are ${\cal Q}$-closed,
and hence contribute to the superconformal index ${\cal I}$.
\begin{align}
\phi_{12}=\sX,\quad
\phi_{13}=\sY,\quad
\phi_{14}=\sZ.
\end{align}
With the first equation in (\ref{deltaphi}) we can show that
$\sX$ and $\sY$ also contribute to the Schur index $\wh{\cal I}$.

These fields and their derivatives
form the irreducible superconformal representation
${\cal B}_1$,
where ${\cal B}_n$ ($n=1,2,\ldots$) are the short representations
of the ${\cal N}=4$ superconformal algebra
denoted in \cite{Dolan:2002zh} by ${\cal B}^{\frac{1}{2},\frac{1}{2}}_{[0,n,0](0,0)}$.
The letter index, or, the single-particle index, of the
representation ${\cal B}_1$ is
\begin{align}
\sI_{{\cal B}_1}(q,y,u,v)&=\frac{q\chi_{(1,0)}(u,v)-(q^2y+qy^{-1})-q^2\chi_{(0,1)}(u,v)+2q^3}{(1-q^2y)(1-qy^{-1})},
\label{iforb1}
\end{align}
where
$\chi_{(r_1,r_2)}$ is the $SU(3)$ character of the representation with the Dynkin labels $(r_1,r_2)$ defined so that
\begin{align}
\chi_{(1,0)}=u+\frac{v}{u}+\frac{1}{v},\quad
\chi_{(0,1)}=\frac{1}{u}+\frac{u}{v}+v.
\label{su3fund}
\end{align}
See also Appendix \ref{characters.sec}.

Another way to obtain the single-particle index
(\ref{iforb1})
is to consider the Maxwell theory
in $\RR\times\bm{S}^3$
and perform the canonical quantization.
The action of the ${\cal N}=4$ $U(1)$ vector multiplet
in $\RR\times\bm{S}^3$ is
\begin{align}
{\cal L}_{{\cal N}=4}
=&-\frac{1}{4}F_{\mu\nu}F^{\mu\nu}
-(\ol\lambda^I\gamma^\mu\partial_\mu\lambda_I)
-\frac{1}{2}(|\partial_\mu\sX|^2+|\partial_\mu\sY|^2+|\partial_\mu\sZ|^2)
\nonumber\\&
-\frac{1}{2L^2}(|\sX|^2+|\sY|^2+|\sZ|^2)
\label{n4action}
\end{align}
where the potential terms in the second line are
the conformal coupling of the scalar fields
to the background curvature $R=6/L^2$.
$L$ is the radius of $\bm{S}^3$.
This action is invariant under the ${\cal N}=4$
supersymmetry transformations
\begin{align}
\delta A_\mu
&=
-(\ol\epsilon^I\gamma_\mu\lambda_I)
-(\epsilon_I\gamma_\mu\ol\lambda^I),\nonumber\\
\delta\lambda_I
&=\frac{1}{2}F_{\mu\nu}\gamma^{\mu\nu}\epsilon_I
+(\partial_\mu\phi_{IJ})\gamma^\mu\ol\epsilon^J
-2\kappa^J\phi_{IJ},\nonumber\\
\delta\ol\lambda^I
&=\frac{1}{2}F_{\mu\nu}\gamma^{\mu\nu}\ol\epsilon^I
+(\partial_\mu\phi^{*IJ})\gamma^\mu\epsilon_J
-2\ol\kappa_J\phi^{*IJ},\nonumber\\
\delta\phi_{IJ}
&=2((\epsilon_I\lambda_J)-(\epsilon_J\lambda_I))
+2\epsilon_{IJKL}(\ol\epsilon^K\ol\lambda^L),
\nonumber\\
\delta\phi^{*IJ}
&=2((\ol\epsilon^I\ol\lambda^J)-(\ol\epsilon^J\ol\lambda^I))
+2\epsilon^{IJKL}(\epsilon_K\lambda_L),
\label{susytransf}
\end{align}
where $\epsilon_I$, $\ol\epsilon^I$,
$\ol\kappa_I$, and $\kappa^I$ are spinors
satisfying the Killing spinor equations
\begin{align}
D_\mu\epsilon_I=\gamma_\mu\ol\kappa_I,\quad
D_\mu\ol\epsilon^I=\gamma_\mu\kappa^I.
\label{killings}
\end{align}

\subsection{${\cal N}=4$ $U(N)$ theory}

The index of the SYM with an arbitrary gauge group $G$
can be calculated by the localization formula
\begin{align}
{\cal I}_G(q,y,u,v)=\int d\mu \Pexp\left( \sI_{{\cal B}_1}(q,y,u,v)\chi_{\rm adj}(g)\right),
\label{locfm}
\end{align}
where $\int d\mu$ is the integration over $g\in G$ with the Haar measure
normalized by $\int d\mu 1=1$ and $\chi_{\rm adj}(g)$ is the character of the
adjoint representation of $G$.
Although we can in principle calculate the index
by this formula for an arbitrary gauge group $G$,
it becomes difficult to carry out the integral as the rank of $G$ increases.

For $U(N)$ gauge group
the integral in the large $N$ limit
was evaluated in \cite{Kinney:2005ej}
by using the saddle pint technique.
It was also confirmed in \cite{Kinney:2005ej} that
the result agrees with the index of Kaluza-Klein modes
in the $AdS_5\times\bm{S}^5$ background.
Namely, the superconformal index of the $U(N)$ ${\cal N}=4$ SYM in the large $N$ limit
is given by
\begin{align}
{\cal I}_{U(\infty)}=\Pexp\sI^{\rm KK}.
\end{align}
where $\sI^{\rm KK}$ is
the single-particle index for the Kaluza-Klein modes in $AdS_5\times\bm{S}^5$.
The Kaluza-Klein modes in $AdS_5\times\bm{S}^5$ form the superconformal
representation \cite{Kim:1985ez,Gunaydin:1984fk}
\begin{align}
\bigoplus_{n=1}^\infty{\cal B}_n,
\end{align}
and 
$\sI^{\rm KK}$ is the sum of the indices of ${\cal B}_n$:
\begin{align}
\sI^{\rm KK}=\sum_{n=1}^\infty\sI_{{\cal B}_n}
&=\frac{1}{1-qu}+\frac{1}{1-qv/u}+\frac{1}{1-q/v}-\frac{1}{1-q^2y}-\frac{1}{1-q/y}-1
\label{ikk1}
\end{align}

\subsection{S-fold projection}\label{sfold.sec}
As is mentioned in Section \ref{bps.sec}
the S-fold group $\ZZ_k$ is generated by
\begin{align}
\exp\left(\frac{2\pi i}{k}\left(S-\frac{A}{2}\right)\right),
\label{zkoperator}
\end{align}
where $S$ is an element of the Cartan subalgebra of $SU(4)_R$ such that $\exp(\pi iS)$
is the generating element of the center $\ZZ_4$ of $SU(4)_R$.
A choice of $S$ breaks $SU(4)_R$ into $SU(3)\times U(1)$,
and the four supercharges $\ol Q_{1,2,3,4}$ split into an $SU(3)$ triplet and an $SU(3)$ singlet.
For $k\geq 3$ the singlet supercharge is projected out, and the ${\cal N}=3$ supersymmetry is
realized.
For the definition of the superconformal index we want to keep $\ol Q_1$ unbroken,
and we have three choices for the eliminated supercharge: $\ol Q_2$, $\ol Q_3$, or $\ol Q_4$.
We denote $S$ corresponding to $\ol Q_2$, $\ol Q_3$, or $\ol Q_4$ by $S_I$, $S_{II}$, and $S_{III}$ respectively. 
They are given by
\begin{align}
S_I&=\tfrac{1}{2}(R^1{}_1-3R^2{}_2+R^3{}_3+R^4{}_4)=-R_x+R_y+R_z,\nonumber\\
S_{II}&=\tfrac{1}{2}(R^1{}_1+R^2{}_2-3R^3{}_3+R^4{}_4)=R_x-R_y+R_z,\nonumber\\
S_{III}&=\tfrac{1}{2}(R^1{}_1+R^2{}_2+R^3{}_3-3R^4{}_4)=R_x+R_y-R_z.
\label{sss}
\end{align}
When we discuss the Schur index, we also need to keep $Q^4$ preserved, and
only $S_I$ and $S_{II}$ are acceptable.

In the case of the orientifold with $k=2$
the $\ZZ_2$ generator (\ref{zkoperator}) is the same for all three choices for $S$,
and there is no difference among them.

Of course three $S$ in (\ref{sss}) are related by the unbroken $SU(3)$ symmetry and
essentially equivalent.
In section \ref{bps.sec} we adopted $S_I$ generating the S-fold action (\ref{omegampp}),
and worked in three different patches.
What we want to do here is the same.
However, working in different patches, or,
with three different wrapped D3-brane configurations, is troublesome
because different configurations preserve
different subgroups of the superconformal group,
and
we need to use different bases of superconformal generators
depending on the configuration we analyze.
To avoid this trouble
we fix a working patch,
and use different S-foldings.
After calculation we translate the results
to those for fixed S-folding by using $SU(3)$ transformations.
We choose the patch corresponding to the D3-brane configuration $X=0$ as the working patch.

The single-particle index for the Kaluza-Klein modes
in $AdS_5\times\bm{S}^5/\ZZ_k$ is obtained from that of $AdS_5\times\bm{S}^5$
by the $\ZZ_k$ projection \cite{Imamura:2016abe}.
For this purpose we first refine the single-particle index
by inserting $\eta^{S-\frac{A}{2}}$ in the definition of the single-particle index.

\begin{align}
\sI(q,y,u,v,\eta)=
\tr[(-1)^F\ol x^{\ol\Delta}
q^{H+J^{(+)1}{}_1+J^{(-)\dot1}{}_{\dot1}}y^{2J^{(+)}}u^{R^2{}_2-R^3{}_3}v^{R^3{}_3-R^4{}_4}
\eta^{S-\frac{1}{2}A}].
\label{defscdefined}
\end{align}

The refined single-particle index of the vector multiplet
is
\begin{align}
\sI_{{\cal B}_1}
&=\frac{
q\eta\chi'_{(1,0)}
-q^2\eta^{-1}\chi'_{(0,1)}
-q^2(y+q^{-1}y^{-1})\eta
+q^3(\eta+\eta^{-1})
}{(1-q^2y)(1-qy^{-1})},
\label{ib1refined}
\end{align}
where $\chi'_{(1,0)}$ and $\chi'_{(0,1)}$
depend on the choice of $S$.
For each choice they are given by
\begin{align}
(S&=S_I)&
\chi'_{(1,0)}&=u\eta^{-2}+u^{-1}v+v^{-1},&
\chi'_{(0,1)}&=u^{-1}\eta^2+uv^{-1}+v,\nonumber\\
(S&=S_{II})&
\chi'_{(1,0)}&=u+u^{-1}v\eta^{-2}+v^{-1},&
\chi'_{(0,1)}&=u^{-1}+uv^{-1}\eta^2+v,\nonumber\\
(S&=S_{III})&
\chi'_{(1,0)}&=u+u^{-1}v+v^{-1}\eta^{-2},&
\chi'_{(0,1)}&=u^{-1}+uv^{-1}+v\eta^2.
\label{chiprime}
\end{align}
These satisfy the relations
\begin{align}
\sigma_{23}\chi'^{(S_I)}_{(r_1,r_2)}=\chi'^{(S_I)}_{(r_1,r_2)},\quad
\sigma_{123}\chi'^{(S_I)}_{(r_1,r_2)}=\chi'^{(S_{II})}_{(r_1,r_2)},\quad
\sigma_{123}^2\chi'^{(S_I)}_{(r_1,r_2)}=\chi'^{(S_{III})}_{(r_1,r_2)},
\end{align}
where the permutation operators act on $(u,\frac{v}{u},\frac{1}{v})$,
the three terms in $\chi_{(1,0)}(u,v)$, instead of $(x,y,z)$.
For $\eta=\pm1$ $\chi'_{(1,0)}$ and $\chi'_{(0,1)}$ reduce to the
$SU(3)$ characters (\ref{su3fund}).

$\sI_{{\cal B}_1}$ splits into two parts with different $\eta$ dependence.
For example, for $S=S_I$,
\begin{align}
\sI^{(S_I)}_{{\cal B}_1}
&=
\frac{
\begin{array}{l}
q(v^{-1}+u^{-1}v)
-q^2u^{-1}
\\
-q^2(y+q^{-1}y^{-1})
+q^3
\end{array}
}{(1-q^2y)(1-qy^{-1})}\eta
+\frac{
\begin{array}{l}
q u
-q^2(v+uv^{-1})
\\
+q^3
\end{array}
}{(1-q^2y)(1-qy^{-1})}
\eta^{-1}.
\end{align}
This splitting implies that the ${\cal N}=4$ vector multiplet consists of
two irreducible ${\cal N}=3$ superconformal multiplets.

The refined single-particle index for Kaluza-Klein modes is
\begin{align}
\sI^{\rm KK}
&=\frac{
\begin{array}{l}
(q\eta-q^4\eta^2)\chi'_{(1,0)}
-q^2(1+\eta^{-1})\chi'_{(0,1)}\\
+q^3(1+\eta^{-1}+2\eta+\eta^2)+q^6\eta\\
-(q^2\eta+q^5(1+\eta))(y+q^{-1}y^{-1})\\
+q^4(y+q^{-1}y^{-1})\chi'_{(0,1)}
\end{array}
}{(1-q^2y)(1-qy^{-1})}
\Pexp(q\chi'_{(1,0)}\eta)
\label{refinedkk}
\end{align}
The bulk single-particle index for $\ZZ_k$ S-fold is defined by
${\cal P}_k\sI^{\rm KK}$ where ${\cal P}_k$ is the projection operator defined by
\begin{align}
{\cal P}_kf
=\frac{1}{k}\sum_{i=0}^{k-1}f|_{\eta=\omega_k^i}.
\end{align}
The bulk contribution to the index ${\cal I}$, which is the same as the large $N$ limit, is
\begin{align}
{\cal I}^{\rm KK}\equiv\Pexp({\cal P}_k\sI^{\rm KK}).
\end{align}
For S-fold theories with non-trivial discrete torsion,
$S(k,N,1)$, there is no correction from wrapped D3-branes
and ${\cal I}^{\rm KK}$ gives
the full index up to the correction of ${\cal O}(q^{kN})$,
which we are not interested in.
\begin{align}
{\cal I}_{S(k,N,1)}\approx
{\cal I}^{\rm AdS}_{S(k,N,1)}
&={\cal I}^{\rm KK}.
\label{igravsokk}
\end{align}
We use the notation ${\cal I}^{\rm AdS}_{S(k,N,p)}$ for the
results on the AdS side.
In the next section we will give a formula
for ${\cal I}^{\rm AdS}_{S(k,N,0)}$
by taking account of wrapped D3-branes
with $|m|=1$.

\section{D3-brane analysis}\label{d3.sec}
In this section we analyze the fluctuations of a wrapped D3-brane.
We take the same strategy as in section \ref{bps.sec}.
Namely, we first calculate $\sI_C^{\rm D3}$,
the refined single-particle index of the fluctuation of a D3-brane
over $C$.
$C$ is one of the three loci $X=0$, $Y=0$, and $Z=0$.
And then, we calculate the multi-particle index
by the relation
\begin{align}
{\cal I}_C^{\rm D3}=(\sI_C^{\rm gr})^N\Pexp({\cal P}_k\sI^{\rm D3}_C)
\end{align}
where
$(\sI_C^{\rm gr})^N$ is the classical contribution of the wrapped D3-brane to the index.
For each $C$ it is given by
\begin{align}
\sI^{\rm gr}_{X=0}=qu,\quad
\sI^{\rm gr}_{Y=0}=q\frac{v}{u},\quad
\sI^{\rm gr}_{Z=0}=q\frac{1}{v}.
\label{igrxyz2}
\end{align} 
These are essentially the same as
the factors in (\ref{igrxyz}).
It is of course possible to directly derive them from
the D3-brane action.
The energy of wrapped D3-brane $E$ is the product of
the D3-brane tension $T_{\rm D3}$ and the volume of the three-cycle
${\rm Vol}(S^3/\mathbb{Z}_k)=2\pi L^3/k$.
We can easily show
\begin{align}
LE=T_{\rm D3}\times\frac{2\pi L^4}{k}=N,
\end{align}
and this does not change even if
we treat the D3-brane collective motions
quantum-mechanically \cite{Berenstein:2002ke}.
We can also obtain the R-charges
satisfying the BPS relation (\ref{ppp})
by analysing the coupling of the worldvolume to
the background R-R flux \cite{McGreevy:2000cw}.
We obtain the classical contribution $(\sI^{\rm gr}_C)^N$ to the index
by substituting these quantities to the
defining equation (\ref{defsc}).

Similarly to (\ref{zxyz}),
the total contribution of wrapped D3-branes with $|m|=1$ is given by
\begin{align}
{\cal I}^{\rm D3}_{X=0}
+{\cal I}^{\rm D3}_{Y=0}
+{\cal I}^{\rm D3}_{Z=0}.
\end{align}

\subsection{Unbroken subalgebra}
As we noted in subsection \ref{sfold.sec}
we carry out the analysis
using a D3-brane wrapped around the locus $X=0$,
which corresponds to the
Pfaffian-like operator ${\cal O}_X$ carrying
the quantum numbers $(R_x,R_y,R_z)=(N,0,0)$.
Let us first determine
the subgroup of the superconformal group
which is preserved by the
state with ${\cal O}_X$ insertion.

The insertion of ${\cal O}_X$ breaks the R symmetry $SU(4)_R$
to $SU(2)\times SU(2)\times U(1)$.
As a result an $SU(4)_R$ quartet like the gaugino $\lambda_I$ ($I=1,2,3,4$)
is split up into two doublets $\lambda_i$ ($i=3,4$) and $\lambda_{\ol i}$ ($\ol i=1,2$).
We denote the $SU(2)$ acting on the former and the latter by $SU(2)_R^{(+)}$
and $SU(2)_R^{(-)}$, respectively.
We define generators of these two $SU(2)$ groups as follows.
\begin{align}
R^{(+)i}{}_j=R^i{}_j-\frac{1}{2}\delta^i_jR^k{}_k,\quad
R^{(-)\ol i}{}_{\ol j}=R^{\ol i}{}_{\ol j}-\frac{1}{2}\delta^{\ol i}_{\ol j}R^{\ol k}{}_{\ol k}.
\end{align}
We denote the $U(1)$ factor by $U(1)_R$, and define the generator $R$ by
\begin{align}
R=\frac{1}{2}(-R^i{}_i+R^{\ol i}{}_{\ol i}).
\end{align}
The eigenvalues of $R$ for the component fields are shown in Table \ref{rqn.tbl}.
\begin{table}[htb]
\caption{$U(1)_R$ quantum numbers of the component fields are shown.}\label{rqn.tbl}
\centering
\begin{tabular}{ccccccc}
\hline
\hline
    & $\sX$  & $\sY$ & $\sZ$ & $A_\mu$ & $\lambda_i$ & $\lambda_{\ol i}$ \\
\hline
$R$ & $+1$ & $0$ & $0$ & $0$     & $-\frac{1}{2}$ & $+\frac{1}{2}$ \\
\hline
\end{tabular}
\end{table}

The insertion of ${\cal O}_X$ at the origin in the spacetime
breaks the conformal symmetry $SU(2,2)$ to $U(1)_H\times SU(2)_J^{(+)}\times SU(2)_J^{(-)}$
where $U(1)_H$, $SU(2)_J^{(+)}$, and $SU(2)_J^{(-)}$ are groups
generated by $H$, $J^{(+)a}{}_b$, and $J^{(-)\dot a}{}_{\dot b}$, respectively.
Although the conformal boosts $K^a{}_{\dot b}$ annihilate the conformal primary operator ${\cal O}_X$ inserted at the origin,
we do not include them in the list of unbroken generators because they do not act linearly on excitations.

In summary, the unbroken bosonic symmetry is generated by the following generators.
\begin{align}
H,\quad
J^{(+)a}{}_b,\quad
J^{(-)\dot a}{}_{\dot b},\quad
R^{(+)i}{}_j,\quad
R^{(-)\ol i}{}_{\ol j},\quad
R.
\label{bosonicgens}
\end{align}

The generators of the preserved supersymmetry are obtained
by picking up $16$ out of $32$ so that the anti-commutators
among them contain only generators in (\ref{bosonicgens}).
There are two possibilities.
One possible set of unbroken supercharges is the operators with $H-R=0$,
and the other is the operators with $H+R=0$.
With our convention of the supersymmetry transformation,
the supercharges that annihilates the operator ${\cal O}_X$ are those with $H-R=0$.
\begin{align}
S_i^a,\quad
Q_a^i,\quad
\ol S^{\ol i}_{\dot a},\quad
\ol Q_{\ol i}^{\dot a}.
\label{unbroken16}
\end{align}
It is important that
both ${\cal Q}$ and ${\cal Q}'$ in (\ref{qqp}) are contained in (\ref{unbroken16}).
The non-vanishing anti-commutators among supercharges in (\ref{unbroken16}) are
\begin{align}
\{S_i^a,Q_b^j\}&=\tfrac{1}{2}\delta^a_b\delta_i^j(H-R)+\delta^j_iJ^{(+)a}{}_b+\delta^a_bR^{(+)j}{}_i,\nonumber\\
\{\ol Q_{\ol i}^{\dot a},\ol S_{\dot b}^{\ol j}\}&=\tfrac{1}{2}\delta^{\dot a}_{\dot b}\delta_{\ol i}^{\ol j}(H-R)
-\delta^{\ol j}_{\ol i}J^{(-)\dot a}{}_{\dot b}
-\delta^{\dot a}_{\dot b}R^{(-)\ol j}{}_{\ol i}.
\end{align}
The bosonic generators (\ref{bosonicgens}) and the fermionic ones
(\ref{unbroken16}) generate
two copies of $SU(2|2)$ with
the common central generator $H-R$.
This is the group used in \cite{Hofman:2006xt}
to derive the dispersion relation of magnons in the spin chain associated with
the ${\cal N}=4$ SYM.

\subsection{D3-brane action}\label{d3action.sec}
In principle, the supersymmetric action of fields on the D3-brane
can be directly read off from the supersymmetric D3-brane action,
which is the sum of the Born-Infeld and the Chern-Simons actions
in the background superspace \cite{Bergshoeff:1996tu}.
Instead, we take an easier way.
We first read off the bosonic part of the action
from the bosonic Born-Indeld and Chern-Simons actions
and complete it so that it becomes invariant under
the unbroken supersymmetry generated by (\ref{unbroken16}).

The field theory on the D3-brane wrapped over $X=0$
is the four-dimensional ${\cal N}=4$ supersymmetric
theory just like the boundary CFT.
We use the same notation for the fields living on the wrapped D3-brane.
We should emphasize that they have no direct relation
to the fields in the boundary theory appearing the previous sections.

Bosonic fields on the D3-brane are one $U(1)$ gauge field $A_\mu$
and three complex scalar fields $\sX$, $\sY$, and $\sZ$.

For the gauge field we obtain the kinetic term from the Born-Infeld action,
and there are no other terms.
The scalar fields corresponds to six directions transverse to the D3-brane worldvolume.
We assume that $\sX$ is associated with the fluctuation in $\bm{S}^5$,
and the other two, $\sY$ and $\sZ$, correspond to the four directions in $AdS_5$.
The quadratic Lagrangian of $\sX$ has been already given in (\ref{bics}).
The Lagrangian of $\sY$ and $\sZ$ read off from the Born-Infeld action is
\begin{align}
{\cal L}=-\frac{1}{2}|\partial_\mu\sY|^2-\frac{1}{2}|\partial_\mu\sZ|^2-\frac{1}{2L^2}(|\sY|^2+|\sZ|^2).
\end{align}
The potential terms for $\sY$ and $\sZ$ agree with the curvature couplings in (\ref{n4action}),
and hence the bosonic part of the D3-brane Lagrangian differs from the bosonic part of the ${\cal N}=4$
supersymmetric Lagrangian (\ref{n4action}) only by the terms shown below.
\begin{align}
{\cal L}_{\rm D3}^{({\rm bos})}
={\cal L}_{{\cal N}=4}^{({\rm bos})}+\frac{2}{L^2}|\sX|^2+\frac{i}{L}(\sX^*\dot\sX-\dot\sX^*\sX).
\label{ld3}
\end{align}

Let us complete the Lagrangian (\ref{ld3}) so that
it becomes invariant under (\ref{unbroken16}).
Notice that the additional terms in (\ref{ld3})
can be absorbed by
the field redefinition
\begin{align}
\sX'=e^{-\frac{2it}{L}}\sX.
\label{zrepl}
\end{align}
We regard the phase factor in
(\ref{zrepl}) as the time-dependent $U(1)_R$ rotation,
and we replace all fields by
\begin{align}
\Phi'=e^{-\frac{2it}{L}R}\Phi.
\label{phir}
\end{align}
Note that the only bosonic field carrying the non-vanishing $U(1)_R$ charge
is $\sX$.

The supersymmetric completion is easily obtained by
this field redefinition from the ${\cal N}=4$ supersymmetric action on $\bm{S}^3\times\RR$.
\begin{align}
{\cal L}_{\rm D3}={\cal L}'_{{\cal N}=4},
\end{align}
where ${\cal L}'_{{\cal N}=4}$ is the action (\ref{n4action}) with all fields
replaced by the primed ones defined by (\ref{phir}).
${\cal L}_{\rm D3}$ in terms of the original variables is
\begin{align}
{\cal L}_{\rm D3}
=&-\frac{1}{4}F_{\mu\nu}F^{\mu\nu}-(\ol\lambda^I\gamma^\mu\partial_\mu\lambda_I)
-\frac{1}{2}(|\partial_\mu\sX|^2+|\partial_\mu\sY|^2+|\partial_\mu\sZ|^2)
\nonumber\\
&+\frac{1}{2L^2}(3|\sX|^2-|\sY|^2-|\sZ|^2)
+\frac{i}{L}(\sX^*\dot\sX-\dot\sX^*\sX)
+\frac{i}{L}(\ol\lambda^i\gamma^0\lambda_i-\ol\lambda^{\ol i}\gamma^0\lambda_{\ol i}).
\end{align}

Before
ending this subsection, let us comment on the Killing spinor equations.
Let $\epsilon_I'$ and $\ol\epsilon'^I$ be the Killing spinors
on the wrapped D3-brane
satisfying the Killing spinor equations (\ref{killings}).
The Lagrangian ${\cal L}'_{{\cal N}=4}$ is invariant under the
supersymmetry transformation (\ref{susytransf})
with all fields and killing spinors replaced by primed ones.
Let $\epsilon_I$ and $\ol\epsilon^I$ be the spinors related to
the primed ones by the phase rotation (\ref{phir}).
Because Killing spinors get transformed under the $U(1)_R$ rotation (\ref{phir})
$\epsilon_I$ and $\ol\epsilon^I$
satisfy the equations
different from the Killing spinor equations (\ref{killings}):
\begin{align}
D_i\epsilon_I=\gamma_i\ol\kappa_I,\quad
D_0\epsilon_I=-\gamma_0\ol\kappa_I,\quad
D_i\ol\epsilon^I=\gamma_i\kappa^I,\quad
D_0\ol\epsilon^I=-\gamma_0\kappa^I.
\label{killingm}
\end{align}
In fact, the Killing spinor equations satisfied by
supersymmetry parameters on the D3-brane
worldvolume are not (\ref{killings})
but (\ref{killingm}).
The reason is as follows.

The Killing spinor equations in the gravity background arise from the
requirement of vanishing of the supersymmetry transformation of the gravitino.
\begin{align}
0=\delta\psi_M=D_M\xi-\frac{ig_s}{16\cdot 5!}F_{N_1\cdots N_5}\Gamma^{N_1\cdots N_5}\Gamma_M\xi.
\end{align}
In the $AdS_5\times\bm{S}^5$ background this reduces to
\begin{align}
D_M\xi=\frac{i}{2L}\Gamma_{AdS}\Gamma_M\xi
\end{align}
where $\Gamma_{AdS}$ is the product of five ten-dimensional Dirac matrices along the AdS directions.
For the $AdS_5$ component $\psi_\mu$ and $\bm{S}^5$ component $\psi_\alpha$ this can be rewritten as
\begin{align}
D_\alpha\xi=\Gamma_\alpha\left(-\frac{i}{2L}\Gamma_{AdS}\xi\right),\quad
D_\mu\xi=-\Gamma_\mu\left(-\frac{i}{2L}\Gamma_{AdS}\xi\right).
\label{dadm}
\end{align}
Two equations in (\ref{dadm}) have the opposite signatures on the right hand side,
and have a similar structure to (\ref{killingm}).
The wrapped D3-brane breaks half of supersymmetry.
By decomposing the unbroken part of $\xi$ into eight two-component spinors $\epsilon_I$
and $\ol\epsilon^I$, we obtain the equations in (\ref{killingm}).

\subsection{Superconformal symmetry on D3}
The action ${\cal L}_{{\cal N}=4}'$
is invariant under the ${\cal N}=4$ superconformal algebra.
We should not confuse this algebra with the algebra of the
boundary CFT.
Let us denote the generators of this ${\cal N}=4$ supersymmetry of ${\cal L}'_{{\cal N}=4}$
by primed ones.
\begin{align}
H',\quad
J'^{(+)a}{}_b,\quad
J'^{(-)\dot a}{}_{\dot b},\quad
P'^{\dot a}{}_b,\quad
K'^a{}_{\dot b},\quad
R'^I{}_J,\quad
Q'^I_a,\quad
S'^a_I,\quad
\ol Q'^{\dot a}_I,\quad
\ol S'^I_{\dot a}.
\label{fulld3sym}
\end{align}
Although this is the symmetry
of the quadratic action ${\cal L}'_{{\cal N}=4}$,
the full D3-brane action is invariant under a subgroup
of this algebra.
In particular, the R-symmetry $su(4)_R'$ mixing the three scalars
$\sY'$, $\sZ'$, and $\sX'$ is broken to
$su(2)'^{(+)}_R\times su(2)'^{(-)}_R\times u(1)'_R$
because $\sX'$ and $(\sY',\sZ')$ describe fluctuations along $\bm{S}^5$ and $AdS_5$ directions,
respectively, and there are no symmetries among them.
By the same reason, $P'^{\dot a}{}_b$ and
$K'^a{}_{\dot b}$ are not true symmetries.
We have the bosonic symmetry generated by
\begin{align}
H',\quad
R',\quad
J'^{(+)a}{}_b,\quad
J'^{(-)\dot a}{}_{\dot b},\quad
R'^{(+)i}{}_j,\quad
R'^{(-)\ol i}{}_{\ol j}.
\label{d3bosonicgen}
\end{align}
For consistency only a half of the $32$ supercharges can be
true symmetries of the D3-brane system.
Again, we have two possibilities, and the choice of one from them
is a matter of convention.
We adopt the choice such that supercharges with $H'-R'=0$
\begin{align}
S'^a_i,\quad
Q'^i_a,\quad
\ol S'^{\ol i}_{\dot a},\quad
\ol Q'^{\dot a}_{\ol i}
\label{unbroken162}
\end{align}
are unbroken.

Let us establish the correspondence between
the generators (\ref{bosonicgens}) and (\ref{unbroken16}) of the
boundary CFT and
the generators
(\ref{d3bosonicgen}) and (\ref{unbroken162}) for the
wrapped D3-brane.

$su(2)'^{(+)}_J\times su(2)'^{(-)}_J$ is the isometry group
of the D3-brane worldvolume.
In the context of the boundary CFT, this corresponds to
the R-symmetry group $su(2)_R^{(+)}\times su(2)_R^{(-)}$.
Conversely,
$su(2)'^{(+)}_R\times su(2)'^{(-)}_R$ is
the isometry in the $AdS_5$, and identified with
$su(2)^{(+)}_J\times su(2)^{(-)}_J$.
We assume the following identifications.
\begin{align}
J^{(+)a}{}_b=R'^{(+)i}{}_j,\quad
J^{(-)\dot a}{}_{\dot b}=R'^{(-)\ol i}{}_{\ol j},\quad
R^{(+)i}{}_j=J'^{(+)a}{}_b,\quad
R^{(-)\ol i}{}_{\ol j}=J'^{(-)\dot a}{}_{\dot b}.
\label{su2map}
\end{align}
These are not the unique choice.
We have some ambiguity associated with the automorphism of
each algebra.
We adopt (\ref{su2map}) for later convenience.

On the two sides of each relation in (\ref{su2map})
generators have different indices,
and an appropriate translation
should be understood.
For example, in the first relation
$a=1,2$ and $b=1,2$ on the left hand side
should be replaced on the right hand side by
$i=3,4$ and $j=3,4$, respectively.

The field redefinition
(\ref{phir}) implies the following relation for the Hamiltonians.
\begin{align}
H'=H-2R.
\label{handhr}
\end{align}
Because unbroken supercharges (\ref{unbroken16}) carry $R=H$,
the relation $H'=-H$ holds for the unbroken supercharges.
This means that
the supercharges 
$(Q,\ol Q)$ with $H=+1/2$ and $(S,\ol S)$ with $H=-1/2$
should correspond to
$(S',\ol S')$ with $H'=-1/2$ and $(Q',\ol Q')$ with $H'=+1/2$, respectively.
The consistency to the relations
in (\ref{su2map}) fixes the correspondence to be
\begin{align}
Q^i_a=S'^a_i,\quad
S^a_i=Q'^i_a,\quad
\ol Q^{\dot a}_{\ol i}=\ol S'^{\ol i}_{\dot a},\quad
\ol S^{\ol i}_{\dot a}=\ol Q'^{\dot a}_{\ol i}.
\label{qqrelation}
\end{align}
For consistency to the relations in (\ref{qqrelation})
\begin{align}
R'=-R,\quad
A'=-A.
\label{rryy}
\end{align}

\subsection{D3-brane contribution to the index}
Now we are ready to calculate the single-particle index
for excitations
on a D3-brane wrapped on $\bm{S}^3/\ZZ_k$.
This is obtained from the refined single-particle index
\begin{align}
\sI^{\rm D3}(q,y,u,v,\eta)
=\tr[(-1)^F\ol x^{\ol\Delta}
q^{H+J^{(+)1}{}_1+J^{(-)\dot1}{}_{\dot1}}y^{2J^{(+)1}{}_1}u^{R^2{}_2-R^3{}_3}v^{R^3{}_3-R^4{}_4}
\eta^{S-\frac{A}{2}}].
\label{sid3}
\end{align}
by the $\ZZ_k$ projection.
The trace in (\ref{sid3}) is taken over excitations
on the D3-brane wrapped on $\bm{S}^3$.
The single-particle index for $\bm{S}^3/\ZZ_k$ is
${\cal P}_k\sI^{\rm D3}$.

As we explained in subsection \ref{d3action.sec}
the theory on the D3-brane is isomorphic to the
standard ${\cal N}=4$ Maxwell theory via the field redefinition (\ref{phir}).
Therefore, the refined index (\ref{sid3}) can be rewritten
by using the operator relations (\ref{su2map}),
(\ref{handhr}), (\ref{qqrelation}), and (\ref{rryy}) as
\begin{align}
\sI^{\rm D3}(q,y,u,v,\eta)
&=
\tr_{{\cal N}=4}[(-1)^F\ol x^{\ol\Delta'}
q^{H'-2R'+R'^{(+)3}{}_3+R'^{(-)1}{}_1}
y^{2R'^{(+)3}{}_3}\times
\nonumber\\&\hspace{8em}
u^{-R'-J'^{(+)1}{}_1-J'^{(-)\dot1}{}_{\dot1}}
v^{2J'^{(+)1}{}_1}
\eta^{S+\frac{1}{2}A'}],
\label{id3eq3}
\end{align}
where $S$ is one of
\begin{align}
S_I&
=2R^{(-)1}{}_1-R
=2J'^{(-)\dot 1}{}_{\dot 1}+R'
,\nonumber\\
S_{II}&
=-2R^{(+)3}{}_3+R
=-2J'^{(+)1}{}_1-R'
,\nonumber\\
S_{III}&
=2R^{(+)3}{}_3+R
=2J'^{(+)1}{}_1-R'
.
\label{s123def}
\end{align}
In (\ref{id3eq3}) the trace is taken over the Fock space
of the standard ${\cal N}=4$ theory
in $\RR\times\bm{S}^3$.
Because the Cartan generators appearing in (\ref{id3eq3})
are the same as those appearing in the
definition of the refined index,
we can rewrite (\ref{id3eq3}) in the same form as the
definition of the index (\ref{defscdefined}) by a variable change of fugacities.
\begin{align}
\sI^{\rm D3}(q,y,u,v,\eta)
&=
\tr_{{\cal N}=4}
[(-1)^F\ol x'^{\ol\Delta'}
q'^{H'+J'^{(+)1}{}_1+J'^{(-)\dot1}{}_{\dot1}}
y'^{2J^{(+)1}{}_1}\times
\nonumber\\&\hspace{8em}
u'^{R'-R'^{(+)3}{}_3-R'^{(-)1}{}_1}
v'^{2R'^{(+)3}{}_3}
\eta'^{S'-\frac{1}{2}A'}]
\nonumber\\
&=\sI_{{\cal N}=4}(q',y',u',v',\eta'),
\label{id3eq4}
\end{align}
where $S'$ is one of $S_I'$, $S_{II}'$, and $S_{III}'$ defined
by
\begin{align}
S'_I&=2R'^{(-)1}{}_1-R',&
S'_{II}&=-2R'^{(+)3}{}_3+R',&
S'_{III}&=2R'^{(+)3}{}_3+R'.
\label{sp123def}
\end{align}
The variable changes to rewrite
(\ref{id3eq3}) to (\ref{id3eq4}) for three choices of $S$
are
\begin{align}
(S&=S_I) & (S&=S_{II}) & (S&=S_{III}) \nonumber\\
\ol x'&=\ol xq^{\frac{1}{3}}u^{\frac{1}{3}}\eta^{-\frac{2}{3}} &
\ol x'&=\ol xq^{\frac{1}{3}}u^{\frac{1}{3}} &
\ol x'&=\ol xq^{\frac{1}{3}}u^{\frac{1}{3}} \nonumber\\
q'&=q^{\frac{2}{3}}u^{-\frac{1}{3}}\eta^{\frac{2}{3}} &
q'&=q^{\frac{2}{3}}u^{-\frac{1}{3}} &
q'&=q^{\frac{2}{3}}u^{-\frac{1}{3}} \nonumber\\
y'&=q^{-\frac{1}{3}}u^{-\frac{1}{3}}v\eta^{-\frac{1}{3}} &
y'&=q^{-\frac{1}{3}}u^{-\frac{1}{3}}v\eta^{-1} &
y'&=q^{-\frac{1}{3}}u^{-\frac{1}{3}}v\eta \nonumber\\
u'&=q^{-\frac{5}{3}}u^{-\frac{2}{3}}\eta^{-\frac{2}{3}} &
u'&=q^{-\frac{5}{3}}u^{-\frac{2}{3}} &
u'&=q^{-\frac{5}{3}}u^{-\frac{2}{3}} \nonumber\\
v'&=q^{-\frac{1}{3}}yu^{-\frac{1}{3}}\eta^{-\frac{1}{3}} &
v'&=q^{-\frac{1}{3}}yu^{-\frac{1}{3}}\eta^{-1} &
v'&=q^{-\frac{1}{3}}yu^{-\frac{1}{3}}\eta \nonumber\\
\eta'&=\eta^{-1} &
\eta'&=\eta^{-1} &
\eta'&=\eta^{-1}
\end{align}

By applying these variable changes to $\sI_{{\cal B}_1}$ in (\ref{ib1refined}) we obtain
\begin{align}
\sI^{(S_I){\rm D3}}_{X=0}&=
\frac{
\frac{\eta}{qu}
-q(y+\frac{1}{qy})\frac{\eta^2}{u}
-q(\frac{1}{v}+\frac{v}{u})
+q^2(y+\frac{1}{qy})
+q^2\frac{1}{u}(\eta+\eta^3)
-q^3\eta
}{(1-q\frac{\eta}{v})(1-q\frac{v\eta}{u})},
\nonumber\\
\sI^{(S_{II}){\rm D3}}_{X=0}&=
\frac{
\frac{1}{qu\eta}
-q(y+\frac{1}{qy})\frac{1}{u}
-q(\frac{1}{v}+\frac{v}{u\eta^2})
+q^2(y+\frac{1}{qy})
+q^2\frac{1}{u}(\frac{1}{\eta}+\eta)
-q^3\eta
}{(1-q\frac{\eta}{v})(1-q\frac{v}{u\eta})},
\nonumber\\
\sI^{(S_{III}){\rm D3}}_{X=0}&=
\frac{
\frac{1}{qu\eta}
-q(y+\frac{1}{qy})\frac{1}{u}
-q(\frac{1}{v\eta^2}+\frac{v}{u})
+q^2(y+\frac{1}{qy})
+q^2\frac{1}{u}(\frac{1}{\eta}+\eta)
-q^3\eta
}{(1-q\frac{1}{v\eta})(1-q\frac{v\eta}{u})}.
\end{align}
In the equations above we explicitly show that these are indices for the
fluctuations around the brane configuration $X=0$.

Now, let us translate these results
to the indices for different patches with a fixed S-folding with $S_I$.
We take $\sI^{(S_I){\rm D3}}_{X=0}$ as is.
$\sI^{(S_I){\rm D3}}_{Y=0}$ and
$\sI^{(S_I){\rm D3}}_{Z=0}$
are obtained from
$\sI^{(S_{II}){\rm D3}}_{X=0}$
and
$\sI^{(S_{III}){\rm D3}}_{X=0}$ by the Weyl reflections:
\begin{align}
\sI^{(S_I){\rm D3}}_{Y=0}=\sigma_{12}\sI^{(S_{II}){\rm D3}}_{X=0},\quad
\sI^{(S_I){\rm D3}}_{Z=0}=\sigma_{13}\sI^{(S_{III}){\rm D3}}_{X=0}.
\end{align}
The results are
\begin{align}
\sI^{(S_I){\rm D3}}_{X=0}&=
\frac{
\frac{\eta}{qu}
-q(y+\frac{1}{qy})\frac{\eta^2}{u}
-q(\frac{1}{v}+\frac{v}{u})
+q^2(y+\frac{1}{qy})
+q^2\frac{1}{u}(\eta+\eta^3)
-q^3\eta
}{(1-q\frac{\eta}{v})(1-q\frac{v\eta}{u})},
\nonumber\\
\sI^{(S_I){\rm D3}}_{Y=0}&=
\frac{
\frac{u}{qv\eta}
-q(y+\frac{1}{qy})\frac{u}{v}
-q(\frac{1}{v}+\frac{u}{\eta^2})
+q^2(y+\frac{1}{qy})
+q^2\frac{u}{v}(\frac{1}{\eta}+\eta)
-q^3\eta
}{(1-q\frac{\eta}{v})(1-q\frac{u}{\eta})},
\nonumber\\
\sI^{(S_I){\rm D3}}_{Z=0}&=
\frac{
\frac{v}{q\eta}
-q(y+\frac{1}{qy})v
-q(\frac{u}{\eta^2}+\frac{v}{u})
+q^2(y+\frac{1}{qy})
+q^2v(\frac{1}{\eta}+\eta)
-q^3\eta
}{(1-q\frac{u}{\eta})(1-q\frac{v\eta}{u})}.
\label{threeid3}
\end{align}
The multi-particle indices for three loci are given by
\begin{align}
{\cal I}^{(S_I)\rm D3}_{X=0}&=q^Nu^N\Pexp({\cal P}_k\sI_{X=0}^{(S_I)\rm D3}),\nonumber\\
{\cal I}^{(S_I)\rm D3}_{Y=0}&=q^N\frac{v^N}{u^N}\Pexp({\cal P}_k\sI_{Y=0}^{(S_I)\rm D3}),\nonumber\\
{\cal I}^{(S_I)\rm D3}_{Z=0}&=q^N\frac{1}{v^N}\Pexp({\cal P}_k\sI_{Z=0}^{(S_I)\rm D3}).
\label{threecali}
\end{align}
These are transformed under the Weyl reflection $\sigma_{23}$
as follows.
\begin{align}
\sigma_{23}{\cal I}^{(S_I){\rm D3}}_{X=0}={\cal I}^{(S_I){\rm D3}}_{X=0},\quad
\sigma_{23}{\cal I}^{(S_I){\rm D3}}_{Y=0}={\cal I}^{(S_I){\rm D3}}_{Z=0}.
\end{align}
In addition, in the orientifold case with $k=2$ these are related by
\begin{align}
\sigma_{123}{\cal I}^{(S_I){\rm D3}}_{X=0}={\cal I}^{(S_I){\rm D3}}_{Y=0},\quad
\sigma^2_{123}{\cal I}^{(S_I){\rm D3}}_{X=0}={\cal I}^{(S_I){\rm D3}}_{Z=0}.
\end{align}
By combining the contributions from Kaluza-Klein modes and wrapped D3-branes
we obtain
\begin{align}
{\cal I}_{S(k,N,0)}\approx
{\cal I}^{\rm AdS}_{S(k,N,0)}
&\equiv{\cal I}^{\rm KK}(1+{\cal I}_{X=0}^{(S_I)\rm D3}+{\cal I}_{Y=0}^{(S_I)\rm D3}+{\cal I}_{Z=0}^{(S_I)\rm D3}).
\label{igravso}
\end{align}
In the following we always use $S=S_I$ without explicit indication.

\subsection{Schur limit}\label{subsec.Schur}
The Schur limit is defined by setting $v=y$ in the
defining equation (\ref{defsc}) of the superconformal index.
As a result the Cartan generators appearing as the exponents of
the fugacities become commutative
with, in addition to ${\cal Q}$, the second supercharge ${\cal Q}'$,
and operators with $\Delta>0$ decouples.
The resulting index, the Schur index,
is a function of two variables.
\begin{align}
{\cal I}(q,y,u,y)
&=
\tr[(-1)^F
(y^{-\frac{3}{4}})^\Delta
(\ol xy^{\frac{1}{4}})^{\ol\Delta}
(qy^{\frac{1}{2}})^{H+J^{(+)1}{}_1+J^{(-)\dot1}{}_{\dot1}}
(uy^{-\frac{1}{2}})^{R^2{}_2-R^3{}_3}
]
\nonumber\\
&=
\wh{\cal I}(qy^{\frac{1}{2}},uy^{-\frac{1}{2}})
\end{align}
We redefine the variables $q$ and $u$ to absorb $y$
and obtain the expression
\begin{align}
\wh{\cal I}(q,u)=\tr[(-1)^Fx^\Delta\ol x^{\ol\Delta}q^{H+J^{(+)1}{}_1+J^{(-)\dot1}{}_{\dot1}}u^{R^2{}_2-R^3{}_3}].
\label{defsch}
\end{align}
Practically, we can set $y=v=1$ to obtain the Schur index from the
superconformal index.

\begin{figure}[htb]
\centering
\includegraphics[width=100mm, bb=86 420 376 480]{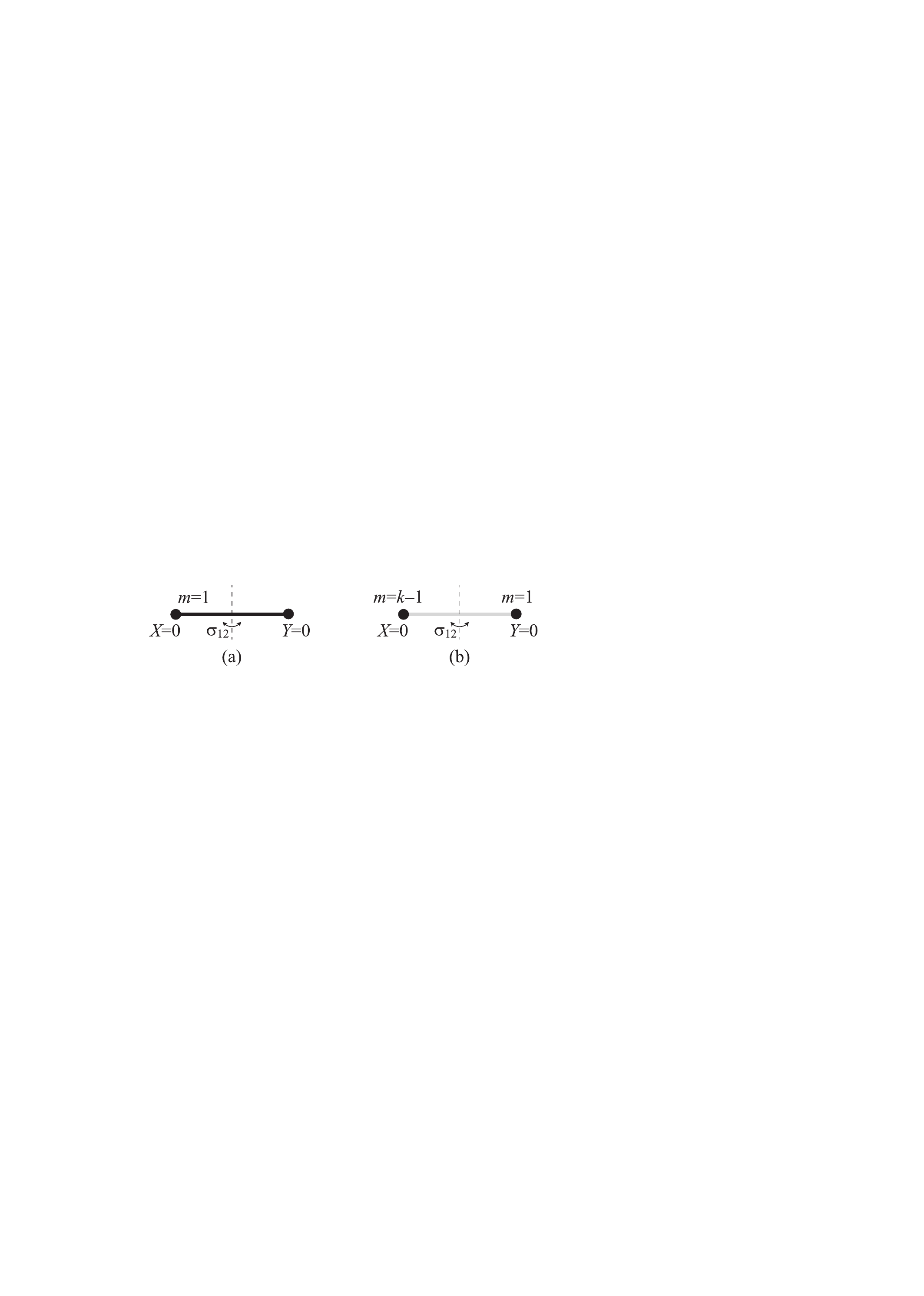}
\caption{The spaces of ground state configurations
contributing to the Schur index are shown.
(a) is the space for $k=2$ and (b) is that for $k\geq3$.}\label{configspace2.eps}
\end{figure}

Among three scalar fields $\sX$, $\sY$, and $\sZ$
the last one, $\sZ$ does not contribute to the Schur index,
and neither does the corresponding D3-brane configuration $Z=0$.
As the result the configuration space reduces as shown in Figure \ref{configspace2.eps}.
In the orientifold case with $k=2$ the space
of ground state configurations contributing to the
Schur index is
$\CC\bm{P}^1$ including $X=0$ and $Y=0$, while in the $k\geq3$ case
the space consists of two points
corresponding to $X=0$ and $Y=0$.

The Schur limit of the refined single-particle index of Kaluza-Klein modes (\ref{refinedkk}) is
\begin{align}
\wh\sI^{{\rm KK}}&=
\frac{
(u-q)q\eta^{-1}
-q^2(1-q^2)
+(u^{-1}-q)q\eta
}{(1-q^2)(1-q\frac{u}{\eta})(1-q\frac{\eta}{u})},
\label{sfschur}
\end{align}
and the corresponding multi-particle index is
$\wh{\cal I}^{\rm KK}=\Pexp({\cal P}_k\wh\sI^{\rm KK})$.
This is expected to give the Schur index of the theories $S(k,N,1)$
up to the error of order $q^{kN}$.
\begin{align}
\wh{\cal I}_{S(k,N,1)}
\approx\wh{\cal I}^{\rm AdS}_{S(k,N,1)}
&\equiv\wh{\cal I}^{\rm KK},\quad
(\mbox{error})\lesssim{\cal O}(q^{kN}).
\label{schkk}
\end{align}

The indices
$\sI^{{\rm D3}}_{X=0}$ and
$\sI^{{\rm D3}}_{Y=0}$
in (\ref{threeid3})
become in the Schur limit
\begin{align}
\wh\sI^{{\rm D3}}_{X=0}
&=\frac{\frac{\eta}{qu}-q\frac{1}{u}(1+\eta^2)+q^2}{1-q\frac{\eta}{u}},&
\wh\sI^{{\rm D3}}_{Y=0}
&=\frac{\frac{u}{q\eta}-qu(1+\frac{1}{\eta^2})+q^2}{1-q\frac{u}{\eta}}.
\label{twoid3}
\end{align}
The corresponding multi-particle indices are
\begin{align}
\wh{\cal I}_{X=0}^{\rm D3}=q^Nu^N\Pexp({\cal P}_k\wh\sI^{(S_I){\rm D3}}_{X=0}),\quad
\wh{\cal I}_{Y=0}^{\rm D3}=q^N\frac{1}{u^N}\Pexp({\cal P}_k\wh\sI^{(S_I){\rm D3}}_{Y=0}).
\end{align}

As is mentioned above the configuration $Z=0$ breaks the supersymmetry
${\cal Q}'$ used in the definition of
the Schur index, and
the multi-particle index ${\cal I}^{(S_I)\rm D3}_{Z=0}$ defined in
(\ref{threecali}) vanishes in the Schur limit.
This is explicitly shown as follows.
On the worldvolume wrapped on $Z=0$
there is a fermion zero-mode
associated with the breaking of ${\cal Q}'$.
This mode corresponds to the term $-v/y$ appearing 
in the Taylor expansion of $\sI_{Z=0}^{(S_I)\rm D3}$.
This term is $\eta$-independent and is not projected out by ${\cal P}_k$.
The plethystic exponential of this term
(in an appropriate parameter region) is
$\Pexp(-v/y)=1-v/y$ and this factor vanishes
in the Schur limit $y=v$.

As is shown in Figure \ref{configspace2.eps}
the configuration space is symmetric under $\sigma_{12}$.
Correspondingly, the indices above satisfy
\begin{align}
\sigma_{12}\wh{\cal I}^{{\rm KK}}
=\wh{\cal I}^{{\rm KK}},\quad
\sigma_{12}\wh{\cal I}_{X=0}^{{\rm D3}}
=\wh{\cal I}_{Y=0}^{{\rm D3}}.
\end{align}

By combining the contributions from Kaluza-Klein modes and wrapped D3-branes
we obtain the final result
\begin{align}
\wh{\cal I}_{S(k,N,0)}
\approx
\wh{\cal I}^{\rm AdS}_{S(k,N,0)}
&\equiv\wh{\cal I}^{\rm KK}(1+\wh{\cal I}_{X=0}^{\rm D3}+\wh{\cal I}_{Y=0}^{\rm D3}).
\label{whiads}
\end{align}

\section{Comparisons to known results}\label{checks.sec}
In this section we calculate the superconformal indices
and the Schur indices
of S-fold theories by using
formulas (\ref{igravso}) and (\ref{whiads})
and compare them to the known results obtained by localization.

Unlike the case of the BPS partition function
for which we know the exact formula (\ref{ximdef}), we have no way to derive
a bound for the error of our formula.
Physically, it is plausible that the error is
$\sim{\cal O}(q^{2N})$ because at this order the sectors with the wrapping number $|m|=2$ start to contribute
to the index.
In addition, the Kaluza-Klein analysis becomes invalid at the order $\sim{\cal O}(q^{kN})$
due to the upper bound of the angular momentum of sphere giants,
and this gives $\sim{\cal O}(q^{2N})$ correction in the $k=2$ case.
Main purpose of this section is to confirm this expectation.
Namely, we want to show
\begin{align}
{\cal I}_{S(k,N,0)}\approx{\cal I}_{S(k,N,0)}^{\rm AdS},\quad
(\mbox{error})\lesssim{\cal O}(q^{2N}),
\label{sserr}
\end{align}
where ${\cal I}_{S(k,N,0)}$ is the superconformal index of the theory
$S(k,N,0)$ and ${\cal I}^{\rm AdS}_{S(k,N,0)}$ is the
result of the calculation on the gravity side using the
formula given in the previous section.

\subsection{Orientifold}
First let us confirm (\ref{igravsokk}) for the bulk contribution.
The single particle index is obtained from (\ref{refinedkk}) by the $\ZZ_2$ projection.
The superconformal index of the bulk Kaluza-Klein modes is
${\cal I}^{\rm KK}=\Pexp({\cal P}_2\sI^{\rm KK})$
and its explicit form is shown in (\ref{eq245}).
As is shown in (\ref{igravsokk}) this gives ${\cal I}^{\rm AdS}_{S(2,N,1)}$.
The comparison to the results of localization shows
\begin{align}
{\cal I}_{S(2,N,1)}
\approx{\cal I}_{S(2,N,1)}^{\rm AdS},\quad
(\mbox{error})\lesssim{\cal O}(q^{2N+2}).
\end{align}
We confirmed this for $N\leq 3$.

To obtain ${\cal I}^{\rm AdS}_{S(2,N,0)}$ we need to take account of the wrapped D3-brane contribution.
The single-particle index obtained by the $\ZZ_2$ projection from
(\ref{threeid3}) is
\begin{align}
{\cal P}_2\sI^{\rm D3}_{X=0}
&=\frac{1}{uv}+\frac{v}{u^2}
-q^{\frac{1}{2}}\frac{1}{u}\chi^J_1
-q(\frac{1}{v}+\frac{v}{u})
+\cdots.
\end{align}
Existence of the two zero-mode terms
indicates the degeneracy of the wrapped D3-brane ground states.
The corresponding multi-particle superconformal index is
\begin{align}
{\cal I}^{\rm D3}_{X=0}
=q^Nu^N\Pexp({\cal P}_2\sI^{\rm D3}_{X=0})
=
\frac{(\mbox{num})}{(1-\frac{1}{uv})(1-\frac{v}{u^2})}
\label{eq174}
\end{align}
with the numerator
\begin{align}
(\mbox{num})
&=u^Nq^N
-\chi^J_1u^{N-1}q^{N+\frac{1}{2}}
+(u^{N-2}-u^N\ol\chi_1)q^{N+1}
\nonumber\\&
+\chi^J_1(u^N+u^{N-1}\ol\chi_1)q^{N+\frac{3}{2}}
+(u^{N-1}\ol\chi_3-u^{N-2}\ol\chi_1-u^{N-1}\chi^J_2)q^{N+2}
\nonumber\\&
+\chi^J_1(-u^{N-2}\ol\chi_3-u^{N-1}\ol\chi_2-u^N\ol\chi_1)q^{N+\frac{5}{2}}
\nonumber\\&
+(\chi^J_2(u^{N-2}\ol\chi_2+u^{N-1}\ol\chi_1+u^N)
\nonumber\\&\hspace{3em}
    +u^{N-3}(\ol\chi_3+1)-u^{N-1}\ol\chi_4+2u^{N-1}\ol\chi_1-u^N\ol\chi_3)q^{N+3}
\nonumber\\&
+(-\chi^J_3u^{N-1}+\chi^J_1(u^{N-2}\ol\chi_4-u^{N-2}\ol\chi_1+3u^{N-1}\ol\chi_3+u^N\ol\chi_2))q^{N+\frac{7}{2}}
\nonumber\\&
+{\cal O}(q^{N+4}).
\label{patch1so}
\end{align}
$\ol\chi_n$ and
$\chi_n^J$ are
the $U(2)$ character and the spin character defined by
\begin{align}
\ol\chi_n=\ol\chi_n(\tfrac{1}{v},\tfrac{v}{u}),\quad
\chi^J_n=\chi_n(\wt y),\quad
\wt y=yq^{\frac{1}{2}}.
\end{align}
As in the case of the BPS partition function,
${\cal I}^{\rm D3}_1$, the contribution of D3-brane with $m=1$,
is the sum of three contributions
${\cal I}^{\rm D3}_{X=0}$,
${\cal I}^{\rm D3}_{Y=0}$, and
${\cal I}^{\rm D3}_{Z=0}$.
It is the Weyl completion of ${\cal I}^{\rm D3}_{X=0}$,
and obtained
by the replacement
\begin{align}
u^\ell\ol\chi_n
\rightarrow \chi_{(\ell-n,n)}.
\end{align}
in the numerator (\ref{patch1so}) of ${\cal I}^{\rm D3}_{X=0}$.
As the result we obtain
\begin{align}
&{\cal I}^{\rm D3}_1={\cal I}^{\rm D3}_{X=0}+{\cal I}^{\rm D3}_{Y=0}+{\cal I}^{\rm D3}_{Z=0}
\nonumber\\&
=q^N\chi_{(N,0)}
-q^{N+\frac{1}{2}}\chi^J_1\chi_{(N-1,0)}
+q^{N+1}(\chi_{(N-2,0)}-\chi_{(N-1,1)})
\nonumber\\&
+q^{N+\frac{3}{2}}\chi_1^J(\chi_{(N,0)}+\chi_{(N-2,1)})
+q^{N+2}(\chi_{(N-4,3)}-\chi_{(N-3,1)}-\chi_2^J\chi_{(N-1,0)})
\nonumber\\&
+q^{N+\frac{5}{2}}\chi^J_1(-\chi_{(N-5,3)}-\chi_{(N-3,2)}-\chi_{(N-1,1)})
\nonumber\\&
+q^{N+3}(\chi^J_2(\chi_{(N-4,2)}+\chi_{(N-2,1)}+\chi_{(N,0)})
\nonumber\\&
\hspace{4em} +\chi_{(N-6,3)}-\chi_{(N-5,4)}+\chi_{(N-3,0)}+2\chi_{(N-2,1)}-\chi_{(N-3,3)})
\nonumber\\&
+q^{N+\frac{7}{2}}(-\chi^J_3\chi_{(N-1,0)}
    +\chi^J_1(\chi_{(N-6,4)}-\chi_{(N-3,1)}+3\chi_{(N-4,3)}+\chi_{N-2,2}))
\nonumber\\&
+q^{N+4}(\chi_{(n-8,6)}-\chi_{(n-7,4)}+\chi_{(n-6,5)}-2\chi_{(n-5,3)}
+\chi_{(n-4,4)}-2\chi_{(n-3,2)}
\nonumber\\&
-2\chi_{(n-2,0)}-\chi_{(n-1,1)}
+\chi^J_2(-2\chi_{(n-5,3)}-\chi_{(n-4,1)}
-2\chi_{(n-3,2)}-\chi_{(n-1,1)}))
\nonumber\\&
+{\cal O}(q^{N+\frac{9}{2}}).
\end{align}
We obtain ${\cal I}^{\rm AdS}_{S(2,N,0)}$
by substituting this into (\ref{igravso}).
The results for $N=1,2,3$ are shown in Appendix \ref{resultsads.sec}.
The comparison to the results of localization
shows
\begin{align}
{\cal I}_{S(2,N,0)}={\cal I}^{\rm AdS}_{S(2,N,0)}+{\cal O}(q^{2N+1}).
\end{align}
We confirmed this for $N\leq 3$.

We also 
consider the Schur index.
Of course once we obtain the superconformal index it is obtained by
taking the Schur limit.
However, we calculate it separately
because it is easier than the derivation via the superconformal index.

The Schur index of the Kaluza-Klein modes is
$\wh{\cal I}^{\rm AdS}_{S(2,N,1)}=\wh{\cal I}^{\rm KK}=\Pexp({\cal P}_2\wh\sI^{\rm KK})$
and the explicit form is shown in (\ref{eq254}).
The comparison to the results of localization $\wh{\cal I}_{S(2,N,1)}=\wh{\cal I}_{SO(2N+1)}$ shows
\begin{align}
\wh{\cal I}_{S(2,N,1)}=\wh{\cal I}^{\rm AdS}_{S(2,N,1)}+{\cal O}(q^{2N+2}).
\end{align}
We confirmed this for $N\leq 3$.

The D3-brane single-particle Schur index is obtained from
(\ref{twoid3}) by the $\ZZ_2$ projection.
\begin{align}
{\cal P}_2\wh\sI^{\rm D3}_{X=0}
=\frac{u^{-2}-2qu^{-1}+q^2}{1-q^2u^{-2}}
=u^{-2}-2qu^{-1}+q^2+q^2u^{-4}+\cdots.
\end{align}
Corresponding to the one-dimensional configuration space
this includes one zero mode term $u^{-2}$.
The plethystic exponential gives
\begin{align}
\wh{\cal I}^{\rm D3}_{X=0}
&=q^Nu^N\Pexp({\cal P}_2\wh\sI^{\rm D3}_{X=0})
=\frac{(\mbox{num})}{1-u^{-2}}
\label{schid3x0}
\end{align}
with the numerator
\begin{align}
(\mbox{num})
&=
u^Nq^N
-2u^{N-1}q^{N+1}
+(u^N+u^{N-2}+u^{N-4})q^{N+2}
\nonumber\\&
-2(u^{N-1}+u^{N-3}+u^{N-5})q^{N+3}
\nonumber\\&
+(u^N+2u^{N-2}+5u^{N-4}+2u^{N-6}+u^{N-8})q^{N+4}
\nonumber\\&
-2(u^{N-1}+2u^{N-3}+3u^{N-5}+2u^{N-7}+u^{N-9})q^{N+5}
+{\cal O}(q^{N+6}).
\label{numwhi}
\end{align}
The result from a D3-brane on $Y=0$ is obtained from this by
the Weyl reflection $\sigma_{12}u=u^{-1}$.
$\wh {\cal I}^{\rm D3}_1$, the sum of $\wh{\cal I}^{\rm D3}_{X=0}$ and $\wh{\cal I}^{\rm D3}_{Y=0}$,
is the Weyl completion of $\wh{\cal I}^{\rm D3}_{X=0}$, and
obtained from (\ref{numwhi}) by
applying the replacement
\begin{align}
u^n\rightarrow \chi_n(u).
\end{align}
The result is
\begin{align}
\wh{\cal I}^{\rm D3}_1
=&
q^N\chi_N
-2q^{N+1}\chi_{N-1}
 +q^{N+2}(\chi_N+\chi_{N-2}+\chi_{N-4})
\nonumber\\&
-2q^{N+3}(\chi_{N-1}+\chi_{N-3}+\chi_{N-5})
\nonumber\\&
 +q^{N+4}(\chi_N+2\chi_{N-2}+5\chi_{N-4}+2\chi_{N-6}+\chi_{N-8})
\nonumber\\&
-2q^{N+5}(\chi_{N-1}+2\chi_{N-3}+3\chi_{N-5}+2\chi_{N-7}+\chi_{N-9})
\nonumber\\&
 +q^{N+6}(\chi_N+2\chi_{N-2}+7\chi_{N-4}+8\chi_{N-6}+7\chi_{N-8}+2\chi_{N-10}+\chi_{N-12})
\nonumber\\&
-2q^{N+7}(\chi_{N-1}+2\chi_{N-3}+5\chi_{N-5}+6\chi_{N-7}+5\chi_{N-9}+2\chi_{N-11}+\chi_{N-13})
\nonumber\\&
+{\cal O}(q^{N+8}),
\label{eq186}
\end{align}
where $\chi_n=\chi_n(u)$ is the $SU(2)$ character.
$\wh{\cal I}_{S(2,N,0)}^{\rm AdS}$
is obtained by substituting (\ref{eq186})
into (\ref{schkk}).
See Appendix \ref{resultsads.sec}
for the results for $N=1,2,3$.
The comparison to the results of localization $\wh{\cal I}_{SO(2N)}=\wh{\cal I}^{\rm AdS}_{S(2,N,0)}$ shows
\begin{align}
\wh{\cal I}_{S(2,N,0)}
=
\wh{\cal I}_{(2,N,0)}^{\rm AdS}
+{\cal O}(q^{2N+2}).
\end{align}
We confirmed this for $N\leq 3$.

\subsection{S-folds with $k\geq 3$}
In the case of $k\geq3$ the ground state configuration space consists of
two components: the non-degenerate component and the $\CC\bm{P}^1$ component (See (b) in Figure \ref{configspace.eps}).

The D3-brane configuration in the non-degenerate component has the
wrapping number $m=k-1$ and hence
${\cal I}^{\rm D3}_{k-1}={\cal I}^{\rm D3}_{X=0}$.
Let us consider the $k=3$ case as an example.
The single-particle index for a D3-brane over $X=0$ is
\begin{align}
{\cal P}_3\sI^{\rm D3}_{X=0}
=q(\frac{1}{u}\ol\chi_2-\ol\chi_1)
+q^{\frac{3}{2}}\chi_1^J(-\frac{1}{u}\ol\chi_1+1)
+q^2\frac{1}{u}
+{\cal O}(q^4).
\end{align}
This does not include zero-mode terms independent of $q$.
The corresponding multi-particle index is
\begin{align}
{\cal I}^{\rm D3}_2
&=q^Nu^N\Pexp({\cal P}_3\sI_{X=0}^{\rm D3})
\nonumber\\
&=q^Nu^N
+q^{N+1}(u^{N-1}\ol\chi _2-u^N\ol\chi_1)
+q^{N+\frac{3}{2}}\chi^J_1(-u^{N-1}\ol\chi_1+u^N)
\nonumber \\&
+q^{N+2}(u^{N-2}\ol\chi _4-u^{N-1}\ol\chi _3-u^{N-2}\ol\chi_1+u^{N-4}+2u^{N-1})
+{\cal O}(q^{N+\frac{5}{2}})
\label{impk3m2}
\end{align}

Two D3-brane configurations $Y=0$ and $Z=0$ in the $\CC\bm{P}^1$ component
have the same wrapping number $m=1$,
and the corresponding contribution
${\cal I}_{Y=0}^{\rm D3}$ and ${\cal I}_{Z=0}^{\rm D3}$ are summed up to
${\cal I}^{\rm D3}_1$.
The single particle index for $Y=0$ is
\begin{align}
{\cal P}_3\sI^{\rm D3}_{Y=0}
&=\frac{u}{v^2}
-q^{\frac{1}{2}}\chi^J_1\frac{u}{v}
+q(-\frac{1}{v}+\frac{u^3}{v})\nonumber \\
&+q^{\frac{3}{2}}\chi^J_1
+q^2(-u^2+\frac{u^2}{v^3})
-q^{\frac{5}{2}}\chi^J_1\frac{u^2}{v^2}
+{\cal O}(q^3).
\end{align}
This has one zero-mode term corresponding to the one-dimensional configuration space $\CC\bm{P}^1$.
The corresponding multi-particle index is
\begin{align}
{\cal I}^{\rm D3}_{Y=0}
=q^N\left (\frac{v}{u}\right )^N\Pexp({\cal P}_3\sI^{\rm D3}_{Y=0})
=\frac{(\mbox{num})}{1-\frac{u}{v^2}},
\end{align}
with the numerator
\begin{align}
(\mbox{num})
&=(\frac{v}{u})^Nq^N-\chi^J_1(\frac{v}{u})^{N-1}q^{N+\frac{1}{2}}
\nonumber\\&
+((\frac{v}{u})^{N-2}+u^2(\frac{v}{u})^{N-1}-u^{-1}(\frac{v}{u})^{N-1})q^{N+1}+{\cal O}(q^{N+\frac{3}{2}}).
\label{s3num}
\end{align}
We obtain ${\cal I}^{\rm D3}_1={\cal I}^{\rm D3}_{Y=0}+{\cal I}^{\rm D3}_{Z=0}$
as the $SU(2)$ completion of ${\cal I}^{\rm D3}_{Y=0}$
from the numerator (\ref{s3num}) by the replacement
\begin{align}
(\frac{v}{u})^n
\rightarrow
\ol\chi_n(\frac{v}{u},\frac{1}{v}).
\end{align}
The result is
\begin{align}
{\cal I}^{\rm D3}_1
&=q^N\ol\chi _N-q^{N+\frac{1}{2}}\chi ^J_1\ol\chi _{N-1}
+q^{N+1}(\ol\chi_{N-2}+u^2\ol\chi_{N-1}-u^{-1}\ol\chi_{N-1})
\nonumber\\&
+q^{N+\frac{3}{2}}\chi^J_1((\frac{1}{u}-u^2)\ol\chi_{N-2}+\ol\chi_N)
\nonumber\\&
+q^{N+2}(u^2\ol\chi_{N-3}+(u^4-u)\ol\chi _{N-2}-\ol\chi_{N-1}-u^2\ol\chi_N-\chi^J_2\chi_{N-1})
+{\cal O}(q^{N+\frac{5}{2}}).
\label{impd31}
\end{align}
By substituting
(\ref{impk3m2}), and (\ref{impd31})
into (\ref{igravso})
we obtain ${\cal I}^{\rm AdS}_{S(3,N,0)}$.
See Appendix \ref{resultsads.sec} for the results for $N=1,2,3$.
Results for $k=4,6$ are also shown there.

In the case of the Schur index
the configuration space consists of two points related by $\sigma_{12}$ (See (b) in Figure \ref{configspace2.eps}).
Let us take the $k=3$ case as an example.
The configuration $X=0$ has wrapping number $m=k-1=2$ and
the corresponding index is
\begin{align}
\wh{\cal I}^{\rm D3}_2&=\wh{\cal I}^{\rm D3}_{X=0}
\nonumber\\&
=q^Nu^N
+q^{N+1}(u^{N-3}-u^{N-1})
+q^{N+2}(u^{N-6}-u^{N-4}-u^{N-2}+u^N)
\nonumber\\&
+{\cal O}(q^{N+3}),
\end{align}
Thanks to the $\ZZ_2$ symmetry $Y=0$ contribution is obtained by
\begin{align}
\wh{\cal I}^{\rm D3}_1&=\sigma_{12}\wh{\cal I}^{\rm D3}_{k-1}.
\end{align}
Substituting these
together with the Kaluza-Klein contribution
$\wh{\cal I}^{\rm KK}=\Pexp({\cal P}_3\wh\sI^{\rm KK})$
into (\ref{whiads}), we obtain
$\wh{\cal I}_{S(3,N,0)}^{\rm AdS}$.
The results for $N=1,2,3$ are shown in Appendix \ref{resultsads.sec}.
The results for $k=4,6$ are also shown there.

At present we have no technique to directly calculate the index of S-fold theories with $k\geq 3$.
Fortunately, it is known that in the S-fold theories with $N=1$ or $2$ and $p=0$ the supersymmetry
is also enhanced to ${\cal N}=4$,
and they are dual to ${\cal N}=4$ supersymmetric gauge theories \cite{Nishinaka:2016hbw,Aharony:2016kai}.
The gauge group for each case is shown in Table \ref{enhance.tbl}.
\begin{table}[htb]
\caption{The gauge groups of the supersymmetry enhanced theories.}\label{enhance.tbl}
\centering
\begin{tabular}{ccccc}
\hline
\hline
    & $S(1,k,0)$ & $S(2,3,0)$ & $S(2,4,0)$ & $S(6,2,0)$ \\
\hline
$G$ & $U(1)$ & $SU(3)$ & $SO(5)$ & $G_2$ \\
\hline
\end{tabular}
\end{table}
By using this duality, we confirm our results for $N=1$ and $2$
are correct up to the expected order of $q$.

If the supersymmetry is enhanced to ${\cal N}=4$,
the manifest $R$ symmetry is enhanced from $SU(2)$ to $SU(3)$, and
the superconformal index must be written in terms of $SU(3)$ characters.
For example, in ${\cal I}^{\rm AdS}_{S(3,2,0)}$ shown in Appendix \ref{resultsads.sec}
is expressed by $SU(3)$ characters up to $q^{\frac{9}{2}}$ terms.
This fact strongly suggests the symmetry enhancement,
and actually we find the agreement with the
index of ${\cal N}=4$ SYM with the gauge group $SU(3)$, the expected dual theory,
up to $q^{\frac{9}{2}}$ terms.

This kind of relations holds for all theories in which the supersymmetry enhancement is expected.
The results of the comparison are shown below.
\begin{align}
{\cal I}^{\rm AdS}_{S(3,1,0)}
&={\cal I}_{U(1)}+{\cal O}(q^3),&
\wh{\cal I}^{\rm AdS}_{S(3,1,0)}
&=\wh{\cal I}_{U(1)}+{\cal O}(q^3)
,\nonumber\\
{\cal I}^{\rm AdS}_{S(4,1,0)}
&={\cal I}_{U(1)}+{\cal O}(q^2),&
\wh{\cal I}^{\rm AdS}_{S(4,1,0)}
&=\wh{\cal I}_{U(1)}+{\cal O}(q^2)
,\nonumber\\
{\cal I}^{\rm AdS}_{S(6,1,0)}
&={\cal I}_{U(1)}+{\cal O}(q^2),&
\wh{\cal I}^{\rm AdS}_{S(6,1,0)}
&=\wh{\cal I}_{U(1)}+{\cal O}(q^2)
,\nonumber\\
{\cal I}^{\rm AdS}_{S(3,2,0)}
&={\cal I}_{SU(3)}+{\cal O}(q^5),&
\wh{\cal I}^{\rm AdS}_{S(3,2,0)}
&=\wh{\cal I}_{SU(3)}+{\cal O}(q^5)
,\nonumber\\
{\cal I}^{\rm AdS}_{S(4,2,0)}
&={\cal I}_{SO(5)}+{\cal O}(q^4),&
\wh{\cal I}^{\rm AdS}_{S(4,2,0)}
&=\wh{\cal I}_{SO(5)}+{\cal O}(q^4)
,\nonumber\\
{\cal I}^{\rm AdS}_{S(6,2,0)}
&={\cal I}_{G_2}+{\cal O}(q^4),&
\wh{\cal I}^{\rm AdS}_{S(6,2,0)}
&=\wh{\cal I}_{G_2}+{\cal O}(q^4).
\end{align}

All these results are consistent with the relation (\ref{sserr}).

\subsection{$U(N)$ SYM}
Encouraged by the success for $k\geq 2$,
let us apply our prescription to the $k=1$ case, $U(N)$ SYM.
The superconformal index in the large $N$ limit is
${\cal I}_{S(1,\infty,p)}^{\rm AdS}=\Pexp({\cal P}_1\sI^{\rm KK})=\Pexp\sI^{\rm KK}$.
See (\ref{eq244}) for the explicit expression.
We want to calculate the correction due to wrapped D3-branes.
Unlike the $k\geq2$ case the internal space $\bm{S}^5$ does not
have topologically non-trivial cycles.
Correspondingly, the single-particle index includes the tachyonic term $\propto q^{-1}$.
\begin{align}
{\cal P}_1\sI_{X=0}^{\rm D3}
=\frac{1}{uq}+u^{-1}\chi_1-u^{-1}\chi_Jq^{\frac{1}{2}}+(-\chi_1+u^{-1}\chi_2)q+\cdots .
\end{align}
Usually we assume $|q|<1$ for convergence, and then the plethystic exponential
of the tachyonic term diverges.
To avoid this problem we formally rewrite
the plethystic exponential of the tachyonic term
by
\begin{align}
\Pexp\frac{1}{qu}=\frac{1}{1-\frac{1}{qu}}=-\frac{qu}{1-qu}=-qu\Pexp(qu).
\end{align}
At the first equality we assumed sufficiently large $q$,
and after the analytic continuation to small $q$
we rewrote the expression with $\Pexp$ again.
Then the corresponding multi-particle index is
\begin{align}
{\cal I}_{X=0}^{\rm D3}=-\frac{q^{N+1}u^{N+1}}{(1-\frac{1}{uv})(1-\frac{v}{u^2})}\Pexp
\left(\frac{
\begin{array}{l}
-u^{-1}\chi^Jq^{\frac{1}{2}}+(u-\chi_1+u^{-1}\chi_2)q\\
+\chi^J_1q^{\frac{3}{2}}+(2u^{-1}-(u^{-2}+u)\chi_1)q^2
\end{array}
}
{(1-\frac{1}{v}q)(1-\frac{v}{u}q)}\right).
\end{align}
${\cal I}_{Y=0}^{\rm D3}$ and ${\cal I}_{Z=0}^{\rm D3}$ are obtained from ${\cal I}_{X=0}^{\rm D3}$
by the Weyl reflections, and the sum of them gives the Weyl
completion of ${\cal I}_{X=0}^{\rm D3}$.
The final results for $N=1,2,3$ are shown in (\ref{sciads1}) in Appendix \ref{resultsads.sec}.
Surprisingly, they agree with the results of the localization up to order $q^{2N+\frac{7}{2}}$ terms.

The Schur index is calculated in a similar way.
$\wh{\cal I}_{X=0}^{\rm D3}$ is given by
\begin{align}
\wh{\cal I}_{X=0}^{\rm D3}=-\frac{q^{N+1}u^{N+1}}{1-\frac{1}{u^2}}\Pexp\left(\frac{qu(1-\frac{1}{u^2})^2}{1-\frac{q}{u}}\right ).
\label{shiunx}
\end{align}
Let us first consider the $u\rightarrow 1$ limit, in which the following
analytic formula was derived in \cite{Bourdier:2015wda}:
\begin{align}
\wh I_{U(N)}(q,u=1)=
\wh I_{U(\infty)}(q,u=1)\sum_{n=0}^\infty\frac{(N+n-1)!(N+2n)}{N!n!}q^{nN+n^2}.
\label{bdfformula}
\end{align}
In the $u\rightarrow1$ limit the plethystic exponential in 
(\ref{shiunx}) becomes $1$
and
the prefactor gives $-q^{N+1}u^{N+1}\chi_{N+1}|_{u=1}=-(N+2)q^{N+1}$ after the Weyl completion.
This correctly reproduces the $n=1$ term in (\ref{bdfformula}).
For $u\neq1$ the results for $N=1,2,3$ are shown in (\ref{schurads1}) in Appendix \ref{resultsads.sec},
and agree with the results of the localization up to order $q^{2N+3}$ terms.

\section{Conclusions and discussion}\label{conc.sec}
We calculated finite $N$ corrections to the superconformal indices
of S-fold theories by using AdS/CFT correspondence.
Among several sources of the correction,
we focused on the contribution of wrapped D3-branes with the wrapping number $m=\pm1\mod k$,
which give the corrections of order $q^N$.
We derived the formula
\begin{align}
{\cal I}_{S(k,N,0)}
\approx&
\Pexp({\cal P}_k\sI^{\rm KK})
\Bigg(1+q^Nu^N\Pexp({\cal P}_k\sI_{X=0}^{\rm D3})
\nonumber\\&
\hspace{6em}
+q^N\frac{v^N}{u^N}\Pexp({\cal P}_k\sI_{Y=0}^{\rm D3})
+q^N\frac{1}{v^N}\Pexp({\cal P}_k\sI_{Z=0}^{\rm D3})
\Bigg).
\label{theformula}
\end{align}
The refined single-particle superconformal index
$\sI^{\rm KK}$ for the bulk Kaluza-Klein modes
and $\sI_{X=0}^{\rm D3}$, $\sI_{Y=0}^{\rm D3}$, and $\sI_{Z=0}^{\rm D3}$ for
the fluctuations on D3-branes wrapped over the indicated loci
are universal in the sense that they are independent of
$k$ and $N$.
We also derived the formula
\begin{align}
\wh{\cal I}_{S(k,N,0)}
\approx&
\Pexp({\cal P}_k\wh\sI^{\rm KK})
\Bigg(1+q^Nu^N\Pexp({\cal P}_k\wh\sI_{X=0}^{\rm D3})
+q^N\frac{1}{u^N}\Pexp({\cal P}_k\wh\sI_{Y=0}^{\rm D3})
\Bigg)
\label{theschformula}
\end{align}
for the Schur index.

The order of corrections from other sources such as multiple wrapping of D3-branes
and the upper bound of the Kaluza-Klein angular momentum of giant gravitons are around $2N$ or more.
Therefore, our results are expected to give the correct answer up to error terms of ${\cal O}(q^{2N})$.
In the case of the orientifold ($k=2$) we compared our results
to the results of localization of $SO(2N)$ SYM,
and we found the agreement in the expected range of the $q$ expansion.
For $k\geq3$, we have no method of the direct calculation of the index.
However, for $N=1,2$ we used the duality to the ${\cal N}=4$ SYM, and
we compared our results to the indices of the dual SYMs.
Again, we found the agreement in the expected range of order.

We also applied the formulas (\ref{theformula}) and (\ref{theschformula})
to the $k=1$ case.
Although there is no non-trivial
three-cycles to be wrapped by D3-branes,
the formulas correctly give the ${\cal O}(q^N)$ corrections to the
index of the $U(N)$ SYM.

These results are summarized in Table \ref{errorssc.tbl} and Table \ref{errorsschur.tbl}.
\begin{table}[htb]
\caption{The order of the term that does not match in the superconformal index is shown for each S-fold
theory that has ${\cal N}=4$ supersymmetry.
${\cal O}(q^n)$ means the term of order $q^n$ does not agree
while ${\cal O}(q^8)$ means we calculated the index up to the
$q^8$ term and we did not find disagreement.}\label{errorssc.tbl}
\centering
\begin{tabular}{ccccccc}
\hline
\hline
$k_p$ & $1_0$ & $2_0$ & $2_1$ & $3_0$ & $4_0$ & $6_0$ \\
\hline
$N=1$ &${\cal O}(q^6)$ &${\cal O}(q^3)$ & ${\cal O}(q^4)$ & ${\cal O}(q^3)$ & ${\cal O}(q^2)$ & ${\cal O}(q^2)$ \\
$N=2$ &${\cal O}(q^8)$ &${\cal O}(q^5)$ & ${\cal O}(q^6)$ & ${\cal O}(q^5)$ & ${\cal O}(q^4)$ & ${\cal O}(q^4)$ \\
$N=3$ &${\cal O}(q^{10})$ &${\cal O}(q^7)$ & ${\cal O}(q^8)$ &        -        &        -        &        -        \\
\hline
\end{tabular}
\end{table}
\begin{table}[htb]
\caption{The order of the term that does not match in the Schur index.}\label{errorsschur.tbl}
\centering
\begin{tabular}{ccccccc}
\hline
\hline
$k_p$ & $1_0$ & $2_0$ & $2_1$ & $3_0$ & $4_0$ & $6_0$ \\
\hline
$N=1$ &${\cal O}(q^6)$ &${\cal O}(q^4)$ & ${\cal O}(q^4)$ & ${\cal O}(q^3)$ & ${\cal O}(q^2)$ & ${\cal O}(q^2)$ \\
$N=2$ &${\cal O}(q^8)$ &${\cal O}(q^5)$ & ${\cal O}(q^6)$ & ${\cal O}(q^5)$ & ${\cal O}(q^4)$ & ${\cal O}(q^4)$ \\
$N=3$ &${\cal O}(q^{10})$ &${\cal O}(q^8)$ & ${\cal O}(q^8)$ &        -        &        -        &        -        \\
\hline
\end{tabular}
\end{table}
As is shown in the tables
in all cases the error terms are of order ${\cal O}(q^{2N})$ or higher.

In this paper we focused only on the corrections starting from ${\cal O}(q^N)$.
To reproduce the corrections of ${\cal O}(q^{2N})$ or higher
we need to take account of multiple wrapping of D3-branes.
This is not so easy as the single wrapping.
A naive expectation is that a part of the contribution of D3-branes
with wrapping number $m$ may be calculated by using the $U(|m|)$ SYM on the worldvolume instead of $U(1)$.
However, it is not clear to what extent this works because
the description in terms of $U(|m|)$ SYM is justified only when all the worldvolumes
of wrapped D3-branes are close to each other,
and in general this is not the case.
Furthermore, even for single wrapping,
the description in terms of the field theory would break down
for large fluctuations comparable to the AdS radius.
It is also necessary to take account of the correction to the Kaluza-Klein contribution
due to giant gravitons.
Namely, for a large angular momentum the Kaluza-klein particles
should be treated as D3-branes puffed up by the interaction with
the background RR flux.

In the case of the BPS partition function the exact partition function
for finite $N$ can be obtained by geometric quantization
of D3-branes in $\bm{S}^5/\ZZ_k$ \cite{Biswas:2006tj,Arai:2018utu}.
It would be very interesting to see what happens
when we extend this analysis to super D3-branes by including the gauge fields and fermion fields on the D3-branes.

\section*{Acknowledgments}
The authors are grateful to Shota~Fujiwara and Tatsuya~Mori for wonderful discussions.
The authors also thank Hirotaka~Kato for collaboration at the early stage of this work.
The work of Y.~I. was 
partially supported by Grand-in-Aid for Scientific Research (C) (No.15K05044),
Ministry of Education, Science and Culture, Japan.

\appendix
\section{Superconformal algebra}\label{algebra.sec}
The anti-commutators among the fermionic generators are shown in (\ref{n4algebra}).
In this appendix we show the other non-vanishing commutators.

The $SU(4)_R$ generators $R^I{}_J$, which satisfy $R^I{}_I=0$,
act on generators with lower and upper
R-indices
as follows.
\begin{align}
[R^I{}_J,\phi_K]&=\delta_K^I\phi_J-\frac{1}{4}\delta^I_J\phi_K,&
[R^I{}_J,\phi^K]&=-\delta^K_I\phi^I+\frac{1}{4}\delta^I_J\phi_K.
\end{align}
The commutator of two $SU(4)_R$ generators is
\begin{align}
[R^I{}_J,R^K{}_L]=
\delta_L^IR^K{}_J-\delta_J^KR^I{}_L.
\end{align}

$J^a{}_b$ and $J^{\dot a}{}_{\dot b}$
are generators of
$SU(2)^{(+)}_J$ and $SU(2)^{(-)}_J$, respectively,
and satisfy similar relations.
They are traceless: $J^a{}_a=J^{\dot a}{}_{\dot a}=0$.
Rules for their action on generators
with lower and upper spinor indices are
\begin{align}
[J^a{}_b,\phi_c]&=\delta^a_c\phi_b-\tfrac{1}{2}\delta^a_b\phi_c,&
[J^{\dot a}{}_{\dot b},\phi_{\dot c}]
&=\delta_{\dot c}^{\dot a}\phi_{\dot b}-\tfrac{1}{2}\delta^{\dot a}_{\dot b}\phi_{\dot c},\nonumber\\
[J^a{}_b,\phi^c]&=-\delta_b^c\phi^a+\tfrac{1}{2}\delta^a_b\phi^c,&
[J^{\dot a}{}_{\dot b},\phi^{\dot c}]
&=-\delta^{\dot c}_{\dot b}\phi^{\dot a}
    +\tfrac{1}{2}\delta^{\dot a}_{\dot b}\phi^{\dot c}.
\end{align}
Commutation relations among $J$ are
\begin{align}
[J^a{}_b,J^c{}_d]
&=\delta^a_dJ^c{}_b-\delta^c_bJ^a{}_d,&
[J^{\dot a}{}_{\dot b},J^{\dot c}{}_{\dot d}]
&=\delta^{\dot a}_{\dot d}J^{\dot c}{}_{\dot b}-\delta^{\dot c}_{\dot b}J^{\dot a}{}_{\dot d}.
\end{align}
Non-vanishing commutators of $H$ and bosonic generators are
\begin{align}
[H,K^a{}_{\dot b}]&=-K^a{}_{\dot b},&
[H,P^{\dot a}{}_b]&=P^{\dot a}{}_b.
\end{align}
The commutators of $H$ and fermionic generators are
\begin{align}
[H,S^a_I]&=-\tfrac{1}{2}S^a_I,&
[H,\ol Q^{\dot a}_I]&=\tfrac{1}{2}\ol Q^{\dot a}_I,&
[H,Q_a^I]&=\tfrac{1}{2}Q_a^I,&
[H,\ol S_{\dot a}^I]&=-\tfrac{1}{2}\ol S_{\dot a}^I.
\end{align}
The commutator between
the momenta $P^{\dot a}{}_b$ and the conformal boosts $K^a{}_{\dot b}$ is
\begin{align}
[K^a{}_{\dot b},P^{\dot c}{}_d]
&=-\delta^a_dJ^{(-)\dot c}{}_{\dot b}
+\delta^{\dot c}_{\dot b}J^{(+)a}{}_d
+\delta^{\dot c}_{\dot b}\delta^a_dH.
\end{align}
Non-vanishing generators including $K$ or $P$ and fermionic generators
are
\begin{align}
[K^a{}_{\dot b},\ol Q^{\dot c}_I]&=\delta^{\dot c}_{\dot b}S^a_I,&
[P^{\dot a}{}_b,S^c_I]&=-\delta^c_b\ol Q^{\dot a}_I,
\nonumber\\
[K^a{}_{\dot b},Q^I_c]&=\delta^a_c\ol S^I_{\dot b},&
[P^{\dot a}{}_b,\ol S^I_{\dot c}]&=-\delta^a_cQ^I_b,
\end{align}

In addition to the generators of superconformal algebra,
we also introduce the $U(1)_A$ generator $A$ acting on the
fermionic generators as follows.
\begin{align}
[A,Q^I_a]=-Q^I_a,\quad
[A,\ol Q_I^{\dot a}]=\ol Q_I^{\dot a},\quad
[A,S_I^a]=S_I^a,\quad
[A,\ol S^I_{\dot a}]=-\ol S^I_{\dot a}.
\end{align}

The Hermiticity of the generators are
\begin{align}
&(J^{(+)a}{}_b)^\dagger=J^{(+)b}{}_a,\quad
(J^{(-)\dot a}{}_{\dot b})^\dagger=J^{(-)\dot b}{}_{\dot a},\quad
(R^I{}_J)^\dagger=R^J{}_I,\nonumber\\
&
H^\dagger=H,\quad
A^\dagger=A,\nonumber\\
&
(K^a{}_{\dot b})^\dagger=P^{\dot b}{}_a,\quad
(S^a_I)^\dagger
=Q^I_a,\quad
(\ol S^I_{\dot a})^\dagger=\ol Q^{\dot a}_I.
\end{align}

\section{Characters}\label{characters.sec}
We define $U(2)$ characters $\ol\chi_n(a,b)$
by
\begin{align}
\ol\chi_n(a,b)=\frac{a^{n+1}-b^{n+1}}{a-b}.
\end{align}
For a few small representations this gives
\begin{align}
\ol\chi_0(a,b)=1,\quad
\ol\chi_1(a,b)=a+b,\quad
\ol\chi_2(a,b)=a^2+ab+b^2.
\end{align}
$U(3)$ characters $\ol\chi_{(r_1,r_2)}(a,b,c)$ are defined by
\begin{align}
\ol\chi_{(r_1,r_2)}(a,b,c)
=\left|\begin{array}{ccc}
a^{r_1+1} & 1 & a^{-r_2-1} \\
b^{r_1+1} & 1 & b^{-r_2-1} \\
c^{r_1+1} & 1 & c^{-r_2-1}
\end{array}\right|
\Bigg/
\left|\begin{array}{ccc}
a & 1 & a^{-1} \\
b & 1 & b^{-1} \\
c & 1 & c^{-1}
\end{array}\right|.
\end{align}
For a few small representations
\begin{align}
\ol\chi_{(0,0)}(a,b,c)=1,\quad
\ol\chi_{(1,0)}(a,b,c)=a+b+c,\quad
\ol\chi_{(0,1)}(a,b,c)=\frac{1}{a}+\frac{1}{b}+\frac{1}{c}.
\end{align}
The $SU(2)$ characters $\chi_n(u)$ and the $SU(3)$ characters $\chi_{(r_1,r_2)}$
are defined by
\begin{align}
\chi_n(u)=\ol\chi_n(u,\frac{1}{u}),\quad
\chi_{(r_1,r_2)}(u,v)=\ol\chi_{(r_1,r_2)}(u,\frac{v}{u},\frac{1}{v})
\end{align}

\section{Differentials in $\CC\bm{P}^d$}
In this appendix we consider some vector bundles
over $\CC\bm{P}^d$
appearing in subsection \ref{global.sec}.
The $SU(d+1)$ action on the base manifold $\CC\bm{P}^d$
is extended to the vector bundles in a natural way,
and we can define the $SU(d+1)$ character as the trace
over holomorphic sections
of the bundle.
We refer to such a character associated with
a vector bundle ${\cal E}$ simply as the character of ${\cal E}$,
and denote it by $\chi({\cal E})$.

We first consider the line bundle
\begin{align}
{\cal E}_0^{(N)}={\cal O}(N).
\end{align}
Let $\alpha^\mu$ ($\mu=0,1,\ldots,d$) be homogeneous coordinates
of $\CC\bm{P}^d$.
A holomorphic section $\Psi$ is expressed in terms of $\alpha^\mu$ as
a homogeneous polynomial of degree $N$.
\begin{align}
\Psi=\Psi^{(N)}(\alpha^\mu).
\end{align}
Therefore, the associated $SU(d+1)$ character is simply given by
\begin{align}
\chi({\cal E}_0^{(N)})=\chi_N ,
\label{appeq1}
\end{align}
where $\chi_n$ in this appendix
represents the $SU(d+1)$ character of the representation $\Sym^n(\mbox{fund})$.

Let $\Psi$ be an $n$-differential with weight $N-1$.
This means $\Psi$ is a section of the vector bundle
\begin{align}
{\cal E}^{(N-1)}_n={\cal O}(N-1)\otimes\Sym^n(T^*\CC\bm{P}^d).
\end{align}
In terms of the homogeneous coordinates $\alpha^\mu$ ($\mu=0,1,\ldots,d$),
$\Psi$ is expressed as
\begin{align}
\Psi=\Psi^{(N-1)}_{\mu_1\cdots\mu_n}(\alpha^\mu)\{d\alpha^{\mu_1}\cdots d\alpha^{\mu_n}\},
\label{psi1}
\end{align}
where each component $\Psi^{(N-1)}_{\mu_1\cdots\mu_n}(\alpha^\mu)$ is a homogeneous polynomial
of degree $N-1$.
We use $\{\cdots\}$ to emphasize the symmetrization.
The coordinates $\alpha^\mu$ and the differentials $d\alpha^\mu$ both
belong to the fundamental representation of $SU(d+1)$.
If the components were all independent the $SU(3)$ character associated with
(\ref{psi1}) would be
$\chi_{N-1}\chi_n$.
This is, however, not correct.
For the differential to be well-defined on the projective space $\CC\bm{P}^d$
$\Psi$ must be transformed homogeneously
under the rescaling of the homogeneous coordinates $\alpha^\mu\rightarrow\lambda\alpha^\mu$.
The parameter $\lambda$ may not be a constant,
and this homogeneity condition requires
the $(n-1)$-differential
\begin{align}
\delta\Psi=n\Psi^{(N-1)}_{\mu_1\cdots\mu_n}(\alpha^\mu)\alpha^{\mu_1}\{d\alpha^{\mu_2}\cdots d\alpha^{\mu_n}\}
\end{align}
must vanish.
We introduced $\delta$-operator which acts on
the homogeneous coordinates and the differentials
as
\begin{align}
\delta\alpha^\mu=0,\quad
\delta(d\alpha^\mu)=\alpha^\mu.
\end{align}
The character associated with the constraints $\delta\Psi$ is
$\chi_N\chi_{n-1}$.
Therefore, the character of ${\cal E}_n^{(N-1)}$ is
\begin{align}
\chi({\cal E}_n^{(N-1)})=
\chi_{N-1}\chi_n-\chi_N\chi_{n-1}.
\label{chichi}
\end{align}
For this to be correct, the homogeneity conditions should be all independent.
This is the case for sufficiently large $N$, but fails for small $N$.
For example, in the $d=2$ case the dimension
\begin{align}
\dim{\cal E}_n^{(N-1)}
=\frac{1}{2}N(N+1)(N-n),
\label{dimchichi}
\end{align}
is negative for $N<n$, and
(\ref{chichi}) cannot be correct.
Although we have not proved it regorously
(\ref{chichi}) seems to give the correct character
when $N\geq n$ as far as we have numerically checked.
A similar restriction applies to other examples below, too.
Namely, the formulas given below are correct only when $N$ is not so small.

Next, let us consider symmetric $n$-$n$-differentials with weight $N-2$,
sections of the vector bundle
\begin{align}
{\cal E}^{(N-2)}_{\{n,n\}}
={\cal O}(N-2)\otimes\Sym^2\left(\Sym^n(T^*\CC\bm{P}^2)\right).
\label{symnn}
\end{align}
A section can be expressed as
\begin{align}
\Psi=\Psi^{(N-2)}_{\mu_1\cdots\mu_n\nu_1\cdots\nu_n}
\{\{d\alpha^{\mu_1}\cdots d\alpha^{\mu_n}\}
\{d\alpha^{\nu_1}\cdots d\alpha^{\nu_n}\}\}.
\end{align}
or, schematically,
\begin{align}
\Psi=\Psi^{(N-2)}\{\{n\}\{n\}\}.
\end{align}
The corresponding character is
$\chi_{N-2}(\chi_n^2+\chi_n^{(2)})/2$.
($\chi_n^{(2)}$ is defined as $\chi_n$ with all fugacities
replaced by the squares of them.)
We must impose the homogeneity condition $\delta\Psi=0$.
$f=\delta\Psi$ is an $(n-1)$-$n$-differential
with weight $N-1$.
Namely, it is schematically given by
\begin{align}
f=f^{(N-1)}\{n-1\}\{n\}.
\label{eq150}
\end{align}
The corresponding character is
$\chi_{N-1}\chi_{n-1}\chi_n$.
This time,
the constraint (\ref{eq150})
are not independent even if $N$ is large.
If $f$ is a general $(n-1)$-$n$-differential,
$g\equiv\delta f$ consists of three parts;
\begin{align}
g
=g_1^{(N)}\{n-2\}\{n\}
+g_2^{(N)}\{\{n-1\}\{n-1\}\}
+g_3^{(N)}[\{n-1\}\{n-1\}],
\end{align}
where $[\cdots]$ in the last term represents the anti-symmetric product.
However, if $f$ is given by $f=\delta\Psi$ the third
term does not exist.
Therefore, we need to subtract the corresponding character
$\chi_N(\chi_{n-1}^2-\chi_{n-1}^{(2)})/2$
from the character of the
constraint.
Combining these, we obtain the character
\begin{align}
\chi({\cal E}^{(N-2)}_{\{n,n\}})=
\chi_{N-2}\frac{\chi_n^2+\chi_n^{(2)}}{2}
-\chi_{N-1}\chi_n\chi_{n-1}
+\chi_N\frac{\chi_{n-1}^2-\chi_{n-1}^{(2)}}{2}
\label{two1}
\end{align}
for the vector bundle (\ref{symnn}).
Again, this formula gives correct character for sufficiently large $N$,
and fails for small $N$.

\section{Results of localization}
In this appendix we show the results of the calculation
using the localization formula (\ref{locfm}).
For each gauge group
we show the terms that are needed to determine the order of the error
of the corresponding result on the gravity side.
If the order of the error is higher than $q^7$
we only show terms
up to $q^7$.

\subsection{Superconformal index}
We use the notations
$\chi^J_n=\chi_n(\wt y)$ and 
$\chi_{(r_1,r_2)}=\chi_{(r_1,r_2)}(u,v)$.

\begin{align}
{\cal I}_{U(1)}
&=1
+\chi_{(1,0)}q
-\chi^J_1q^{\frac{3}{2}}
+(\chi_{(2,0)}-\chi_{(0,1)})q^2
+(\chi_{(3,0)}-\chi_{(1,1)}+1-\chi^J_2)q^3
\nonumber\\&
+\chi^J_1\chi_{(0,1)}q^{\frac{7}{2}}
+(\chi_{(4,0)}-\chi_{(2,1)}+\chi_{(1,0)}-\chi^J_2\chi_{(1,0)})q^4
\nonumber\\&
+(\chi^J_2\chi_{(0,1)}+\chi_{(2,0)}-\chi_{(3,1)}+\chi_{(5,0)})q^5
+(-\chi^J_1\chi_{(1,0)}-\chi^J_3\chi_{(1,0)})q^{\frac{11}{2}}
\nonumber\\&
+(\chi^J_2\chi_{(1,1)}+\chi_{(3,0)}-\chi_{(4,1)}+\chi_{(6,0)})q^6
+{\cal O}(q^{\frac{13}{2}})
\nonumber\\
{\cal I}_{U(2)}
&=1
+\chi_{(1,0)}q
-\chi^J_1q^{\frac{3}{2}}
+(-\chi_{(0,1)}+2\chi_{(2,0)})q^2
-\chi^J_1\chi_{(1,0)}q^{\frac{5}{2}}
\nonumber\\&
+(2-\chi^J_2-\chi_{(1,1)}+2\chi_{(3,0)})q^3
+\chi^J_1(\chi_{(0,1)}-\chi_{(2,0)})q^{\frac{7}{2}}
\nonumber\\&
+(-\chi^J_2\chi_{(1,0)}+\chi_{(1,0)}-2\chi_{(2,1)}+3\chi_{(4,0)})q^4
+\chi^J_1(1+2\chi_{(1,1)}-\chi_{(3,0)})q^{\frac{9}{2}}
\nonumber\\&
+(-2\chi^J_2\chi_{(2,0)}-2\chi_{(0,1)}+\chi_{(2,0)}-2\chi_{(3,1)}+3\chi_{(5,0)})q^5
\nonumber\\&
+\chi^J_1(\chi_{(1,0)}+2\chi_{(2,1)}-\chi_{(4,0)})q^{\frac{11}{2}}
\nonumber\\&
+(\chi^J_2(\chi_{(1,1)}-\chi_{(3,0)})
    -\chi_{(1,1)}+2\chi_{(3,0)}-3\chi_{(4,1)}+4\chi_{(6,0)})q^6
\nonumber\\&
+(-\chi^J_3\chi_{(2,1)}+\chi^J_1(-3\chi_{(0,1)}-\chi_{(1,2)}+2\chi_{(3,1)}-\chi_{(5,0)}))q^{\frac{13}{2}}
\nonumber\\&
+(\chi^J_2(3\chi_{(0,2)}+2\chi_{(1,0)}+2\chi_{(2,1)}-2\chi_{(4,0)})
     +4\chi_{(1,0)}+2\chi_{(4,0)}
\nonumber\\&\hspace{3em}
          -3\chi_{(5,1)}+4\chi_{(7,0)})q^7
+{\cal O}(q^{\frac{15}{2}})
\nonumber\\
{\cal I}_{U(3)}
&=1
+\chi_{(1,0)}q
-\chi^J_1q^{\frac{3}{2}}
+(-\chi_{(0,1)}+2\chi_{(2,0)})q^2
-\chi^J_1\chi_{(1,0)}q^{\frac{5}{2}}
\nonumber\\&
+(2-\chi^J_2-\chi_{(1,1)}+3\chi_{(3,0)})q^3
+\chi^J_1(\chi_{(0,1)}-2\chi_{(2,0)})q^{\frac{7}{2}}
\nonumber\\&
+(-\chi^J_2\chi_{(1,0)}+\chi_{(0,2)}+2\chi_{(1,0)}-2\chi_{(2,1)}+4\chi_{(4,0)})q^4
\nonumber\\&
+\chi^J_1(1+\chi_{(1,1)}-2\chi_{(3,0)})q^{\frac{9}{2}}
\nonumber\\&
+(\chi^J_2(\chi_{(0,1)}-2\chi_{(2,0)})
     -2\chi_{(0,1)}+2\chi_{(2,0)}-2\chi_{(3,1)}+5\chi_{(5,0)})q^5
\nonumber\\&
+\chi^J_1(\chi_{(1,0)}+3\chi_{(2,1)}-3\chi_{(4,0)})q^{\frac{11}{2}}
\nonumber\\&
+(-\chi_{(0,3)}-4\chi_{(1,1)}+\chi_{(2,2)}+3\chi_{(3,0)}-3\chi^J_2\chi_{(3,0)}-4\chi_{(4,1)}+7\chi_{(6,0)})q^6
\nonumber\\&
+\chi^J_1(\chi_{(0,1)}+\chi_{(1,2)}+3\chi_{(2,0)}+3\chi_{(3,1)}-3\chi_{(5,0)})q^{\frac{13}{2}}
\nonumber\\&
+(\chi^J_2(-\chi_{(0,2)}-\chi_{(1,0)}+\chi_{(2,1)}-4\chi_{(4,0)})
   -\chi_{(0,2)}+\chi_{(1,0)}-5\chi_{(2,1)}
\nonumber\\&\hspace{3em}
        +3\chi_{(4,0)}-4\chi_{(5,1)}+8\chi_{(7,0)})q^7
+{\cal O}(q^{\frac{15}{2}})
\label{sciun}
\end{align}

\begin{align}
{\cal I}_{SO(2)}&={\cal I}_{U(1)}
\nonumber\\
{\cal I}_{SO(4)}
&=1
+2\chi_{(2,0)}q^2
-2\chi^J_1\chi_{(1,0)}q^{\frac{5}{2}}
+(2-2\chi_{(1,1)})q^3
+2\chi^J_1(\chi_{(0,1)}+\chi_{(2,0)})q^{\frac{7}{2}}
\nonumber\\&
+(-2\chi^J_2\chi_{(1,0)}
 +\chi_{(0,2)}+\chi_{(2,1)}+3\chi_{(4,0)})q^4
-4\chi^J_1(\chi_{(1,1)}+\chi_{(3,0)})q^{\frac{9}{2}}
\nonumber\\&
+(\chi^J_2(3\chi_{(0,1)}+3\chi_{(2,0)})
 +\chi_{(0,1)}-2\chi_{(1,2)}+\chi_{(2,0)}-4\chi_{(3,1)})q^5
+{\cal O}(q^{\frac{11}{2}})
\nonumber\\
{\cal I}_{SO(6)}
&=1
+\chi_{(2,0)}q^2
-\chi^J_1\chi_{(1,0)}q^{\frac{5}{2}}
+(1-\chi_{(1,1)}+\chi_{(3,0)})q^3
+\chi^J_1\chi_{(0,1)}q^{\frac{7}{2}}
\nonumber\\&
+(\chi_{(0,2)}+\chi_{(1,0)}-\chi^J_2\chi_{(1,0)}-\chi_{(2,1)}+2\chi_{(4,0)})q^4
+\chi^J_1(-\chi_{(1,1)}-\chi_{(3,0)})q^{\frac{9}{2}}
\nonumber\\&
+(-\chi_{(0,1)}+2\chi^J_2\chi_{(0,1)}+2\chi_{(2,0)}-\chi_{(3,1)}+\chi_{(5,0)})q^5
\nonumber\\&
+(\chi^J_1(\chi_{(0,2)}+\chi_{(1,0)}+\chi_{(2,1)})-\chi^J_3\chi_{(1,0)})q^{\frac{11}{2}}
\nonumber\\&
+(1-2\chi_{(1,1)}+\chi_{(3,0)}-\chi_{(4,1)}+3\chi_{(6,0)}
  +\chi^J_2(-1-2\chi_{(1,1)}-\chi_{(3,0)}))q^6
\nonumber\\&
+(\chi^J_1(-\chi_{(0,1)}+\chi_{(2,0)}-2\chi_{(5,0)})
 +\chi^J_3(2\chi_{(0,1)}+\chi_{(2,0)}))q^{\frac{13}{2}}
\nonumber\\&
+(-\chi^J_4\chi_{(1,0)}
  +\chi^J_2(2\chi_{(0,2)}+2\chi_{(1,0)}+\chi_{(2,1)})
\nonumber\\&\hspace{3em}
  +\chi_{(1,0)}+\chi_{(4,0)}-2\chi_{(5,1)}+2\chi_{(7,0)}
 )q^7
+{\cal O}(q^{\frac{15}{2}})
\label{sciso2n}
\end{align}

\begin{align}
{\cal I}_{SO(3)}
&=1
+\chi_{(2,0)}q^2
-\chi^J_1\chi_{(1,0)}q^{\frac{5}{2}}
+(1-\chi_{(1,1)})q^3
+\chi^J_1(\chi_{(0,1)}+\chi_{(2,0)})q^{\frac{7}{2}}
\nonumber\\&
+(-\chi^J_2\chi_{(1,0)}+\chi_{(4,0)})q^4
+{\cal O}(q^{\frac{9}{2}})
\nonumber\\
{\cal I}_{SO(5)}
&=1
+\chi_{(2,0)}q^2
-\chi^J_1\chi_{(1,0)}q^{\frac{5}{2}}
+(-\chi_{(1,1)}+1)q^3
+\chi^J_1(\chi_{(2,0)}+\chi_{(0,1)})q^{\frac{7}{2}}
\nonumber\\&
+(2\chi_{(4,0)}+\chi_{(0,2)}-\chi^J_2\chi_{(1,0)})q^4
-\chi^J_1(-2\chi_{(1,1)}-2\chi_{(3,0)})q^{\frac{9}{2}}
\nonumber\\&
+(2\chi^J_2\chi_{(0,1)}-\chi_{(0,1)}-\chi_{(1,2)}+\chi_{(2,0)}+\chi_{(5,0)})q^5
\nonumber\\&
+(-\chi^J_3\chi_{(1,0)}
      +\chi^J_1(2\chi_{(0,2)}+\chi_{(1,0)}+4\chi_{(2,1)}+2\chi_{(4,0)}))q^{\frac{11}{2}}
\nonumber\\&
+(\chi^J_2(-1-4\chi_{(1,1)}-3\chi_{(3,0)})
    -1+\chi_{(0,3)}-3\chi_{(1,1)}+\chi_{(2,2)}+\chi_{(4,1)}
\nonumber\\&\hspace{3em}
         +2\chi_{(6,0)})q^6
+{\cal O}(q^{\frac{13}{2}})
\nonumber\\
{\cal I}_{SO(7)}
&=1
+\chi_{(2,0)}q^2
-\chi^J_1\chi_{(1,0)}q^{\frac{5}{2}}
+(1-\chi_{(1,1)})q^3
+\chi^J_1(\chi_{(0,1)}+\chi_{(2,0)})q^{\frac{7}{2}}
\nonumber\\&
+(-\chi^J_2\chi_{(1,0)}+\chi_{(0,2)}+2\chi_{(4,0)})q^4
+\chi^J_1(-2\chi_{(1,1)}-2\chi_{(3,0)})q^{\frac{9}{2}}
\nonumber\\&
+(\chi^J_2(2\chi_{(0,1)}+\chi_{(2,0)})
 -\chi_{(1,2)}+2\chi_{(2,0)}-2\chi_{(3,1)})q^5
\nonumber\\&
+(-\chi^J_3\chi_{(1,0)}
 +\chi^J_1(2\chi_{(0,2)}+\chi_{(1,0)}+4\chi_{2,1)}+2\chi_{(4,0)}))q^{\frac{11}{2}}
\nonumber\\&
+(\chi^J_2(-1-4\chi_{(1,1)}-3\chi_{(3,0)})
 +\chi_{(0,3)}-3\chi_{(1,1)}+2\chi_{(2,2)}+\chi_{(4,1)}+3\chi_{(6,0)})q^6
\nonumber\\&
+(\chi^J_3(2\chi_{(0,1)}+2\chi_{(2,0)})
 +\chi^J_1(-\chi_{(0,1)}-5\chi_{(1,2)}-\chi_{(2,0)}-6\chi_{(3,1)}-4\chi_{(5,0)}))q^{\frac{13}{2}}
\nonumber\\&
+(\chi^J_4(-\chi_{(1,0)})
 +\chi^J_2(5\chi_{(0,2)}+4\chi_{(1,0)}+8\chi_{(2,1)}+4\chi_{(4,0)})
 +3\chi_{(0,2)}+4\chi_{(1,0)}
\nonumber\\&\hspace{2em}
        -2\chi_{(1,3)}+4\chi_{(2,1)}-3\chi_{(3,2)}+3\chi_{(4,0)}-4\chi_{(5,1)})q^7
+{\cal O}(q^{\frac{15}{2}})
\label{sciso2np1}
\end{align}

\begin{align}
{\cal I}_{SU(3)}
&=1
+\chi_{(2,0)}q^2
-\chi^J_1\chi_{(1,0)}q^{\frac{5}{2}}
+(\chi_{(3,0)}-\chi_{(1,1)}+1)q^3
+\chi^J_1\chi_{(0,1)}q^{\frac{7}{2}}
\nonumber\\&
+(\chi_{(4,0)}-\chi_{(2,1)}+\chi_{(0,2)}+\chi_{(1,0)}-\chi^J_2\chi_{(1,0)})q^4
-\chi^J_1\chi_{(1,1)}q^{\frac{9}{2}}
\nonumber\\&
+(\chi^J_2(2\chi_{(0,1)}+\chi_{(2,0)})-\chi_{(1,2)}+2\chi_{(2,0)}-2\chi_{(3,1)})q^5
+{\cal O}(q^{\frac{11}{2}}),
\end{align}

\begin{align}
{\cal I}_{G_2}
&=1
+\chi_{(2,0)}q^2
-\chi^J_1\chi_{(1,0)}q^{\frac{5}{2}}
+(-\chi_{(1,1)}+1)q^3
+\chi^J_1(\chi_{(2,0)}+\chi_{(0,1)})q^{\frac{7}{2}}
\nonumber\\&
+(\chi_{(4,0)}+\chi_{(0,2)}-\chi^J_2\chi_{(1,0)})q^4
+\chi^J_1(-2\chi_{(1,1)}-\chi_{(3,0)})q^{\frac{9}{2}}
\nonumber\\&
+(\chi^J_2(2\chi_{(0,1)}+\chi_{(2,0)})-\chi_{(1,2)}+\chi_{(2,0)}-\chi_{(3,1)})q^5
+{\cal O}(q^{\frac{11}{2}})
\end{align}

\subsection{Schur index}\label{subsec.SI}
We use the notation $\chi_n=\chi_n(u)$.
Note that $q$ in the following Schur indices
is denoted by $q^{\frac{1}{2}}$
in the standard reference \cite{Gadde:2011uv}.

\begin{align}
\wh{\cal I}_{U(1)}
&=1
+\chi _1q
+(\chi _2-2)q^2
+(\chi _3-\chi _1)q^3
+(\chi _4-\chi _2)q^4
\nonumber\\&
+(\chi _5-\chi _3-\chi _1)q^5
+(\chi _6-\chi _4+\chi _0)q^6
+{\cal O}(q^7)
\nonumber\\
\wh{\cal I}_{U(2)}
&=1
+\chi_1q
+(-2+2\chi_2)q^2
+(-2\chi_1+2\chi_3)q^3
+(-3\chi_2+3\chi_4)q^4
\nonumber\\&
+(\chi_1-3\chi_3+3\chi_5)q^5
+(-4\chi_4+4\chi_6)q^6
+(-4\chi_5+4\chi_7)q^7
+{\cal O}(q^8)
\nonumber\\
\wh{\cal I}_{U(3)}
&=1
+\chi_1q
+(-2+2\chi_2)q^2
+(-2\chi_1+3\chi_3)q^3
+(1-4\chi_2+4\chi_4)q^4
\nonumber\\&
+(-4\chi_3+5\chi_5)q^5
+(2\chi_2-7\chi_4+7\chi_6)q^6
+(\chi_1-7\chi_5+8\chi_7)q^7
+{\cal O}(q^8)
\label{schurun}
\end{align}

\begin{align}
\wh{\cal I}_{SO(2)}
&=\wh{\cal I}_{U(1)}.
\nonumber\\
\wh{\cal I}_{SO(4)}
&=1
+2\chi _2q^2
-4\chi _1q^3
+3(\chi _4+\chi _2+1)q^4
-8(\chi _3+\chi _1)q^5
+{\cal O}(q^6)
\nonumber\\
\wh{\cal I}_{SO(6)}
&=1
+\chi _2q^2
+(\chi _3-2\chi _1)q^3
+(2\chi _4-\chi _2+2)q^4
\nn \\&
+(\chi _5-2\chi _3-2\chi _1)q^5+(3\chi _6-\chi _4+\chi _2+3)q^6
\nn \\&
+(2\chi _7-4\chi _5-\chi _3-3\chi _1)q^7
+(4\chi _8-2\chi _6+4\chi _4+\chi _2+6)q^8
\nn \\&
+{\cal O}(q^9)
\label{schurso2n}
\end{align}

\begin{align}
\wh{\cal I}_{SO(3)}
&=1
+\chi _2q^2
-2\chi _1q^3
+(1+\chi_2+\chi _4)q^4
+{\cal O}(q^5),\\
\wh{\cal I}_{SO(5)}&=1+\chi _2q^2-2\chi _1q^3+(2\chi _4+\chi _2+2)q^4 -4(\chi _3+\chi _1)q^5\nn \\
&\quad+(2\chi _6+3\chi _4+6\chi _2+5)q^6
+{\cal O}(q^7),\\
\wh{\cal I}_{SO(7)}&=1+\chi _2q^2-2\chi _1q^3+(2\chi _4+\chi _2+2)q^4-4(\chi _3+\chi _1)q^5\nn \\
&\quad +(3\chi _6+3\chi _4+7\chi _2+5)q^6-4(2\chi _5+3\chi _3+3\chi _1)q^7\nn \\
+{\cal O}(q^8),
\label{schurso2np1}
\end{align}

\begin{align}
\wh{\cal I}_{SU(3)}
&=1
+\chi _2q^2
+(\chi _3-2\chi _1)q^3
+(\chi _4-\chi _2+2)q^4
+(\chi _5-3\chi _1)q^5
\nn \\&
+{\cal O}(q^6)
\nonumber\\
\wh{\cal I}_{G_2}
&=1
+\chi _2q^2
-2\chi _1q^3
+(\chi _4+\chi _2+2)q^4
+{\cal O}(q^5).
\end{align}
\section{Results of D3-brane analysis}\label{resultsads.sec}
For a theory which has (manifest and hidden)
the ${\cal N}=4$ supersymmetry we show the terms
up to the order of $q$ at which
we find disagreement with the localization result
provided that
the order of the error term is lower than or equal to $q^7$.
If the error term is higher order than $q^7$
we show only terms up to $q^7$.

For theories with the ${\cal N}=3$ supersymmetry
we show the terms up to the order $q^{2N-\frac{1}{2}}$ for the superconformal index
and $q^{2N-1}$ for the Schur index
which are expected to be correct.

\subsection{Superconformal index}
We use the notations
$\chi^J_n=\chi_n(\wt y)$,
$\ol\chi_n=\ol\chi_n(\frac{v}{u},\frac{1}{v})$,
and
$\chi_{(r_1,r_2)}=\chi_{(r_1,r_2)}(u,v)$.

We first list the results for $S(k,N,1)$ with $k=1,2,3,4,6$
in which only the bulk KK mode contribution is included:
${\cal I}_{S(k,\infty,p)}^{\rm AdS}=\Pexp({\cal P}_k\sI^{\rm KK})$.
\begin{align}
{\cal I}_{S(1,N,1)}^{\rm AdS}
&=1
+\chi_{(1,0)}q
-\chi^J_1q^{\frac{3}{2}}
+(-\chi_{(0,1)}+2\chi_{(2,0)})q^2
-\chi^J_1\chi_{(1,0)}q^{\frac{5}{2}}
\nonumber\\&
+(2-\chi^J_2-\chi_{(1,1)}+3\chi_{(3,0)})q^3
+\chi^J_1(\chi_{(0,1)}-2\chi_{(2,0)})q^{\frac{7}{2}}
\nonumber\\&
+(-\chi^J_2\chi_{(1,0)}+\chi_{(0,2)}+2\chi_{(1,0)}-2\chi_{(2,1)}+5\chi_{(4,0)})q^4
\nonumber\\&
+\chi^J_1(1+\chi_{(1,1)}-3\chi_{(3,0)})q^{\frac{9}{2}}
\nonumber\\&
+(\chi^J_2(\chi_{(0,1)}-2\chi_{(2,0)})-2\chi_{(0,1)}+\chi_{(1,2)}+3\chi_{(2,0)}-2\chi_{(3,1)}+7\chi_{(5,0)})q^5
\nonumber\\&
+\chi^J_1(-\chi_{(0,2)}+\chi_{(1,0)}+2\chi_{(2,1)}-5\chi_{(4,0)})q^{\frac{11}{2}}
\nonumber\\&
+(\chi^J_2(\chi_{(1,1)}-3\chi_{(3,0)})
     +1-\chi_{(0,3)}-3\chi_{(1,1)}+3\chi_{(2,2)}+5\chi_{(3,0)}
\nonumber\\&\hspace{3em}
        -4\chi_{(4,1)}+11\chi_{(6,0)})q^6
+{\cal O}(q^{\frac{13}{2}}).
\label{eq244}
\end{align}

\begin{align}
{\cal I}^{\rm AdS}_{S(2,N,1)}
&=1+\chi_{(2,0)}q^2
-\chi^J_1\chi_{(1,0)}q^\frac{5}{2}
+(1-\chi_{(1,1)})q^3
+\chi_1^J(\chi_{(2,0)}+\chi_{(0,1)})q^{\frac{7}{2}}
\nonumber\\&
+(2\chi_{(4,0)}-\chi_2^J\chi_{(1,0)}+\chi_{(0,2)})q^4
-2\chi^J_1(\chi_{(3,0)}+\chi_{(1,1)})q^{\frac{9}{2}}
\nonumber\\&
+(\chi^J_2(2\chi_{(0,1)}+\chi_{(2,0)})-\chi_{(1,2)}+2\chi_{(2,0)}-2\chi_{(3,1)})q^5
\nonumber\\&
+(-\chi^J_3\chi_{(1,0)}+\chi^J_1(2\chi_{(0,2)}+\chi_{(1,0)}+4\chi_{(2,1)}+2\chi_{(4,0)})q^{\frac{11}{2}}
\nonumber\\&
+(\chi^J_2(-1-4\chi_{(1,1)}-3\chi_{(3,0)})
  +\chi_{(0,3)}-3\chi_{(1,1)}+2\chi_{(2,2)}
\nonumber\\&\hspace{3em}
     +\chi_{(4,1)}+3\chi_{(6,0)})q^6
\nonumber\\&
+(\chi^J_3(2\chi_{(0,1)}+2\chi_{(2,0)})
\nonumber\\&\hspace{3em}
 +\chi^J_1(-\chi_{(0,1)}-5\chi_{(1,2)}-\chi_{(2,0)}-6\chi_{(3,1)}-4\chi_{(5,0)}))q^{\frac{13}{2}}
\nonumber\\&
+(\chi^J_4(-\chi_{(1,0)})
 +\chi^J_2(5\chi_{(0,2)}+4\chi_{(1,0)}+8\chi_{(2,1)}+4\chi_{(4,0)})
 +3\chi_{(0,2)}
\nonumber\\&\hspace{3em}
        +4\chi_{(1,0)}-2\chi_{(1,3)}+4\chi_{(2,1)}-3\chi_{(3,2)}+3\chi_{(4,0)}-4\chi_{(5,1)})q^7
\nonumber\\&
+{\cal O}(q^{\frac{15}{2}})
\label{eq245}
\end{align}

\begin{align}
{\cal I}^{\rm AdS}_{S(3,N,1)}
&=1
+u\ol\chi_1q^2
-u\chi^J_1q^{\frac{5}{2}}
+(-1+u^3-u\ol\chi_2+\ol\chi_3)q^3
+\chi^J_1(2u\ol\chi_1-\ol\chi_2)q^{\frac{7}{2}}
\nonumber\\&
+(-u\chi^J_2-u+2\ol\chi_1-u^3\ol\chi_1-\frac{1}{u}\ol\chi_2+2u^2\ol\chi_2)q^4
\nonumber\\&
+\chi^J_1(-2+u^3+\frac{1}{u}\ol\chi_1-2u^2\ol\chi_1-u\ol\chi_2+\ol\chi_3)q^{\frac{9}{2}}
\nonumber\\&
+(\chi^J_2(2u\ol\chi_1-\ol\chi_2)
 -\frac{1}{u}+3u^2-2u\ol\chi_1+2u^4\ol\chi_1-2u^2\ol\chi_3+2u\ol\chi_4)q^5
\nonumber\\&
+(-u\chi^J_3
  +\chi^J_1(3u-2u^4+\ol\chi_1-u^3\ol\chi_1-\frac{1}{u}\ol\chi_2+5u^2\ol\chi_2-3u\ol\chi_3))q^{\frac{11}{2}}
\nonumber\\&
+(\chi^J_2(-2+u^3+\frac{1}{u}\ol\chi_1-4u^2\ol\chi_1+\ol\chi_3)
  +1-3u^3+2u^6-\frac{1}{u}\ol\chi_1-4u^2\ol\chi_1
\nonumber\\&\hspace{3em}
       +\frac{1}{u^2}\ol\chi_2+7u\ol\chi_2-3u^4\ol\chi_2-4\ol\chi_3
       +4u^3\ol\chi_3-2u\ol\chi_5+2\ol\chi_6)q^6
\nonumber\\&
+{\cal O}(q^{\frac{13}{2}})
\end{align}

\begin{align}
{\cal I}^{\rm AdS}_{S(4,N,1)}
&=1
+u\ol\chi_1q^2
-u\chi^J_1q^{\frac{5}{2}}
+(-1-u\ol\chi_2)q^3
+2u\chi^J_1\ol\chi_1q^{\frac{7}{2}}
\nonumber\\&
+(-u+u^4-u\chi^J_2+\ol\chi_1+2u^2\ol\chi_2+\ol\chi_4)q^4
\nonumber\\&
+\chi^J_1(-2-2u^2\ol\chi_1-u\ol\chi_2-\ol\chi_3)q^{\frac{9}{2}}
\nonumber\\&
+(2u\chi^J_2\ol\chi_1
   +2u^2-2u\ol\chi_1-u^4\ol\chi_1+\ol\chi_2-\frac{1}{u}\ol\chi_3-2u^2\ol\chi_3)q^5
\nonumber\\&
+(-u\chi^J_3+\chi^J_1(3u+u^4+\ol\chi_1+\frac{1}{u}\ol\chi_2+5u^2\ol\chi_2+\ol\chi_4))q^{\frac{11}{2}}
\nonumber\\&
+(\chi^J_2(-2-4u^2\ol\chi_1-u\ol\chi_2-\ol\chi_3)
  +u^3-\frac{1}{u}\ol\chi_1-4u^2\ol\chi_1+2u^5\ol\chi_1
\nonumber\\&\hspace{3em}
               +4u\ol\chi_2+3u^2\ol\chi_3+2u\ol\chi_5)q^6
+{\cal O}(q^{\frac{13}{2}})
\end{align}

\begin{align}
{\cal I}^{\rm AdS}_{S(6,N,1)}
&=1
+u\ol\chi_1q^2
-u\chi^J_1q^{\frac{5}{2}}
+(-1-u\ol\chi_2)q^3
+2u\chi^J_1\ol\chi_1q^{\frac{7}{2}}
\nonumber\\&
+(-u-\chi^J_2+\ol\chi_1+2u^2\ol\chi_2)q^4
+\chi^J_1(-2-2u^2\ol\chi_1-u\ol\chi_2)q^{\frac{9}{2}}
\nonumber\\&
+(2u^2-2u\ol\chi_1+2u\chi^J_2\ol\chi_1-2u^2\ol\chi_3)q^5
\nonumber\\&
+(-u\chi^J_3+\chi^J_1(3u+\ol\chi_1+5u^2\ol\chi_2))q^{\frac{11}{2}}
\nonumber\\&
+(\chi^J_2(-2-4u^2\ol\chi_1-u\ol\chi_2)
   +u^6-4u^2\ol\chi_1+4u\ol\chi_2+3u^2\ol\chi_3+\ol\chi_6)q^6
\nonumber\\&
+{\cal O}(q^{\frac{13}{2}})
\end{align}

The results for $S(k,N,0)$ which include the wrapped D3-brane
contributions are shown below.
\begin{align}
{\cal I}_{S(1,1,0)}^{\rm AdS}
&=1
+\chi_{(1,0)}q
-\chi^J_1q^{\frac{3}{2}}
+(-\chi_{(0,1)}+\chi_{(2,0)})q^2
+(1-\chi^J_2-\chi_{(1,1)}+\chi_{(3,0)})q^3
\nonumber\\&
+\chi^J_1\chi_{(0,1)}q^{\frac{7}{2}}
+(\chi_{(1,0)}-\chi^J_2\chi_{(1,0)}-\chi_{(2,1)}+\chi_{(4,0)})q^4
\nonumber\\&
+(\chi^J_2\chi_{(0,1)}+\chi_{(2,0)}-\chi_{(3,1)}+\chi_{(5,0)})q^5
+(-\chi^J_1\chi_{(1,0)}-\chi^J_3\chi_{(1,0)})q^{\frac{11}{2}}
\nonumber\\&
+(-2+\chi^J_2\chi_{(1,1)}-2\chi_{(2,2)}+\chi_{(3,0)}-\chi_{(4,1)}-\chi_{(6,0)})q^6
+{\cal O}(q^{\frac{13}{2}}).
\nonumber\\
{\cal I}_{S(1,2,0)}^{\rm AdS}
&=1+\chi_{(1,0)}q^2
-\chi^J_1q^{\frac{3}{2}}
+(-\chi_{(0,1)}+2\chi_{(2,0)})q^2
-\chi^J_1\chi_{(1,0)}q^{\frac{5}{2}}
\nonumber\\&
+(2-\chi^J_2-\chi_{(1,1)}+2\chi_{(3,0)})q^3
+\chi^J_1(\chi_{(0,1)}-\chi_{(2,0)})q^{\frac{7}{2}}
\nonumber\\&
+(-\chi^J_2\chi_{(1,0)}+\chi_{(1,0)}-2\chi_{(2,1)}+3\chi_{(4,0)})q^4
+\chi^J_1(1+2\chi_{(1,1)}-\chi_{(3,0)})q^{\frac{9}{2}}
\nonumber\\&
+(-2\chi^J_2\chi_{(2,0)}-2\chi_{(0,1)}+\chi_{(2,0)}-2\chi_{(3,1)}+3\chi_{(5,0)})q^5
\nonumber\\&
+\chi^J_1(\chi_{(1,0)}+2\chi_{(2,1)}-\chi_{(4,0)})q^{\frac{11}{2}}
\nonumber\\&
+(\chi^J_2(\chi_{(1,1)}-\chi_{(3,0)})
    -\chi_{(1,1)}+2\chi_{(3,0)}-3\chi_{(4,1)}+4\chi_{(6,0)})q^6
\nonumber\\&
+(-\chi^J_3\chi_{(2,1)}+\chi^J_1(-3\chi_{(0,1)}-\chi_{(1,2)}+2\chi_{(3,1)}-\chi_{(5,0)}))q^{\frac{13}{2}}
\nonumber\\&
+(\chi^J_2(3\chi_{(0,2)}+2\chi_{(1,0)}+2\chi_{(2,1)}-2\chi_{(4,0)})
     +4\chi_{(1,0)}+2\chi_{(4,0)}
\nonumber\\&\hspace{3em}
             -3\chi_{(5,1)}+4\chi_{(7,0)})q^7
+{\cal O}(q^{\frac{15}{2}})
\nonumber\\
{\cal I}_{S(1,3,0)}^{\rm AdS}
&=1+\chi_{(1,0)}q^2
-\chi^J_1q^{\frac{3}{2}}
+(-\chi_{(0,1)}+2\chi_{(2,0)})q^2
-\chi^J_1\chi_{(1,0)}q^{\frac{5}{2}}
\nonumber\\&
+(2-\chi^J_2-\chi_{(1,1)}+3\chi_{(3,0)})q^3
+\chi^J_1(\chi_{(0,1)}-2\chi_{(2,0)})q^{\frac{7}{2}}
\nonumber\\&
+(-\chi^J_2\chi_{(1,0)}+\chi_{(0,2)}+2\chi_{(1,0)}-2\chi_{(2,1)}+4\chi_{(4,0)})q^4
\nonumber\\&
+\chi^J_1(1+\chi_{(1,1)}-2\chi_{(3,0)})q^{\frac{9}{2}}
\nonumber\\&
+(\chi^J_2(\chi_{(0,1)}-2\chi_{(2,0)})-2\chi_{(0,1)}+2\chi_{(2,0)}-2\chi_{(3,1)}+5\chi_{(5,0)})q^5
\nonumber\\&
+\chi^J_1(\chi_{(1,0)}+3\chi_{(2,1)}-3\chi_{(4,0)})q^{\frac{11}{2}}
\nonumber\\&
+(-\chi_{(0,3)}-4\chi_{(1,1)}+\chi_{(2,2)}+3\chi_{(3,0)}-3\chi^J_2\chi_{(3,0)}-4\chi_{(4,1)}+7\chi_{(6,0)})q^6
\nonumber\\&
+\chi^J_1(\chi_{(0,1)}+\chi_{(1,2)}+3\chi_{(2,0)}+3\chi_{(3,1)}-3\chi_{(5,0)})q^{\frac{13}{2}}
\nonumber\\&
+(\chi^J_2(-\chi_{(0,2)}-\chi_{(1,0)}+\chi_{(2,1)}-4\chi_{(4,0)})
   -\chi_{(0,2)}+\chi_{(1,0)}-5\chi_{(2,1)}
\nonumber\\&\hspace{3em}
        +3\chi_{(4,0)}-4\chi_{(5,1)}+8\chi_{(7,0)})q^7
+{\cal O}(q^{\frac{15}{2}})
\label{sciads1}
\end{align}

\begin{align}
{\cal I}^{\rm AdS}_{S(2,1,0)}
&=1
+\chi_{(1,0)}q
-\chi^J_1q^{\frac{3}{2}}
+(\chi_{(2,0)}-\chi_{(0,1)})q^2
+(\chi_{(3,0)}-\chi_{(1,1)}-\chi ^J_2+2)q^3
\nonumber \\&
+{\cal O}(q^4).
\nonumber\\
{\cal I}^{\rm AdS}_{S(2,2,0)}
&=1
+2\chi_{(2,0)}q^2
-2\chi^J\chi_{(1,0)}q^{\frac{5}{2}}
-2(\chi_{(1,1)}-1)q^3
+2\chi _1^J(\chi_{(2,0)}+\chi_{(0,1)})q^{\frac{7}{2}}
\nonumber\\&
+(-2\chi^J_2\chi_{(1,0)}+3\chi_{(4,0)}+\chi_{(2,1)}+\chi_{(0,2)})q^4
-4\chi^J_1(\chi_{(3,0)}+\chi_{(1,1)})q^{\frac{9}{2}}
\nonumber\\&
+(\chi_{(0,1)}+3\chi^J_2(\chi_{(0,1)}+\chi_{(2,0)})-2\chi_{(1,2)}+2\chi_{(2,0)}-4\chi_{(3,1)})q^5
\nn \\&
+{\cal O}(q^{\frac{11}{2}})
\nonumber\\
{\cal I}^{\rm AdS}_{S(2,3,0)}
&=1
+\chi_{(2,0)}q^2
-\chi^J_1\chi_{(1,0)}q^{\frac{5}{2}}
+(1-\chi_{(1,1)}+\chi_{(3,0)})q^3
+\chi^J_1\chi_{(0,1)}q^{\frac{7}{2}}
\nonumber\\&
+(\chi_{(0,2)}+\chi_{(1,0)}-\chi^J_2\chi_{(1,0)}-\chi_{(2,1)}+2\chi_{(4,0)})q^4
-\chi^J_1(\chi_{(1,1)}+\chi_{(3,0)})q^{\frac{9}{2}}
\nonumber\\&
+(-\chi_{(1,0)}+2\chi^J_2\chi_{(0,1)}+2\chi_{(2,0)}-\chi_{(3,1)}+\chi_{(5,0)})q^5
\nonumber\\&
+(\chi^J_1(\chi_{(0,2)}+\chi_{(1,0)}+\chi_{(2,1)})-\chi^J_3\chi_{(1,0)})q^{\frac{11}{2}}
\nonumber\\&
+(1-2\chi_{(1,1)}+\chi_{(3,0)}-\chi_{(4,1)}+3\chi_{(6,0)}-\chi^J_2(1+2\chi_{(1,1)}+\chi_{(3,0)}))q^6
\nonumber\\&
+(\chi^J_1(-\chi_{(0,1)}+\chi_{(2,0)}-2\chi_{(5,0)})+\chi^J_3(2\chi_{(0,1)}+\chi_{(2,0)}))q^{\frac{13}{2}}
\nonumber\\&
+(\chi_{(0,2)}+\chi_{(1,0)}+2\chi_{(4,0)}-2\chi_{(5,1)}+2\chi_{(7,0)}
\nonumber\\&\hspace{3em}
+\chi^J_2(2\chi_{(0,2)}+2\chi_{(1,0)}+\chi_{(2,1)})
-\chi^J_4\chi_{(1,0)})q^7
+{\cal O}(q^{\frac{15}{2}})
\label{sciads2}
\end{align}

\begin{align}
{\cal I}^{\rm AdS}_{S(3,1,0)}
&=1
+\underbrace{(u+\ol\chi _1)}_{\chi _{(1,0)}}q
-\chi _1^Jq^{\frac{3}{2}}
+\underbrace{(\ol\chi _2+u^2-u^{-1})}_{\chi _{(2,0)}-\chi _{(0,1)}}q^2
\nonumber\\&
+(1-\chi _2^J-u^{-1}\ol\chi _1+u^{-1}\ol\chi _4)q^3
+{\cal O}(q^{\frac{7}{2}})
\nonumber\\
{\cal I}^{\rm AdS}_{S(3,2,0)}
&=1
+\underbrace{(u^2+u\ol\chi _1+\ol\chi _2)}_{\chi _{(2,0)}}q^2
-\chi _1^J\underbrace{(u+\ol\chi _1)}_{\chi _{(1,0)}}q^{\frac{5}{2}}
+\underbrace{(u^3+\ol\chi _3-u^{-1}\ol\chi _1)}_{\chi _{(3,0)}-\chi _{(1,1)}+1}q^3
\nn \\
&+\chi _1^J\underbrace{(u\ol\chi _1+u^{-1})}_{\chi _{(0,1)}}q^{\frac{7}{2}}
\nonumber\\&
+\underbrace{(u^4+u^2\ol\chi _2-u\chi _2^J+\ol\chi _1-\chi _2^J\ol\chi _1+\ol\chi _4-u^{-1}\ol\chi _2+u^{-2})}_{\chi _{(4,0)}+\chi _{(0,2)}        +\chi _{(1,0)}-\chi _2^J\chi _{(1,0)}-\chi _{(2,1)}}q^4
\nn \\
&-\chi _1^J\underbrace{(1+u^{-1}\ol\chi _1+u^2\ol\chi _1+u\ol\chi _2)}_{\chi _{(1,1)}}q^{\frac{9}{2}}\nn \\
&+(-2u^{-1}+u^2+2u^{-1}\chi _2^J-u^{-2}\ol\chi _1+u^4\ol\chi _1+2u\chi _2^J\ol\chi _1+u^{-3}\ol\chi _2
\nonumber\\&\hspace{3em}
                  +u\ol\chi _4+u^{-1}\ol\chi _6)q^5
+{\cal O}(q^{\frac{11}{2}})
\nonumber\\
{\cal I}^{\rm AdS}_{S(3,3,0)}
&=1
+u\ol\chi_1q^2
-u\chi^J_1q^{\frac{5}{2}}
+(-1+2u^3-u\ol\chi_2+2\ol\chi_3)q^3
\nonumber\\&
+\chi^J_1(2u\ol\chi_1-2\ol\chi_2)q^{\frac{7}{2}}
+(-u-u\chi^J_2+3\ol\chi_1-2u^3\ol\chi_1-2\frac{1}{u}\ol\chi_2+4u^2\ol\chi_2)q^4
\nonumber\\&
+\chi^J_1(-2+2u^3+2\frac{1}{u}\ol\chi_1-4u^2\ol\chi_1-u\ol\chi_2+2\ol\chi_3)q^{\frac{9}{2}}
\nonumber\\&
+(\chi^J_2(2u\ol\chi_1-2\ol\chi_2)
  +6u^2-4u\ol\chi_1+4u^4\ol\chi_1-4u^2\ol\chi_3+4u\ol\chi_4)q^5
\nonumber\\&
+{\cal O}(q^{\frac{11}{2}})
\label{sciads3}
\end{align}

\begin{align}
{\cal I}^{\rm AdS}_{S(4,1,0)}
&=1
+\underbrace{(u+\ol\chi _1)}_{\chi _{(1,0)}}q
-\chi _1^Jq^{\frac{3}{2}}
-u^{-1}q^2
+{\cal O}(q^{\frac{5}{2}})
\nonumber\\
{\cal I}^{\rm AdS}_{S(4,2,0)}
&=1
+\underbrace{(u^2+u\ol\chi _1+\ol\chi _2)}_{\chi _{(2,0)}}q^2
-\chi _1^J\underbrace{(u+\ol\chi _1)}_{\chi _{(1,0)}}q^{\frac{5}{2}}
\nonumber\\&
-\underbrace{(u^{-1}\ol\chi _1+u^2\ol\chi _1+u\ol\chi _2)}_{\chi _{(1,1)}-1}q^3
+\chi _1^J\underbrace{(u^{-1}+u^2+2u\ol\chi _1+\ol\chi _2)}_{\chi _{(0,1)}+\chi _{(2,0)}}q^{\frac{7}{2}}
\nn \\&
+(u^4-u\chi _2^J+\ol\chi _1+2u^3\ol\chi _1-\chi _2^J\ol\chi _1+2u^2\ol\chi _2+2u\ol\chi _3+\ol\chi _4)q^4
+{\cal O}(q^{\frac{9}{2}})
\nonumber\\
{\cal I}_{S(4,3,0)}^{\rm AdS}
&=1
+u\ol\chi_2q^2
-u\chi^Jq^{\frac{5}{2}}
+(-1+u^3-u\ol\chi_2+\ol\chi_3)q^3
+\chi^J_1(2u\ol\chi_1-\ol\chi_2)q^{\frac{7}{2}}
\nonumber\\&
+(-u+u^4-u\chi^J_2+2\ol\chi_1-u^3\ol\chi_1-\frac{1}{u}\ol\chi_2+2u^2\ol\chi_2+\ol\chi_4)q^4
\nonumber\\&
+\chi^J_1(-2+u^3+\frac{1}{u}\ol\chi_1-2u^2\ol\chi_1-u\ol\chi_2)q^{\frac{9}{2}}
\nonumber\\&
+(\chi^J_2(2u\ol\chi_1-\ol\chi_2)
 +3u^2-2u\ol\chi_1+\ol\chi_2+u^3\ol\chi_2-\frac{1}{u}\ol\chi_3-u^2\ol\chi_3+u\ol\chi_4)q^5
\nonumber\\&
+{\cal O}(q^{\frac{11}{2}})
\label{sciads4}
\end{align}

\begin{align}
{\cal I}^{\rm AdS}_{S(6,1,0)}
&=1
+\underbrace{(u+\ol\chi _1)}_{\chi _{(1,0)}}q
-\chi _1^Jq^{\frac{3}{2}}
-u^{-1}q^2
+{\cal O}(q^{\frac{5}{2}})
\nonumber\\
{\cal I}^{\rm AdS}_{S(6,2,0)}
&=1
+\underbrace{(u^2+u\ol\chi _1+\ol\chi _2)}_{\chi _{(2,0)}}q^2
-\chi _1^J\underbrace{(u+\ol\chi _1)}_{\chi _{(1,0)}}q^{\frac{5}{2}}
-\underbrace{(u^{-1}\ol\chi _1+u^2\ol\chi _1+u\ol\chi _2)}_{\chi _{(1,1)}-1}q^3
\nn \\&
+\chi _1^J\underbrace{(u^{-1}+u^2+2u\ol\chi _1+\ol\chi _2)}_{\chi _{(0,1)}+\chi _{(2,0)}}q^{\frac{7}{2}}
\nn \\&
+(-u\chi _2^J+\ol\chi _1+u^3\ol\chi _1-\chi _2^J\ol\chi _1+2u^2\ol\chi _2+u\ol\chi _3)q^4
+{\cal O}(q^{\frac{9}{2}}).
\nonumber\\
{\cal I}_{S(6,3,0)}^{\rm AdS}
&=1
+u\ol\chi_1q^2
-u\chi^J_1q^{\frac{5}{2}}
+(-1+u^3-u\ol\chi_2+\ol\chi_3)q^3
+\chi^J_1(2u\ol\chi_1-\ol\chi_2)q^{\frac{7}{2}}
\nonumber\\&
+(-u-u\chi^J_2+2\ol\chi_1-u^3\ol\chi_1-\frac{1}{u}\ol\chi_2+2u^2\ol\chi_2)q^4
\nonumber\\&
+\chi^J_1(-2+u^3+\frac{1}{u}\ol\chi_1-2u^2\ol\chi_1-u\ol\chi_2+\ol\chi_3)q^{\frac{9}{2}}
\nonumber\\&
+(\chi^J_2(2u\ol\chi_1-\ol\chi_2)
    +3u^2-2u\ol\chi_1+u^4\ol\chi_1-2u^2\ol\chi_3+u\ol\chi_4)q^5
+{\cal O}(q^{\frac{11}{2}})
\label{sciads6}
\end{align}

\subsection{Schur index}
We use the notation $\chi_n=\chi_n(u)$.
Note that $q$ in the following
Schur indices is denoted by $q^{\frac{1}{2}}$
in the standard reference \cite{Gadde:2011uv}.

\begin{align}
\wh{\cal I}_{S(1,\infty,p)}^{\rm AdS}
&=1
+\chi_1q
+(-2+2\chi_2)q^2
+(-2\chi_1+3\chi_3)q^3
+(1-4\chi_2+5\chi_4)q^4
\nonumber\\&
+(\chi_1-5\chi_3+7\chi_5)q^5
+{\cal O}(q^6)
\nonumber\\
\wh{\cal I}_{S(2,\infty,p)}^{\rm AdS}
&=1
+\chi_2q^2
-2\chi_1q^3
+(2+\chi_2+2\chi_4)q^4
+(-4\chi_1-4\chi_3)q^5
\nonumber\\&
+(5+7\chi_2+3\chi_4+3\chi_6)q^6
+(-12\chi_1-12\chi_3-8\chi_5)q^7
+{\cal O}(q^8)
\nonumber\\
\wh{\cal I}_{S(3,\infty,p)}^{\rm AdS}
&=1
+q^2
+(-2\chi_1+\chi_3)q^3
+(5-\chi_2)q^4
+(-6\chi_1+3\chi_3)q^5
+{\cal O}(q^6)
\nonumber\\
\wh{\cal I}_{S(4,\infty,p)}^{\rm AdS}
&=1
+q^2
-2\chi_1q^3
+(4-\chi_2+\chi_4)q^4
+(-2\chi_1-\chi_3)q^5
+{\cal O}(q^6)
\nonumber\\
\wh{\cal I}_{S(6,\infty,p)}^{\rm AdS}
&=1
+q^2
-2\chi_1q^3
+4q^4
-3\chi_1q^5
+{\cal O}(q^6)
\label{eq254}
\end{align}

\begin{align}
\wh{\cal I}_{S(1,1,0)}^{\rm AdS}
&=1
+\chi_1q
+(-2+\chi_2)q^2
+(-\chi_1+\chi_3)q^3
+(-2\chi_2+\chi_4)q^4
\nonumber\\&
+(-\chi_1-\chi_3+\chi_5)q^5
+(2-2\chi_2-\chi_6+\chi_8)q^6
+{\cal O}(q^7),
\nonumber\\
\wh{\cal I}_{S(1,2,0)}^{\rm AdS}
&=1
+\chi_1q
+(-2+2\chi_2)q^2
+(-2\chi_1+2\chi_3)q^3
+(-3\chi_2+3\chi_4)q^4
\nonumber\\&
+(\chi_1-3\chi_3+3\chi_5)t^5
+(-4\chi_4+4\chi_6)q^6
+(-4\chi_5+4\chi_7)q^7
+{\cal O}(q^8),
\nonumber\\
\wh{\cal I}_{S(1,3,0)}^{\rm AdS}
&=1
+\chi_1q
+(-2+2\chi_2)q^2
+(-2\chi_1+3\chi_3)q^3
+(1-4\chi_2+4\chi_4)q^4
\nonumber\\&
+(-4\chi_3+5\chi_5)q^5
+(2\chi_2-7\chi_4+7\chi_6)q^6
+(\chi_1-7\chi_5+8\chi_7)q^7
\nonumber\\&
+{\cal O}(q^8)
\label{schurads1}
\end{align}

\begin{align}
\wh{\cal I}_{S(2,1,0)}^{\rm AdS}
&=1
+\chi_1q
+(-2+\chi_2)q^2
+(-\chi_1+\chi_3)q^3
+(-\chi_2+2\chi_4)q^4
+{\cal O}(q^5)
\nonumber\\
\wh{\cal I}_{S(2,2,0)}^{\rm AdS}
&=1
+2\chi_2q^2
-4\chi_1q^3
+(3+3\chi_2+3\chi_4)q^4
+(-4\chi_1-4\chi_3+2\chi_5)q^5
\nonumber\\&
+{\cal O}(q^6)
\nonumber\\
\wh{\cal I}_{S(2,3,0)}^{\rm AdS}
&=1
+\chi_2q^2
+(-2\chi_1+\chi_3)q^3
+(2-\chi_2+2\chi_4)q^4
\nonumber\\&
+(-2\chi_1-2\chi_3+\chi_5)q^5
+(3+\chi_2-\chi_4+3\chi_6)q^6
\nonumber\\&
+(-3\chi_1-\chi_3-4\chi_5+2\chi_7)q^7
+{\cal O}(q^8)
\label{schurads2}
\end{align}

\begin{align}
\wh{\cal I}_{S(3,1,0)}^{\rm AdS}
&=1
+\chi_1q
+(-2+\chi_2)q^2
+(-\chi_3+\chi_5)q^3
+{\cal O}(q^4)
\nonumber\\
\wh{\cal I}_{S(3,2,0)}^{\rm AdS}
&=1
+\chi_2q^2
+(-2\chi_1+\chi_3)q^3
+(2-\chi_2+\chi_4)q^4
\nonumber\\&
+(-3\chi_1+\chi_3-\chi_5+\chi_7)q^5
+{\cal O}(q^6)
\nonumber\\
\wh{\cal I}_{S(3,3,0)}^{\rm AdS}
&=1
+q^2
+(-3\chi_1+2\chi_3)q^3
+(8-2\chi_2)q^4
+(-11\chi_1+6\chi_3)q^5
\nonumber\\&
+{\cal O}(q^6)
\label{schurads3}
\end{align}

\begin{align}
\wh{\cal I}_{S(4,1,0)}^{\rm AdS}
&=1
+\chi_1q
-q^2
+{\cal O}(q^3)
\nonumber\\
\wh{\cal I}_{S(4,2,0)}^{\rm AdS}
&=1
+\chi_2q^2
-2\chi_1q^3
+(1+2\chi_2+\chi_4)q^4
+{\cal O}(q^5)
\nonumber\\
\wh{\cal I}_{S(4,3,0)}^{\rm AdS}
&=1
+q^2
+(-2\chi_1+\chi_3)q^3
+(5-2\chi_2+\chi_4)q^4
+(-3\chi_1+\chi_3)q^5
+{\cal O}(q^6)
\label{schurads4}
\end{align}

\begin{align}
\wh{\cal I}_{S(6,1,0)}^{\rm AdS}
&=1
+\chi_1q
-q^2
+{\cal O}(q^3)
\nonumber\\
\wh{\cal I}_{S(6,2,0)}^{\rm AdS}
&=1
+\chi_2q^2
-2\chi_1q^3
+(2+2\chi_2)q^4
+{\cal O}(q^5)
\nonumber\\
\wh{\cal I}_{S(6,3,0)}^{\rm AdS}
&=1
+q^2
+(-2\chi_2+\chi_3)q^3
+(5-\chi_2)q^4
+(-5\chi_1+2\chi_3)q^5
+{\cal O}(q^6)
\label{schurads6}
\end{align}

\end{document}